\documentclass[aps,prb,amsmath,amssymb,twocolumn,superscriptaddress,longbibliography,footinbib]{revtex4-2} 

\usepackage{graphicx}% Include figure files
\usepackage{tikz}
\usepackage{mathtools}
\usepackage{dcolumn}% Align table columns on decimal point
\usepackage{bm}% bold math
\usepackage{mathrsfs}
\usepackage[colorlinks=true,linkcolor=blue,citecolor=blue,urlcolor=blue]{hyperref}
\usepackage{siunitx}
\usepackage{xcolor}

\begin{document}

\preprint{APS/123-QED}

\title{Nonuniform grids for Brillouin zone integration and interpolation}
\author{Siyu Chen}
\email{sc2090@cam.ac.uk}
\affiliation{TCM Group, Cavendish Laboratory, University of Cambridge,J. J. Thomson Avenue, Cambridge CB3 0HE, United Kingdom}
\author{Pascal T. Salzbrenner}
\affiliation{Department of Materials Science and Metallurgy, University of Cambridge, 27 Charles Babbage Road, Cambridge CB3 0FS, United Kingdom}
\author{Bartomeu Monserrat}
\email{bm418@cam.ac.uk}
\affiliation{TCM Group, Cavendish Laboratory, University of Cambridge,J. J. Thomson Avenue, Cambridge CB3 0HE, United Kingdom}
\affiliation{Department of Materials Science and Metallurgy, University of Cambridge, 27 Charles Babbage Road, Cambridge CB3 0FS, United Kingdom}

\date{\today}% It is always \today, today,

\begin{abstract}
We present two developments for the numerical integration of a function over the Brillouin zone. First, we introduce a nonuniform grid, which we refer to as the Farey grid, that generalizes regular grids. Second, we introduce symmetry-adapted Voronoi tessellation, a general technique to assign weights to the points in an arbitrary grid. Combining these two developments, we propose a strategy to perform Brillouin zone integration and interpolation that provides a significant computational advantage compared to the usual approach based on regular uniform grids. We demonstrate our methodology in the context of first principles calculations with the study of Kohn anomalies in the phonon dispersions of graphene and MgB$_2$, and in the evaluation of the electron-phonon driven renormalization of the band gaps of diamond and bismuthene. In the phonon calculations, we find speedups by a factor of $3$ to $4$ when using density functional perturbation theory, and by a factor of $6$ to $7$ when using finite differences in conjunction with supercells. As a result, the computational expense between density functional perturbation theory and finite differences becomes comparable. For electron-phonon coupling calculations we find even larger speedups. Finally, we also demonstrate that the Farey grid can be expressed as a combination of the widely used regular grids, which should facilitate the adoption of this methodology. 
\end{abstract}

\maketitle

\section{Introduction}
\label{sec:intro}
Many macroscopic properties of crystalline materials are approximated by averaging the contributions from every unit cell in an infinite crystal. Invoking Bloch's theorem, this real-space average can be equivalently evaluated as an integral of a periodic function of the wavevector $\boldsymbol{k}$ in the Brillouin zone of reciprocal space. In practice, the integral is approximated with a sum over a discrete set of points $\{\boldsymbol{k}_i\}$ and corresponding weights $\{w_i\}$, and several schemes have been proposed to generate \textit{special points} to perform these integrals at small computational cost~\cite{Baldereschi-mean-value-point, Chadi-Cohen-special-points, cunningham1974, Monkhorst-Pack-grids, Fehlner1977, Froyen1989, Moreno1992, Wisesa2016}, with the most popular being that of Monkhorst and Pack~\cite{Monkhorst-Pack-grids}. These grids are a cornerstone of first principles calculations, and they have been used to evaluate Brillouin zone averages of properties related to (quasi-)particles including electrons~\cite{VASP-Original-Paper}, phonons~\cite{alfe2009}, excitons~\cite{marini2009}, and magnons~\cite{Tancogne2020}, as well as their interactions~\cite{sabiryanov1999, ponce2016, antonius2022}. 

One common feature of special points grids is that the $\boldsymbol{k}$-points are uniformly distributed over the Brillouin zone and the corresponding weights $w$ can be easily obtained using the point group symmetries of the system. These regular grids have been tremendously successful in many applications, and a key example is their central role in density functional theory where they are used to evaluate the self-consistent potential. However, some quantities exhibit sharp variations over the Brillouin zone, and the required grid sizes can become computationally prohibitive. Examples include Kohn anomalies in phonon dispersions~\cite{piscanec2004, calandra2010} and electron-phonon coupling matrix elements~\cite{giustino2007, sjakste2015,verdi2015}, which often necessitate sophisticated interpolation methods~\cite{marzari2012, ponce2016, zhou2021, cepellotti2021} to achieve the required grid sizes.  

In this paper, we propose two independent developments that generalize regular $\boldsymbol{k}$-point grids and weights. First, we introduce a nonuniform grid, the Farey grid, which contains regular grids as a subset of its $\boldsymbol{k}$-points. Second, we introduce symmetry-adapted Voronoi tessellation as a general strategy to calculate the weights $w$ of an arbitrary grid in the Brillouin zone and subject to translational and point group symmetries.

Farey grids provide two advantages compared to regular grids. The first advantage is that the grids used in convergence tests, which are discarded when using regular grids, are all incorporated into the Farey grid. This typically implies that convergence is reached using fewer $\boldsymbol{k}$-points, accelerating calculations. The second advantage is restricted to calculations in which the Brillouin zone is sampled exploiting commensurate real space supercells, for example when calculating phonons using finite differences. In this case, converged Farey grids typically include $\boldsymbol{k}$-points that correspond to smaller commensurate real space supercells, again accelerating calculations.

Symmetry-adapted Voronoi tessellation provides a general strategy to assign weights to \textit{any} $\boldsymbol{k}$-point grid. The weights reduce to the standard weights for uniform grids, but are also applicable to nonuniform grids. Although we use symmetry-adapted Voronoi tessellation in conjunction with Farey grids in this paper, other applications could include stochastic grids or physically-inspired tailored irregular grids.

We find that combining Farey grids with symmetry-adapted Voronoi tesselation dramatically reduces the computational cost of first principles calculations, and a computational cost reduction is observed irrespective of whether the Brillouin zone grids are accessed directly or using commensurate real space supercells. For example, using density functional perturbation theory, which directly samples the Brillouin zone grid, the convergence of the Kohn anomalies in the phonon dispersions of graphene and MgB$_2$ is accelerated by factors of three and four, respectively. The same calculations using finite differences with real space supercells are accelerated by factors of six and seven, respectively. The larger acceleration observed in the latter implies that overall computational costs become similar between the two methods, and in the case of graphene, the finite displacement method using supercells is computationally cheaper than density functional perturbation theory. Similar computational cost gains are observed in the evaluation of electron-phonon interactions in diamond and bismuthene.

We also highlight that, while the results reported for graphene, MgB$_2$, and diamond have been reported elsewhere and are here used as a proof-of-principle, bismuthene has not been studied before in this context. One important conclusion revealed by our results is that bismuthene, a known topological insulator at low temperature, will remain a topological insulator up to room temperature. 

The paper is organized as follows. We first introduce the nonuniform Farey grid in Sec.\,\ref{sec:theory} and compare it with the widely used uniform regular grids. In Sec.\,\ref{sec:savt}, we describe the assignment of weights to points on an arbitrary grid by means of symmetry-adapted Voronoi tessellation. Section\,\ref{sec:nnfi} extends the ideas of Farey grids and symmetry-adapted Voronoi tesselation from BZ integration to BZ interpolation. The proposed methods are illustrated in Sec.\,\ref{sec:sctv-action} with two examples: the calculation of Kohn anomalies in graphene and MgB$_2$, and the calculation of electron-phonon driven band gap renormalization in diamond and bismuthene. Finally, a short summary and conclusions are presented in Sec.\,\ref{sec:conclusion}.

\section{Brillouin zone grids}
\label{sec:theory}
Let $f(\boldsymbol{k})$ be a scalar function defined in $\tau$-dimensional reciprocal space and with the same symmetry as the reciprocal lattice of the crystal. A general integral of this function over the Brillouin zone (BZ) reads
\begin{equation}
\label{general integral}
I=\frac{1}{\Omega} \int_{\mathrm{BZ}}  f(\boldsymbol{k})\; \mathrm{d}^{\tau}\boldsymbol{k} ,
\end{equation}
where $\Omega$ is the $\tau$-dimensional volume of the BZ. According to the definition of the Riemann integral, Equation \eqref{general integral} can always be approximated by a weighted sum over a discrete set of points in the BZ:
\begin{equation}
\label{Eq.{I=wf}}
I \approx \frac{1}{W}\sum_{\boldsymbol{k}_i \in \mathrm{BZ}} f\left(\boldsymbol{k}_{i}\right)w_{i},
\end{equation}
where $W=\sum_i w_i$, and $w_i$ is the weight for the $i$-th point. We henceforth refer to the discrete set $\{\boldsymbol{k}_i\}$ as a grid. 

\begin{figure}
\includegraphics[width=1.0\linewidth]{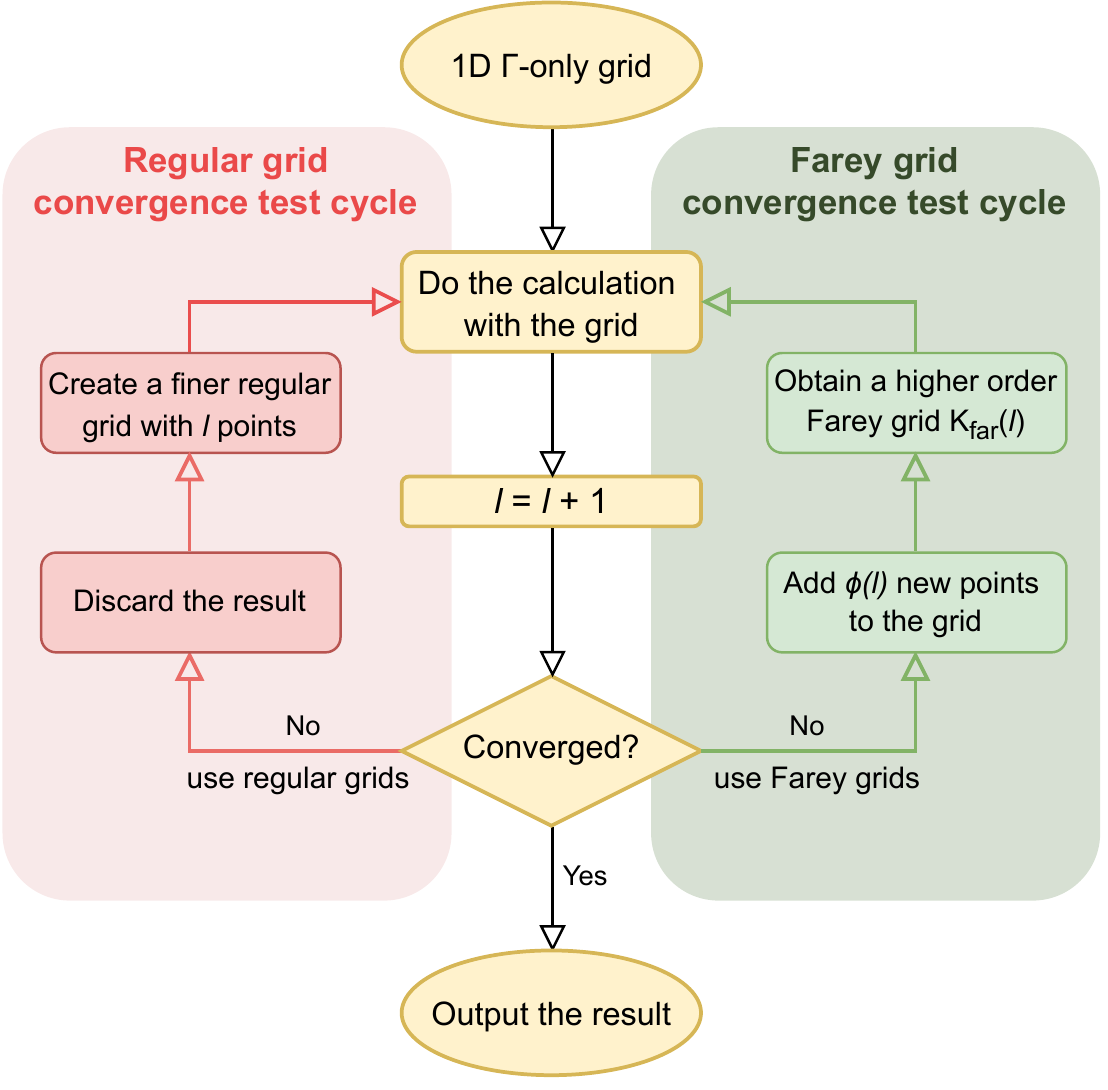}
\caption{(color online). Workflow of the convergence test to obtain the converged sampling density using the regular grid (left, red panel) and the Farey grid (right, green panel).}
\label{fig:workflow}
\end{figure}

\begin{figure*}
\centering
\includegraphics[width=1.0\textwidth]{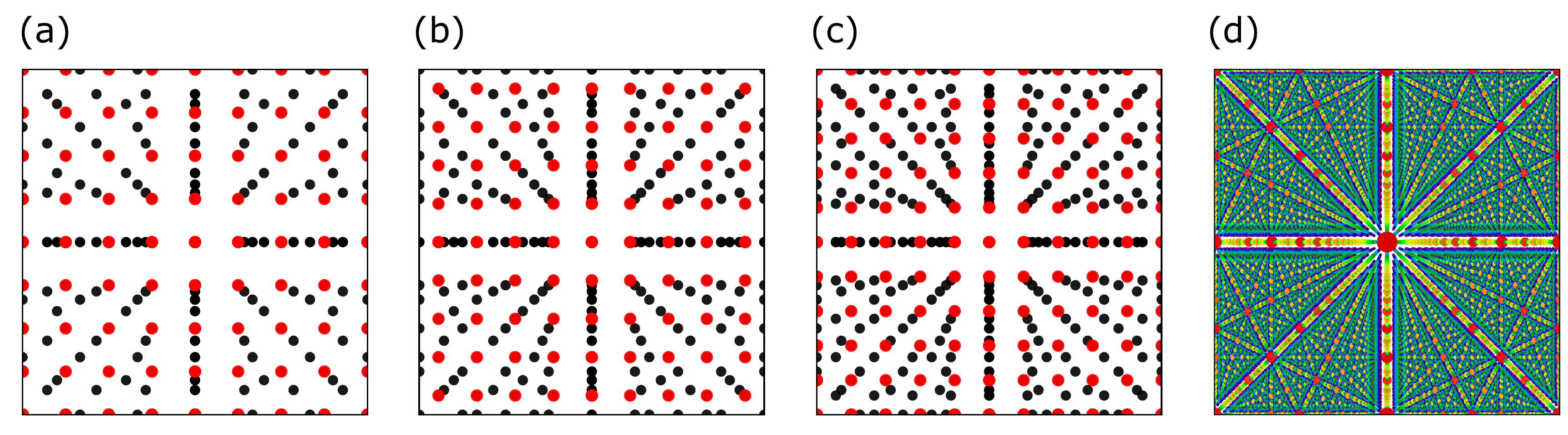}
\caption{(color online). (a-c) 2D Farey grids of order $l=\text{(a) } 8, \text{(b) }9 \text{ and } \text{(c) }10$, where the red dots mark the sampling points of the $l \times l$ regular grid, and the black dots mark the ``easy" sampling points missed by it. (d) The graph of $J(\boldsymbol{k})$ in 2D for $s=1$, where green represents small and red large values. }
\label{fig:farey}
\end{figure*}

\subsection{Regular Brillouin zone grids}
\label{sec:regular}
One possible implementation of Eq.\,\eqref{Eq.{I=wf}} is to find an optimal set $\{(\boldsymbol{k}_i, w_i)\}$ such that as many terms as possible in the Fourier expansion of $f(\boldsymbol{k})$ are integrated exactly, and this is the strategy followed by the various special points methods~\cite{Baldereschi-mean-value-point, Chadi-Cohen-special-points, cunningham1974, Monkhorst-Pack-grids, Fehlner1977, Froyen1989, Moreno1992, Wisesa2016}. It is worth noting that the logic behind these methods is essentially an analogy of Gauss-Legendre quadrature, where Legendre polynomials are replaced by trigonometric polynomials in order to exploit the translational symmetry of $f(\boldsymbol{k})$. 

A regular grid is indeed the optimal grid in this case in terms of maximizing the number of Fourier components that are exactly integrated given a fixed sampling density. This can be rationalized by considering the one-dimensional (1D) case in which the integration evaluated from $n$ equidistant points yields an exact result for any function representable by a Fourier series of $n$ Fourier components or fewer:
\begin{equation}
f(k)=\sum_{i=0}^{n-1} \left[C_{i} \cos \left(\frac{2 \pi i k}{G}\right)+S_{i} \sin \left(\frac{2 \pi i k}{G}\right)\right],
\end{equation}
where $C_{i}$ and $S_{i}$ are the Fourier coefficients for the $i$-th Fourier components and $G$ is the period of $f(k)$. The conclusion holds for higher dimensions and also holds when the Fourier series is symmetrized according to the point group of the crystal. 

In summary, the special points methods (which could be called \textit{Fourier quadrature} methods) invariably lead to regular grids in which the sampling points are uniformly distributed.

\subsection{Computational cost of regular Brillouin zone grids}
In a practical BZ integration using a regular grid, the converged sampling density is not known in advance. The general convergence workflow is depicted in the left panel of Fig.\,\ref{fig:workflow}: sequential calculations are performed with increasing linear grid size $l$ until the integral agrees within some pre-defined tolerance between successive grid sizes. In this process, all intermediate calculations necessary to reach the converged grid size are discarded.

Figures\,\ref{fig:farey}(a-c) depict the regular grids considered in a typical convergence workflow. The red dots in Fig.\,\ref{fig:farey}(a) correspond to a $8\times8$ grid, and the black dots correspond to all the discarded grids of sizes between $1\times1$ and $7\times7$. Figure \ref{fig:farey}(b) depicts the subsequent step in the convergence workflow, with the red dots corresponding to a $9\times9$ grid, and the black dots to all the discarded grids of sizes between $1\times1$ and $8\times8$. Figure \ref{fig:farey}(c) corresponds to the next step, when the sampled grid is of size $10\times10$.

In calculations where the BZ grids are accessed directly, the computational cost of every $\boldsymbol{k}$-point is equivalent. By contrast, in calculations where the BZ grids are accessed using commensurate real space supercells, the smallest supercell commensurate with $\tau$-dimensional wavevector $\boldsymbol{k} = \left({u_1}/{v_1}, \cdots, {u_\tau}/{v_\tau}\right)$, written in reduced fractional coordinates, has a size of $S_{\text{min}} =\mathrm{lcm}\left(v_1, \cdots, v_\tau\right)$~\cite{Tomeu-nondiagonal-supercells}. Based on these insights, we define a reciprocal computational cost function on the $\tau$-dimensional BZ as follows:
\begin{equation}
J(\boldsymbol{k})= \begin{cases}\frac{1}{|S_{\text{min}}|^{s}} & \text { if all entries of } \boldsymbol{k} \text { are rational } \\ \; 0^s & \text { if }\text {any entry of } \boldsymbol{k} \text { is irrational}\end{cases}. \label{eq:cost-function-general}
\end{equation}
In other words, $J(\boldsymbol{k})$ is defined such that the computational cost needed to evaluate the $\boldsymbol{k}$-point is proportional to $J(\boldsymbol{k})^{-1}$. The parameter $s$ characterizes the scaling of the solver used. For example, in calculations where the BZ grids are accessed directly, $s=0$ because each $\boldsymbol{k}$-point has an equivalent cost. As another example, in calculations where the BZ grids are accessed using commensurate real space supercells, and in which density functional theory (DFT) is used, then $s=3$ because the computational cost of DFT scales cubically with system size. The 2D version of $J(\boldsymbol{k})$ with $s=1$ is visualized in Fig.\,\ref{fig:farey}(d), and it exhibits a fractal structure dominated by the lines with rational slopes. Conceptually, Figure \ref{fig:farey}(d) shows how computationally expensive it is to sample a particular point in the BZ.

Figure \ref{fig:farey} conveys two important insights. The first insight is that, irrespective of the method used to sample the BZ, a typical convergence workflow (Fig.\,\ref{fig:workflow}, left panel) requires many calculations that are ultimately discarded, as shown by the black dots in Figs.\,\ref{fig:farey}(a-c). The second insight is that, when the BZ grid is sampled using commensurate real space supercells, the computational cost of accessing a given $\boldsymbol{k}$-point varies significantly across the BZ, as shown in Fig.\,\ref{fig:farey}(d). Both insights suggest that the use of regular grids with standard convergence workflows is computationally wasteful.

\subsection{Farey Brillouin zone grids}
\label{sec:Farey}
Unlike the widely used special points methods and associated regular grids~\cite{Baldereschi-mean-value-point, Chadi-Cohen-special-points, cunningham1974, Monkhorst-Pack-grids, Fehlner1977, Froyen1989, Moreno1992, Wisesa2016}, whose objective is to maximize the number of Fourier components of $f(\boldsymbol{k})$ that are exactly integrated given a fixed sampling density, our starting point is to maximize the sampling density one can achieve with a fixed computational cost. 

\subsubsection{Farey Brillouin zone grids in 1D}
We start with the 1D case to illustrate the key ideas. In 1D, the reciprocal computational cost function defined in Eq.\,\eqref{eq:cost-function-general} becomes:
\begin{equation}
\label{Eq.J(k)}
J(k)= \begin{dcases}\frac{1}{|q|^{s}} & \text { if }|k|=\frac{|p|}{|q|}, \; p, q \in \mathbb{Z}, \; \gcd(p,q)=1 \\ \; 0^s & \text { if } k \text { is irrational }\end{dcases}.
\end{equation}
Note that $-{1}/{2} < k \leq {1}/{2}$ is represented in units of the reciprocal lattice and $|q|$ corresponds to the size of the smallest supercell commensurate with $k$. Also remember that $s$ characterizes the scaling of the solver used, and that overall $J(k)$ is defined such that the computational cost needed to evaluate the $k$-point is proportional to $J(k)^{-1}$.

It is worth noting that when $s=1$, $J(k)$ reduces to the Thomae's function, a famous pathological function with an intriguing fractal structure. Thomae's function is nowhere differentiable, but it has been shown that its local maxima are located strictly at every rational number. Therefore, one can easily deduce that the positions of the top $l\in \mathbb{N}^*$ maxima of $J(k)$ for any $s>0$ are at:
\begin{equation}
\begin{aligned}
&\mathscr{F}\left(l\right) = \\ &\left\{\frac{p}{q} \; \middle| \;   q \in \mathbb{N^*} \leq  l, \; -\frac{q}{2} < p \in \mathbb{Z} \leq \frac{q}{2}, \; \gcd(p,q)=1 \right\}.
\end{aligned}
\end{equation}
It is clear that the most cost-effective scheme to sample the BZ is to follow the points provided by $\mathscr{F}\left(l\right)$. We henceforth refer to this kind of grid generated according to $\mathscr{F}\left(l\right)$ as the 1D \textit{Farey grid} of order $l$, as the grid is isomorphic to the Farey sequence of order $l$ modulo $1$~\cite{niven1991} when its sampling points are arranged in ascending order. 

The 1D Farey grid has two noteworthy properties inherited from the Farey sequence~\cite{ledoan2018}:
\begin{itemize}
\item  For any subinterval $I=\left(\alpha, \beta \right] \subseteq (-1/2,1/2]$, $\left |I \cap \mathscr{F}\left(l\right) \right| /\left |\mathscr{F}\left(l\right) \right| = \left(\beta-\alpha \right) +\mathscr{O}\left(l^{-1}\log l\right)$. 
\item  $\mathscr{F}\left(l\right) \supsetneq \mathscr{F}\left(l-1\right)$ and $\left|\mathscr{F}\left(l\right) - \mathscr{F}\left(l-1\right) \right|=  \phi(l) $, where $\phi(l)$ is Euler's totient function.
\end{itemize}
The first property states that, as $l$ increases, the 1D Farey grid tends to become uniform at the same speed as $\mathscr{O}\left(l^{-1}\log l\right)$ approaches zero. This asymptotically uniform nature guarantees that the 1D Farey grid is an almost unbiased sampling grid. It follows that for a sufficiently large sampling density, the error for the BZ integration calculated with the Farey grid ought to be close to that estimated by Chadi and Cohen for the regular grid~\cite{Chadi-Cohen-special-points}. For the same reason, interpolation between different points on a sufficiently dense Farey grid should also perform as well as that on a regular grid with the same sampling density. Overall, just like for regular grids, by deploying successively finer Farey grids one can achieve increasingly precise BZ integration and interpolation.

The second property states the 1D Farey grid has a compact nested hierarchy. A Farey grid of order $l$ contains all sampling points of any lower-order grid, all the way to the Farey grid of order $1$ (that is, the $\Gamma$-only grid). More specifically, increasing the order $l$ of the Farey grid by $1$, then the new Farey 1D grid or order $l+1$ will have $\phi(l+1)$ more sampling points than the order $l$ grid. The average growth rate of Euler's totient function is known to be approximately constant, namely ${6}/{\pi^2}$. From this linear growth, one can estimate that a 1D Farey grid of order $l$ samples approximately $\sum_{i=1}^l {6i}/{\pi^2} = \mathscr{O}\left(l^{2}\right)$ points in the BZ and the corresponding computational cost is $\sum_{i=1}^l i^s = \mathscr{O}\left(l^{s+1}\right)$. By comparison, a regular grid would require a computational effort of $\mathscr{O}\left(l^{2s}\right)$ to achieve the same sampling density.

As a result of the second property, the Farey grid resolves the computational waste associated with the discarded points in a typical convergence workflow using regular grids (Fig.\,\ref{fig:workflow}, left panel). The true computational cost of a calculation that converges with a regular grid of size $l$ requires the addition of an extra computational cost of $\mathscr{O}\left(l^{2s+1}\right)$. In contrast, due to the nested hierarchical structure of the Farey grid, the corresponding convergence workflow does not add any computational cost (Fig.\,\ref{fig:workflow}, right panel), as the sampled results from the lower-order Farey grids are recycled. Therefore, even considering a practical convergence workflow, the total computational cost of the Farey grid is still $\mathscr{O}\left(l^{s+1}\right)$. 

In this sense, sampling with the Farey grid always outperforms sampling with the regular grid. Consider an example 1D system whose converged results require semi-uniformly sampling $128$ points. In this case, one needs to carry out $128$ loops of the regular grid convergence test in order to determine this result, while only $20$ loops of the Farey grid convergence test will be necessary (as $\sum_{j=1}^{20}\phi(j) = 128$). Importantly, this conclusion holds for any method used to sample the BZ. For methods in which the grid points are accessed directly [$s=0$ in the context of Eq.\,\eqref{Eq.J(k)}], the computational gains arise due to overall sampling fewer $k$-points. For methods in which the grid points are accessed using commensurate real space supercells [$s>0$ in the context of Eq.\,\eqref{Eq.J(k)}], the advantage of the Farey grid becomes more pronounced as convergence is achieved with smaller supercells. The advantage of the Farey grid compounds as the scaling $s$ of the solver becomes larger. 

\subsubsection{Farey Brillouin zone grids in arbitrary dimensions}
We define the $\tau$-dimensional Farey grid of linear order $l$ as follows:
\begin{equation}
\begin{aligned}
&\mathcal{K}^{(\tau)}_\text{far}\left(l\right) = \\ &\bigg\{ \left(\frac{u_1}{v_1}, \cdots, \frac{u_\tau}{v_\tau}\right)\; \bigg| \; \frac{u_i}{v_i} \in \mathscr{F}\left(l\right),\mathrm{lcm}\left(v_1, \cdots, v_\tau\right) \leq l \bigg\}.
\end{aligned}
\end{equation}
In this notation, $\mathcal{K}^{(1)}_\text{far}\left(l\right)=\mathscr{F}\left(l\right)$. Also, in the context of supercell calculations, the condition $\mathrm{lcm}\left(v_1, \cdots, v_n\right) \leq l$ filters out all sampling points whose corresponding smallest commensurate supercells have sizes greater than $l$ such that the sampling density is, as desired, strictly maximized at a fixed computational cost. It is worth noting that this condition preserves the asymptotic uniformity and nested hierarchy which characterized the 1D Farey grids.

Going back to Figs.\,\ref{fig:farey}(a-c), they clearly demonstrate how a Farey grid resolves the wasteful calculations associated with regular grids. For a 2D Farey grid of order $8$, \textit{all} points (red and black dots) in Fig.\,\ref{fig:farey}(a) are included in the evaluation of the BZ integral. By contrast, a regular grid requires calculations on all points, but only the points indicated with red dots are included in the final BZ integral. Figure \ref{fig:farey}(b) depicts the corresponding situation for a Farey grid of linear order $9$, and Fig.\,\ref{fig:farey}(c) for a Farey grid of linear order $10$.

\begin{figure}
\includegraphics[width=1.0\linewidth]{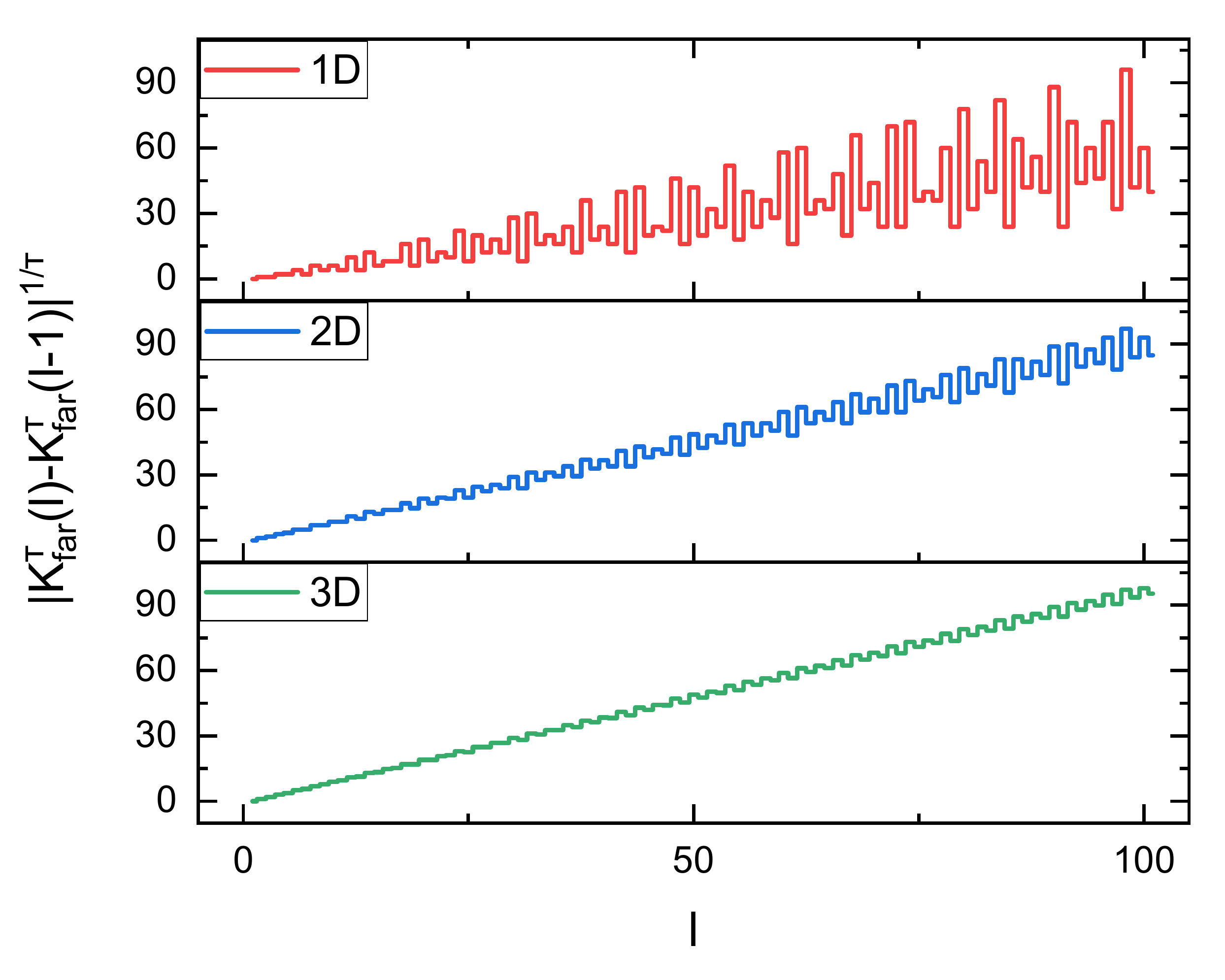}
\caption{(color online). $|\mathcal{K}^{(\tau)}_\text{far}\left(l\right) - \mathcal{K}^{(\tau)}_\text{far}\left(l-1\right)|^{1/\tau}$, which characterizes how many new sampling points are introduced when one increments the $\tau$-dimensional Farey grid, as a function of $l$.}
\label{fig:farey_size}
\end{figure}

The main difference between the higher-dimensional Farey grid compared to 1D is in the growth rate of the grid size. Figure \ref{fig:farey_size} depicts $|\mathcal{K}^{(\tau)}_\text{far}\left(l\right) - \mathcal{K}^{(\tau)}_\text{far}\left(l-1\right)|^{1/\tau}$, which characterizes the number of new sampling points introduced upon increasing the Farey grid linear order, in $\tau=1, 2 \text{ and } 3$ dimensions. It is clear that irrespective of the dimensionality, $|\mathcal{K}^{(\tau)}_\text{far}\left(l\right) - \mathcal{K}^{(\tau)}_\text{far}\left(l-1\right)|^{1/\tau}$, on average, grows linearly. The local fluctuations around the curves are reduced in higher dimensions, with the 3D case approaching a smooth curve with an average slope close to $1$. Overall, a $\tau$-dimensional Farey grid of order $l$ has $\mathscr{O}\left(l^{\tau}\right)$ more sampling points than one of order $l-1$, and it contains $\mathscr{O}\left(l^{\tau+1}\right)$ points in total.

\subsubsection{Relation between Farey and regular grids}
A Farey grid of order $l$ can be treated as a union of regular grids with their linear size increasing step-by-step from $\lfloor l/2 \rfloor +1 $ to $l$:
\begin{equation}
\mathcal{K}^{(\tau)}_\text{far}\left(l\right)=\bigcup_{j=\lfloor l/2 \rfloor +1}^l\mathcal{K}^{(\tau)}_\text{reg}\left(j\right). \label{eq:farey-regular}
\end{equation}
In this expression, 
\begin{equation}
\begin{aligned}
\label{uvw_regular}
\mathcal{K}^{(\tau)}_\text{reg}\left(l\right) = \left\{\left(\frac{u_1}{v_1}, \cdots, \frac{u_3}{v_3}\right)\; \middle| \; \frac{u_i}{v_i} \in \mathscr{A}\left(l\right) \right\}
\end{aligned}
\end{equation}
denotes a $\tau$-dimensional $\Gamma$-centered regular grid generated by a finite set of the arithmetic sequence 
\begin{equation}
\begin{aligned}
\mathscr{A}\left(l\right) = & \left\{\frac{j}{l} \; \middle| \; -\frac{l}{2} < j \in \mathbb{Z}  \leq \frac{l}{2}\right\}
\end{aligned} 
\end{equation}with a common difference of $1/l$. 

It follows from Eq.\,\eqref{eq:farey-regular} that the Farey grid can be constructed by combining the results (insofar as they exist) from consecutive regular grids, and this implies that software implementing the regular grid, including all major DFT codes, can be easily modified to adopt the Farey grid. Indeed, this is the approach taken in the calculations reported in Sec.\,\ref{sec:sctv-action} of this work.

We note that, in practice, it may be more natural to start with a tentative regular grid of linear size $a\in \mathbb{N^*}$ and combine it with increasingly large regular grids, step by step, until a linear size $b\in \mathbb{N^*}$ :
\begin{equation}
\label{HtoA}
\mathcal{K}^{(\tau)}_\text{far}\left(a,b\right)=\bigcup_{j=a}^b\mathcal{K}^{(\tau)}_\text{reg}\left(j\right).
\end{equation}
If $b < 2 (a-1)$, this corresponds to a ``truncated" Farey grid. This kind of Farey grid misses a small number of ``easy" points whose fractional coordinates have a component of $p/q \in \mathscr{H}_\text{miss}\left(a, b\right)$, where
\begin{equation}
\begin{aligned}
&\mathscr{H}_\text{miss}\left(a, b\right) = \\& \left\{\frac{p}{q} \in \mathscr{F}\left(b, 0\right) \; \middle| \; \forall \; a \leq n  \in \mathbb{N^*} \leq b \; \gcd(np,q) \neq q \right\},
\end{aligned}
\end{equation}
but it can still be very useful in practical calculations.

\section{Symmetry-Adapted Voronoi Tessellations}
\label{sec:savt}
\begin{figure*}
\centering
\includegraphics[width=0.99\textwidth]{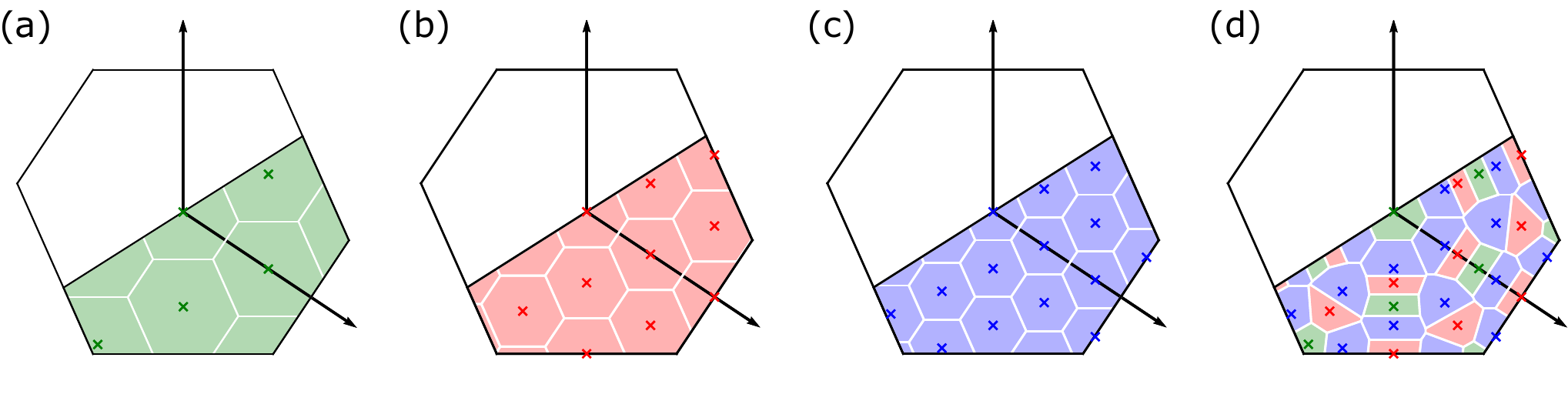}
\caption{(color online). SAVTs of the oblique lattice IBZ for the regular grids (a) $\mathcal{K}^{(2)}_\text{reg}\left(3\right)$, (b) $\mathcal{K}^{(2)}_\text{reg}\left(4\right)$, (c) $\mathcal{K}^{(2)}_\text{reg}\left(5\right)$ and (d) the Farey grid $\mathcal{K}^{(2)}_\text{far}\left(5\right) = \mathcal{K}^{(2)}_\text{reg}\left(3\right) \cup \mathcal{K}^{(2)}_\text{reg}\left(4\right) \cup\mathcal{K}^{(2)}_\text{reg}\left(5\right)$. The grid points in the IBZ of interest (colored) are marked by `x'. The color distinguishes the Voronoi cells generated by the different regular grids.}
\label{fig:SAVT}
\end{figure*}

In this section we consider the assignment of weights to an arbitrary Brillouin zone grid. 

\subsection{Weights for a regular grid}

All the points of a regular grid carry exactly the same weight. This is reasonable, since by definition the grid points are evenly spaced in each independent direction, so there is no reason to think of any sampling point as more important than any other. 

Regular grids allow for an efficient symmetry reduction since a regular grid spanned by the reciprocal basis vectors of the BZ is essentially a scaled version of the reciprocal lattice. This means that all the points of a regular grid can be mapped to the same irreducible BZ (IBZ) by some point group symmetry operation of the crystal. As a result, the BZ integration can be restricted to the IBZ,
\begin{equation}
I \approx  \frac{1}{W}\sum_{\boldsymbol{k}_i \in \mathrm{IBZ}} f\left(\boldsymbol{k}_{i}\right)w_{i},
\end{equation}
by recasting the weights as the symmetry multiplicity of the sampling points in the IBZ. In other words, the weight of a grid point is equal to the cardinality of its orbit in the crystal point group:

\begin{equation}
\label{wight}
w_i  = \left|\mathrm{Orb} \left(\boldsymbol{k}_{i}\right)\right|.
\end{equation}

\subsection{Weights for a general grid}

One may expect that a general grid (for example, the nonuniform Farey grid introduced in Sec.\,\ref{sec:Farey}) can also be reduced by symmetry, as in the case of regular grids. To address this question, we introduce \textit{symmetry-adapted Voronoi tessellation} (SAVT). 

Consider first plain Voronoi tessellation (VT) in Euclidean space. Given a $\tau$-dimensional space $\mathbb{R}^{\tau}$ and a set of distinct points $\mathcal{K} = \left\{\boldsymbol{k}_{i}\right\}$, for each point $\boldsymbol{k}_{i}$, the corresponding Voronoi cell is given by
\begin{equation}
\label{plain_vornoi}
C_{i}(\mathcal{K})=\left\{\boldsymbol{k} \in \mathbb{R}^{\tau}  \mid \;\left\|\boldsymbol{k}-\boldsymbol{k}_{i}\right\|<\left\|\boldsymbol{k}-\boldsymbol{k}_{j}\right\|\; \forall \; i \neq j  \right \},
\end{equation}
where $\|\dots\|$ denotes the Euclidean distance. That is, the Voronoi cell of a given point includes all space closer to it than to any other point in the set. 

To incorporate the crystal point group symmetry $\mathcal{G}$ and the periodicity of the reciprocal lattice $\mathbf{G}$, we change the space metric to extend Eq.\,\eqref{plain_vornoi} to define the SAVT of a bounded domain $\Pi = \text{IBZ}$ as follows:
\begin{equation}
\label{SAVT}
C_{i}(\mathcal{K})=\left\{\boldsymbol{k} \in \Pi \mid \;{}_\mathcal{G}\left\|\boldsymbol{k}-\boldsymbol{k}_{i}\right\|_\mathbf{G} < {}_\mathcal{G} \left\|\boldsymbol{k}-\boldsymbol{k}_{j}\right\|_\mathbf{G} \; \forall \; i \neq j  \right \},
\end{equation}
where
\begin{equation}
\label{SAVT_dis}
{}_\mathcal{G}\|\boldsymbol{k}-\boldsymbol{k}_i\|_{\mathbf{G}} = \min_{ g \in \mathcal{G}, \; \eta \in \mathbf{G} } \left( \|\boldsymbol{k} - g\boldsymbol{k}_i - \eta  \|\right).
\end{equation}
One can easily show that for $i\neq j$, $C_{i} \cap C_{j}=\emptyset$, so the set $\left\{C_{i}\right\}$ defined above is indeed a tessellation of $\Pi$. It is worth further noting that: 

\begin{itemize}
\item There are two types of Voronoi cells in the SAVT, inner cells which are entirely inside the domain and marginal cells which touch the boundary of the domain. The inner cell is the same as the plain Voronoi cell, while the marginal cell can either be composed of just one singly connected polyhedron or several visually disconnected polyhedra.
\item The metric function defined in Eq.\,\eqref{SAVT_dis} guarantees that the SAVT of the IBZ conserves both the point group and translational symmetries of the crystal. Therefore, the corresponding Voronoi diagram allows for a symmetry reduction similar to that used in regular grids. The weight of a sampling point is just equal to the volume of its corresponding symmetry-adapted Voronoi cell:
\begin{equation}
\label{SAVT-weight}
w_i = \left|C_i\right|.
\end{equation}
This is one of the main conclusions of this work.
\end{itemize}

Equation \eqref{SAVT-weight} can be rationalized by considering the 2D example of the SAVTs of the oblique lattice IBZ shown in Fig.\,\ref{fig:SAVT}. From (a) to (d), the SATVs are constructed for the regular grids $\mathcal{K}^{(2)}_\text{reg}\left(3\right)$, $\mathcal{K}^{(2)}_\text{reg}\left(4\right)$ and $\mathcal{K}^{(2)}_\text{reg}\left(5\right)$ as well as the Farey grid $\mathcal{K}^{(2)}_\text{far}\left(5\right)$, which is equivalent to their union. For the regular grids, the SAVT of the IBZ generates a number of Voronoi cells with exactly the shape of the BZ, except for some marginal Voronoi cells that are cut off by the IBZ boundary. It is important to realize that some marginal cells appear not to be associated with any grid point, but they are in fact connected to some other marginal cell that visually possesses a point. These ``point-less" marginal cells, from another perspective, can be treated as parts of a Voronoi cell associated with a symmetry image of the point in an adjacent IBZ which intersects with the IBZ of interest. The volume associated with such a point is the sum of all symmetry-connected volumes within the IBZ of interest. This is the main difference between SAVTs and plain VTs. In the special case when the IBZ boundary edges coincide with mirror planes, SAVTs and VTs become equivalent because the IBZ boundary coincides with the boundaries of all marginal cells. In 2D space, except for the oblique lattice, the IBZ boundaries of the other four Bravais lattices are mirror-like. In 3D space, four out of the fourteen Bravais lattices also possess this kind of mirror-like IBZ boundaries. These are the orthorhombic, tetragonal, rhombohedral, and hexagonal lattices.

It is also worth noting that there is an intimate connection between the geometry of the Voronoi cells and the symmetry of the associated points. In order to show this, we consider the little group $\mathcal{L}_{\boldsymbol{k}}$ of $\mathcal{G}$ for a given $\boldsymbol{k}$-point and the action of the symmetry operator $g \in \mathcal{L}_{\boldsymbol{k}}$ on the corresponding Voronoi cell. For an inner cell whose associated point always has a trivial little group containing the identity operator only, it will, by definition, be simply mapped to itself through $g$. However, for a marginal cell whose associated point is located exactly on the IBZ boundary, thus having a higher symmetry, it will be mapped in its entirety to an adjacent IBZ by $g$ other than the identity operator. In the usual scheme for the regular grid, the weight of the boundary point is generally reduced compared to that of an inner point by $1/|\mathcal{L}_{\boldsymbol{k}}|$. In our geometric picture, because the symmetry operations $g \in \mathcal{L}_{\boldsymbol{k}}$ map points on the IBZ boundary to themselves, images created by the operation of these symmetries can never contribute to the volume of the marginal cell within the IBZ of interest. The volume of marginal cells associated with points on the IBZ boundary is therefore also reduced by $1/|\mathcal{L}_{\boldsymbol{k}}|$ compared to the volume of an inner cell -- that is, its weight, determined from the volume of its marginal cell, is reduced to exactly the same extent as its weight determined by symmetry. This is why the geometric and the symmetry pictures are equivalent when one considers the regular grid. For the more general case, only the geometric approach developed here works.

We emphasize that using the SAVTs to reduce the sampling points to the IBZ and obtain the corresponding weights is not restricted to Farey grids, but is also suitable for any other nonuniform grid. Examples of these might include stochastic grids or manually customized grids which sample some specific regions of the BZ heavily but the remaining regions in a light-weight manner.

Finally, we point out the relation between the SAVT method and the currently used weighting methods for the special points regular grids~\cite{Baldereschi-mean-value-point, Chadi-Cohen-special-points, cunningham1974, Monkhorst-Pack-grids, Fehlner1977, Froyen1989, Moreno1992, Wisesa2016}. As pointed out in Sec.\,\ref{sec:regular}, the logic behind these special points grids is analogous to Gaussian quadrature. The various methods are based on the choice of a small set $\{(\boldsymbol{k}_i, w_i)\}$ such that as many terms as possible in the symmetrized Fourier expansion of $f(\boldsymbol{k})$ are integrated exactly. For this reason, these methods work effectively when the Fourier coefficients converge rapidly to zero, that is, when $f(\boldsymbol{k})$ is sufficiently smooth. When $f(\boldsymbol{k})$ is a rapidly varying function, a better approach would be to expand $f(\boldsymbol{k})$ on a symmetrized basis that is localized in $\boldsymbol{k}$-space
\begin{equation}
\delta_i(\boldsymbol{k}-\boldsymbol{k_i}) = \begin{dcases}\frac{1}{\left|C_i\right|} & \text { if } k \in C_i \\ 0 & \text { if } k \notin C_i\end{dcases}.
\end{equation}
This is precisely what the SAVT method does.

To some extent, the SAVT method can be treated as a generalization of the special point method combined with the tetrahedron method, which was proposed to resolve the BZ integration of metals where $f(\boldsymbol{k})$ usually varies sharply near the Fermi energy~\cite{Kleinman1983, Jepsen1984, Blochl1994}. The method divides the BZ into tetrahedra in such a way that the edges of the tetrahedra have minimal lengths for the given geometry of the Bravais lattice (in order to make the tetrahedra as compact as possible). In fact, this is very close to the idea of Delaunay triangulations. The relation becomes even more clear when one notes that the dual graph of the Delaunay triangulation of a discrete point set is exactly the Voronoi tessellation of the same point set.

\section{Natural neighbor interpolation in the Billouin zone}
\label{sec:nnfi}
Rather than the integral of $f(\boldsymbol{k})$, some calculations require the value of $f$ at some particular point $\boldsymbol{k}^{*}$. The construction of $f(\boldsymbol{k}^{*})$ from the knowledge of $f$ on a $\boldsymbol{k}$-point grid is known as BZ interpolation. In this section we describe a strategy to perform BZ interpolation starting from a nonuniform Farey grid.

We can recast BZ interpolation as an instance of BZ integration using the following integral in terms of the Dirac delta function:
\begin{equation}
\begin{aligned}
\label{bz_integration}
f(\boldsymbol{k}^{*}) &= \int_{\mathrm{BZ}}  f(\boldsymbol{k})\delta(\boldsymbol{k}-\boldsymbol{k}^{*})\; \mathrm{d}^{\tau}\boldsymbol{k}.
\end{aligned}
\end{equation}
BZ interpolation is the name given to evaluating this integral as a weighted sum over a discrete $\boldsymbol{k}$-point grid, which results in the interpolation of $f$ from the grid to $\boldsymbol{k}^{*}$. Due to the locality of the Dirac delta function, the weighted sum for calculating the above integral only needs to involve $\boldsymbol{k}$-points that are located close to the target point $\boldsymbol{k}^{*}$. Based on this, Sibson proposed the so-called natural neighbor interpolation scheme~\cite{sibson1981} to interpolate a function defined in $\mathbb{R}^{\tau}$:
\begin{equation}
\label{sibson inter}
\mathcal{N}[f(\boldsymbol{k}^{*}) | f\left(\mathcal{K}\right)]= \frac{1}{W}\sum_{i=1}^{|\mathcal{K}|} f\left(\boldsymbol{k}_i\right)w_{i}^{*}.
\end{equation}
In this expression, $\mathcal{N}[f(\boldsymbol{k}^{*}) | f\left(\mathcal{K}\right)]$ is the natural neighbor interpolation of $f$ at $\boldsymbol{k}^{*}$ given the values of $f$ at the grid points $\boldsymbol{k}_i \in \mathcal{K}$, and $W=\sum_i w_i^*$  is the normalization factor. According to Sibson's definition, the weight elements $w^*_i$ are equal to the volume of each of the surrounding Voronoi cells which is ``stolen'' when inserting $\boldsymbol{k}^{*}$ into the original Voronoi tessellation, that is
\begin{equation}
\label{sibson weight}
w^*_i=\left| C_{i}(\mathcal{K} \cup \boldsymbol{k}^{*}) \cap C_{i}(\mathcal{K})\right|.
\end{equation}

To perform BZ interpolation starting from arbitrary grids, we extend this interpolation strategy to SAVT. As we are interested in functions $f$ that have the periodicity of the reciprocal lattice, we replace $f\left(\boldsymbol{k}_i\right)$ in Eq.\,\eqref{sibson inter} by a series of Fourier interpolated values at $\boldsymbol{k}^{*}$, that is
\begin{equation}
\label{nnf inter}
\mathcal{N}[f(\boldsymbol{k}^{*}) | f\left(\mathcal{K}\right)]=\frac{1}{W}\sum_{i=1}^{|\mathcal{K}|}   \mathcal{F}[f(\boldsymbol{k}^{*})| f\left(\mathcal{K}_i\right)]w_{i}^{*},
\end{equation}
where $\mathcal{F}[f(\boldsymbol{k}^{*}) | f\left(\mathcal{K}_i\right)]$ denotes the Fourier interpolation of $f$ at $\boldsymbol{k}^{*}$ given the values of $f$ at the grid points $\mathcal{K}_i$. $\mathcal{K}_i \subseteq \mathcal{K}$ is the finest regular grid which contains the $i$-th point $\boldsymbol{k}_i \in \mathcal{K}$, and $w^*_i$ is calculated using SAVT. It is straightforward to see that when $\mathcal{K}$ is itself a regular grid, Equation \eqref{nnf inter} reduces to Fourier interpolation, and when $\mathcal{K}$ is a Farey grid, the decomposition of the grid into a series of regular grids is always feasible according to Eq\,\eqref{HtoA}. As a result, the Fourier interpolation required in Eq.\,\eqref{nnf inter} can always be performed. Henceforth we refer to the interpolation based on Eq.\,\eqref{nnf inter} as \textit{natural neighbor Fourier interpolation}.

In particular, when $\boldsymbol{k}^{*}$ coincides exactly with some grid point $\boldsymbol{k}_i$, the natural neighbor Fourier interpolated value of $f$ is guaranteed to be equal to the actual value, that is $\mathcal{N}[f(\boldsymbol{k}^{*}=\boldsymbol{k}_i) | f\left(\mathcal{K}\right)] = f\left(\boldsymbol{k}_i\right)$. A limitation of this strategy is that $\mathcal{N}[f(\boldsymbol{k}^{*}) | f\left(\mathcal{K}\right)]$ is not differentiable. In other words, the function obtained by natural neighbor Fourier interpolation is smooth everywhere in the BZ except at the locations of the grid points, where it is only piece-wise continuous. To obtain a smooth interpolation of $f$ everywhere, we propose to first map the results evaluated on the nonuniform grid $\mathcal{K}$ to a target regular grid $ \mathcal{K}_{\substack{\scalebox{0.6}{reg} \\ \scalebox{0.6}{tar}}}$ using natural neighbor Fourier interpolation, and then to obtain an interpolation that is smooth everywhere in the BZ through plain Fourier interpolation, as summarized below:
\begin{widetext}
\begin{equation}
 \mathcal{K}_{\substack{\scalebox{0.6}{reg} \\ \scalebox{0.6}{tar}}}	\xrightarrow{{k}^\text{tar}_{i} \in  \mathcal{K}_{\substack{\scalebox{0.6}{reg} \\ \scalebox{0.6}{tar}}}} f( \mathcal{K}_{\substack{\scalebox{0.6}{reg} \\ \scalebox{0.6}{tar}}}) = \{\mathcal{N}[f(\boldsymbol{k}^\text{tar}_i) | f\left(\mathcal{K}\right)]\}  \xrightarrow{\boldsymbol{k}^{*} \in \text{BZ}} f\left(\boldsymbol{k}^{*}\right) = \mathcal{F}[f(\boldsymbol{k}^{*} |f( \mathcal{K}_{\substack{\scalebox{0.6}{reg} \\ \scalebox{0.6}{tar}}})].
\end{equation}
\end{widetext}

The size of the target regular grid is, in principle, a free parameter. However, we suggest the following two options for practical calculations when the nonuniform grid is the Farey grid $\mathcal{K} = \mathcal{K}_\text{far}$: (i)  $| \mathcal{K}_{\substack{\scalebox{0.6}{reg} \\ \scalebox{0.6}{tar}}}| \approx |\mathcal{K}_\text{far}|$, that is, the regular grid is chosen such that the sampling densities of the two grids are almost equal; (ii) $\mathcal{K}_\text{far} \subseteq  \mathcal{K}_{\substack{\scalebox{0.6}{reg} \\ \scalebox{0.6}{tar}}} $, that is, choosing the smallest $ \mathcal{K}_{\substack{\scalebox{0.6}{reg} \\ \scalebox{0.6}{tar}}}$ which contains all grid points of $\mathcal{K}_\text{far} $. As discussed in Sec.\,\ref{sec:Farey}, the Farey grid is an asymptotically uniform grid, so option (i) is usually a sufficiently good choice to access the interpolated function $f$. Indeed, we use option (i) for the calculations reported in Sec.\,\ref{sec:sctv-action} below, in which we match the sampling density along high symmetry lines of interest in the BZ. Option (ii) guarantees that the interpolated curve strictly passes through all the data points on the starting grid, but its evaluation can become computationally costly when the size of the Farey grid is large.

\section{Farey grids and symmetry-adapted Voronoi tessellation in action}
\label{sec:sctv-action}
We now describe how to use Farey grids and SAVT to perform first principles calculations in practice. We focus on two examples that have traditionally proven challenging for calculations based on regular grids: (i) the calculation of phonon dispersions with Kohn anomalies in metallic and semimetallic systems, and (ii) the calculation of electron-phonon coupling (EPC).

Unless otherwise indicated, first-principles calculations are carried out at the DFT level~\cite{DFT-Hohenberg-Kohn, DFT-Kohn-Sham} with the Vienna \textit{ab initio} Simulation Package ({\sc vasp})~\cite{VASP-Original-Paper}. The interaction between ions and valence electrons is modeled with pseudopotentials based on the projector-augmented wave~\cite{VASP-PAW-One, VASP-PAW-Two} method. Some of the calculations, clearly identified below, are performed using the {\sc castep} plane-wave DFT code \cite{CASTEP} using norm-conserving pseudopotentials~\cite{nc_pseudo}.
 
The phonon and EPC calculations are evaluated within the harmonic approximation, and most are preformed using the finite differences (FD) method~\cite{Parlinski1997} in conjunction with nondiagonal supercells~\cite{Tomeu-nondiagonal-supercells}. Some phonon calculations are also performed using linear response in the context of density functional perturbation theory (DFPT)~\cite{baroni2001}.

To follow standard notation, in this section we use $\boldsymbol{q}$ instead of $\boldsymbol{k}$ to label points in the BZ related to phonons.

\subsection{Kohn anomalies in the phonon dispersion}
Phonons are a critical concept for understanding lattice vibration in solids. Accurate phonon dispersions and densities of states are necessary to evaluate properties of solids including thermal expansion, heat capacity, vibrational entropy, and free energy, and also form the starting point for calculations of quantities such as thermal conductivity and electronic conductivity. The fundamental physical quantity for accessing phonons in solids is the so-called dynamical matrix:
\begin{equation}
D_{\alpha \alpha^{\prime}}(\boldsymbol{q}) =  \frac{1}{N \sqrt{m_{\alpha} m_{\alpha^{\prime}}}} \sum_{p \, p^{\prime}} C_{\alpha \alpha^{\prime}}\left(\boldsymbol{\mathbf{R}}_{p}, \boldsymbol{\mathbf{R}}_{p^{\prime}}\right) \mathrm{e}^{i \boldsymbol{q} \cdot\left(\boldsymbol{\mathbf{R}}_{p}-\boldsymbol{\mathbf{R}}_{p^{\prime}}\right)}, \label{eq:dyn-mat}
\end{equation}
whose eigenvalues and eigenvectors correspond to the squared phonon frequencies $\omega^2_{\nu \boldsymbol{q}}$ and phonon polarization vectors $\boldsymbol{\xi}_{\nu \boldsymbol{q}}$, respectively. In Eq.\,\eqref{eq:dyn-mat}, $N$ is the number of primitive cells contained in a supercell subject to Born–von Karman periodic boundary conditions, $m_{\alpha}$ denotes the mass of the $\alpha$-th atom in the primitive cell, and $C_{\alpha \alpha^{\prime}}\left(\boldsymbol{\mathbf{R}}_{p}, \boldsymbol{\mathbf{R}}_{p^{\prime}}\right) $ is the force constant matrix, which is defined as the derivative of the force on atom $\alpha$ in the primitive cell indicated by the Bravais lattice vector $\boldsymbol{\mathbf{R}}_{p}$ with respect to displacement of atom $\alpha^{\prime}$ in the primitive cell indicated by $\boldsymbol{\mathbf{R}}_{p^{\prime}}$.  
 
First principles DFT calculations of phonons can be performed using two different methodologies. The first method is based on linear response theory in the context of DFPT, which is a method that provides direct access to the dynamical matrix on a grid of $\boldsymbol{q}$-points, often called the coarse $\boldsymbol{q}$-point grid. The second method is based on finite differences and it constructs the dynamical matrix on a grid of $\boldsymbol{q}$-points using commensurate real space supercells. In both cases, the dynamical matrix $D_{\alpha \alpha^{\prime}}(\boldsymbol{q})$ is exact at the $\boldsymbol{q}$-points on the grid, and it can be approximated at other $\boldsymbol{q}$-points in the BZ using Fourier interpolation. The size of the $\boldsymbol{q}$-point grid becomes a convergence parameter as the dynamical matrix may deviate significantly from its exact value when the $\boldsymbol{q}$-point grid is not dense enough. 

The changes driven by the displacement of an atom from its equilibrium position are usually screened by electrons, which implies that the matrix of force constants decays with a rapid power law as a function of distance. As a result, relatively small regular $\boldsymbol{q}$-point grids have been very successful at converging phonon calculations. An exception to this rule-of-thumb is provided by Kohn anomalies, which arise at nesting vectors of the Fermi surface of metallic and semimetallic systems, and whose convergence typically requires large regular $\boldsymbol{q}$-point grids that capture the relevant nesting vectors. In the following, we evaluate the phonon dispersions of graphene and MgB$_2$ which host Kohn anomalies. 

\subsubsection{Computational details}
For graphene, the exchange-correlation functional is treated in the local density approximation (LDA)~\cite{Perdew-Zunger-LDA}. A vacuum layer with a thickness of $15$ \AA\@ is used in the calculation to avoid periodic image interactions along the direction perpendicular to the plane. Precise forces are obtained when the energy is converged to within the strict criterion of $10^{-4}$\,meV per cell. This is achieved with a plane-wave energy cutoff of $500$\,eV and by sampling the electronic BZ using a $24 \times 24$  $\Gamma$-centered $\boldsymbol{k}$-point grid.

For MgB$_2$, the exchange-correlation functional is treated in the generalized gradient approximation (GGA) parametrized by Perdew-Burke-Ernzerhoff (PBE)~\cite{PBE-exchange-correlation}. The same energy convergence criterion as that in graphene is achieved with a plane-wave energy cutoff of $500$\,eV and by sampling the electronic BZ using an $18 \times 18 \times 12$ $\Gamma$-centered $\boldsymbol{k}$-point grid.

\subsubsection{Phonon dispersion and Kohn anomaly of graphene}
\begin{figure}
\includegraphics[width=0.98\linewidth]{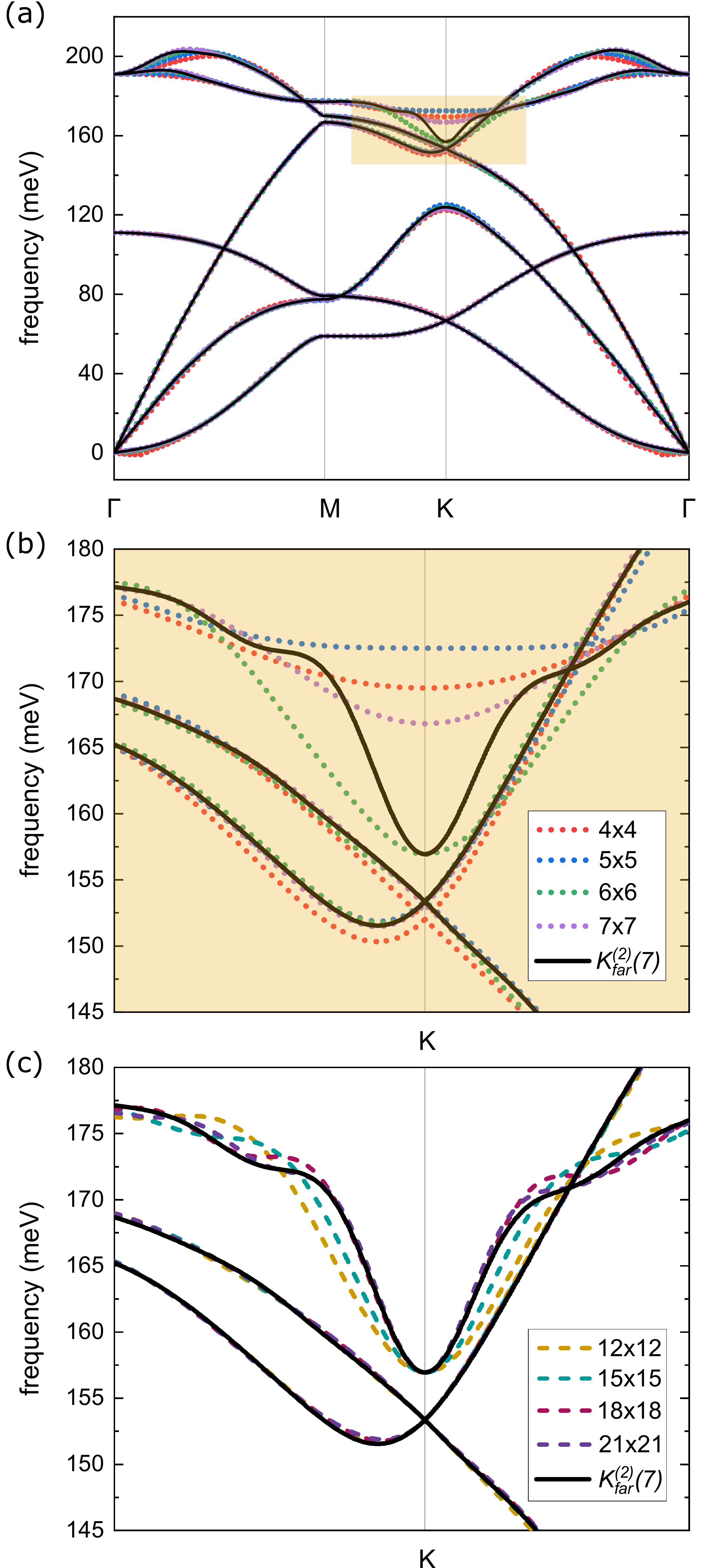}
\caption{(color online). (a) Phonon dispersion of graphene along the high-symmetry path $\Gamma \rightarrow \mathrm{M} \rightarrow \mathrm{K} \rightarrow \Gamma$, calculated with regular grids of different sizes (colored dotted line: $l \times l$ where $l$ ranges from $4$ to $7$) and the Farey grid of order $7$ (black solid line: $\mathcal{K}^{(2)}_{\text{far}}\left(7\right)$). The shaded region highlights the Kohn anomaly located near the K point, whose details are magnified and shown in Figure (b). The phonon dispersion calculated with the Farey grid is also compared to that calculated with even finer regular grids (colored dashed line: $3l \times 3l$ where $l$ ranges from $4$ to $7$) and shown in Figure (c).}
\label{fig:graphene-ph}
\end{figure}

Figure \ref{fig:graphene-ph}(a) shows the phonon dispersion of graphene along the boundary of its IBZ, calculated using regular grids of different sizes as well as the Farey grid combining them. Both acoustic and optical branches rapidly converge with respect to the size of the $\boldsymbol{q}$-point grid in most regions of the high-symmetry path, except for the window in the vicinity of the K point highlighted in yellow shade and magnified in Fig.\,\ref{fig:graphene-ph}(b). The dispersion of the highest frequency optical branch near the K point is very flat when the dynamical matrix is sampled on the $4 \times 4 $, $5 \times 5 $ and $7 \times 7 $ regular $\boldsymbol{q}$-point grids. In contrast, it shows a strong $\boldsymbol{q}$-dependence when the dynamical matrix is instead sampled on the $6\times 6 $ regular $\boldsymbol{q}$-point grid.

The fractional coordinates of the K point are $(1/3, 1/3)$, which implies that of all the regular grids considered, only the $6\times 6 $ grid contains the point $\boldsymbol{q}=\mathrm{K}$. This implies that the dynamical matrix at the K point is only exact for this grid, and is obtained through Fourier interpolation for the other considered grid sizes. In general, the dynamical matrix at the K point will be exact for any regular $\boldsymbol{q}$-point grid of size $3l \times 3l $ for any positive integer $l$, and interpolated for any other grid size. Since Fourier interpolation does not tend to produce peaks or pits that are not already represented by the explicit input data points, it is expected that the regular grids in Fig.\,\ref{fig:graphene-ph}(a) that do not include the K point miss the abrupt change of phonon frequency around that point.

This frequency drop is due to the abrupt change in the screening of lattice vibrations by conduction electrons in metals and semi-metallic systems, which is known as the Kohn anomaly~\cite{kohn1959}. For graphene, this Kohn anomaly has been experimentally observed~\cite{maultzsch2004, mohr2007} and theoretically characterized~\cite{piscanec2004}. However, the vast majority of phonon calculations in the literature report a flat dispersion near the K point~\cite{portal1999, dubay2003, yanagisawa2005, mounet2005, zhang2011, marquina2013, slotman2014, gu2015, diery2018}. Among the above works, Ref.\,\cite{mounet2005} uses the largest regular $\boldsymbol{q}$-point grid of size $16\times 16 $, but the Fourier interpolated frequency at the K point is still significantly overestimated compared to the exact value. In our simulations, we used regular grids up to a size of $20\times 20 $, but the frequency at the K point was still substantially overestimated. These results demonstrate that it is computationally challenging to converge the phonon frequency of the upper branch at the K point with respect to the regular grid size.

However, as we have already noted earlier, it is not actually computationally expensive to capture the exact phonon frequencies at the K point -- a $3 \times 3$ $\boldsymbol{q}$-grid can capture them exactly. As a result, in this example the Farey grid has a clear advantage compared to regular grids: while regular grids alternately sample the K point or not as the grid size grows, Farey grids always guarantee the K point is sampled exactly when the grid order exceeds $l=3$. 

In addition to the phonon frequency itself, the slope of the phonon dispersion curve also has an important physical meaning: it is proportional to the phonon momentum derivative of the dielectric function of the solid. Furthermore, Piscanec and co-workers showed that, in graphene, the slope of the phonon dispersion at the K point, where the Kohn anomaly is located, is proportional to the square of the EPC strength~\cite{piscanec2004}. The dynamical matrix of graphene is nonanalytic at the K point, which leads to a nonzero slope. This nonanalytic behavior cannot be captured with standard phonon calculations and Fourier interpolation, as the dynamical matrix is assumed to be analytically dependent on $\boldsymbol{q}$. In fact, if the dynamical matrix is analytical, symmetry enforces the slope at the K point to always be zero in graphene. This means we cannot directly access the slope at the K point, and instead we consider the slope at points very close to the K point. To be precise, we use $\boldsymbol{q} = \mathrm{K} + \delta \boldsymbol{q}$, where $|\delta \boldsymbol{q}| = 1/10$ of the $\mathrm{M}-\mathrm{K}$ distance. As shown in Fig.\,\ref{fig:graphene-ph}(c), We find that at least an $18\times 18 $ $\boldsymbol{q}$-point grid is necessary to converge this slope with a regular grid. In contrast, the Farey grid of order 7, $\mathcal{K}^{(2)}_\text{far}\left(7\right)$, which has the same sampling density along the high-symmetry path $\Gamma \rightarrow \mathrm{M} \rightarrow \mathrm{K} \rightarrow \Gamma$, leads to the same level of convergence at a much smaller computational cost.

\subsubsection[Phonon dispersion and Kohn anomalies of MgB2]{Phonon dispersion and Kohn anomalies of MgB$_2$}

We carry out a similar phonon calculation for the 3D system MgB$_2$ to further verify the superior performance of the Farey grid. The crystal structure of MgB$_2$ consists of hexagonal graphene-like planes of boron atoms intercalated vertically with magnesium atoms located in the centers of the hexagons. This simple binary compound has an exceptionally high superconducting transition temperature $T_{\text{c}} = 39 $\,K~\cite{nagamatsu2001}. Another important feature is that the phonons of the second highest frequency on the high-symmetry segment from $\Gamma$ to A couple strongly to the electrons~\cite{kong2001}, which leads to a Kohn anomaly as discussed below. These phonons correspond to a transverse mode with $E_{2g}$ symmetry propagating along the vertical direction with the atomic displacement parallel to the horizontal plane.

\begin{figure}
\includegraphics[width=0.98\linewidth]{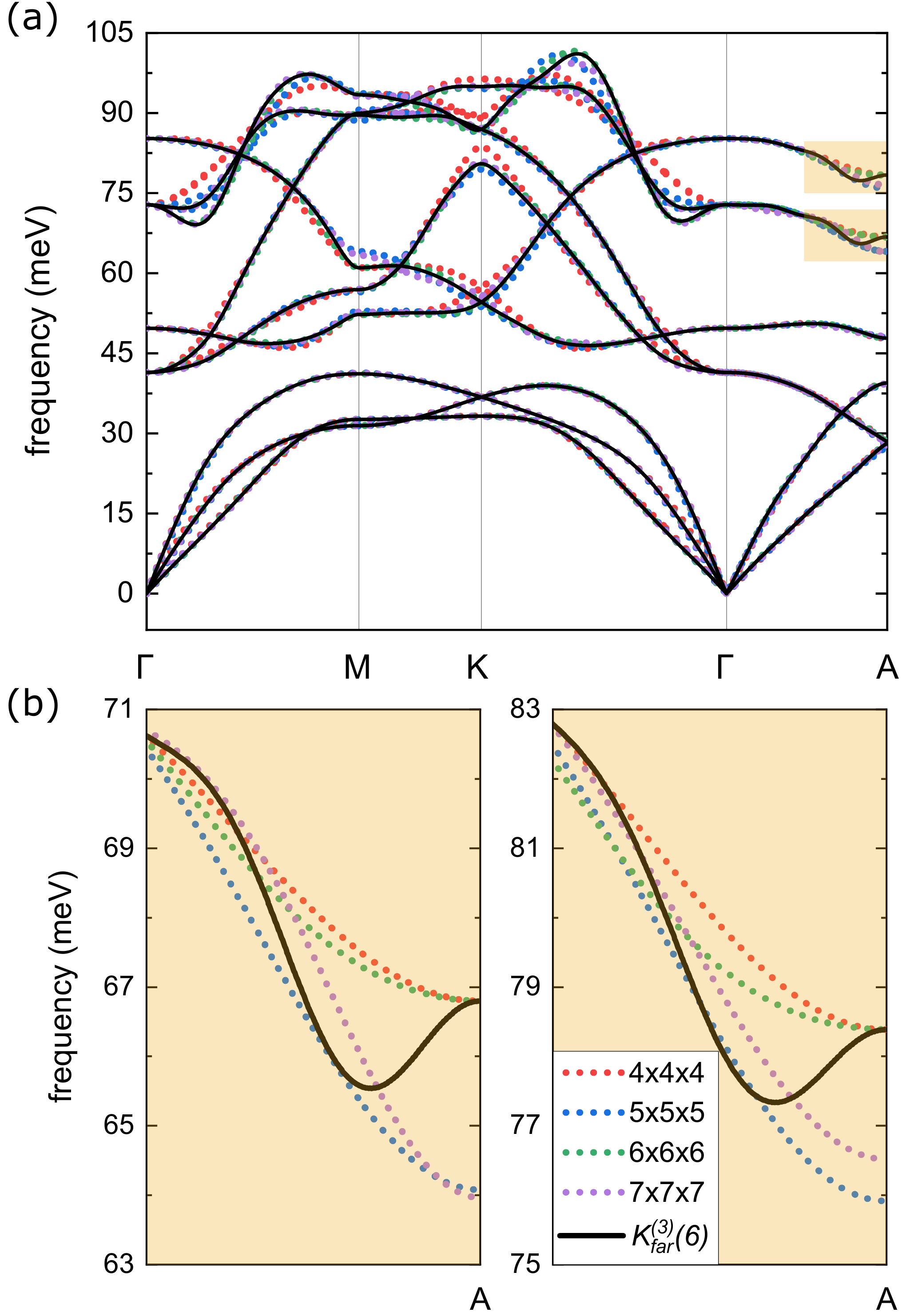}
\caption{(color online). (a) Phonon dispersion of MgB$_2$ along the high-symmetry path $\Gamma \rightarrow \mathrm{M} \rightarrow \mathrm{K} \rightarrow \Gamma \rightarrow \mathrm{A}$, calculated with regular grids of different sizes (colored dotted line: $l \times l \times l$ where $l$ ranges from $4$ to $7$) and the Farey grid of order $6$ (black solid line: $\mathcal{K}^{(3)}_{\text{far}}\left(6\right)$). The shaded region highlights the Kohn anomaly located near the A point, whose details are magnified and shown in Figure (b). The left panel indicates the lower and the right the higher mode.}
\label{fig:mgb2-ph}
\end{figure}

Figure \ref{fig:mgb2-ph} shows the phonon dispersion of MgB$_2$ calculated with the Farey grid $\mathcal{K}^{(3)}_\text{far}\left(6\right)$, as well as with the regular grids whose sizes range from $4 \times 4 \times 4$ to $7 \times 7 \times 7$ for comparison. We observe Kohn anomalies in the $E_{2g}$ optical branch near the $\Gamma$ point (along the $\Gamma-\mathrm{K}$ and $\Gamma-\mathrm{M}$ directions) both in the phonon dispersions calculated with the Farey grid and the regular grids whose sizes exceed $6 \times 6 \times 6$, which is in agreement with previously reported results~\cite{bohnen2001, matteo2007, alarco2014, mackinnon2017}. In addition, we also find Kohn anomalies on the first- and second-highest frequency phonon branches along the $\Gamma-\mathrm{A}$ direction, and at $|\boldsymbol{q}| \approx 9/11$ of the $\Gamma-\mathrm{A}$ distance (highlighted in yellow shade and magnified in Fig.\,\ref{fig:mgb2-ph}(b)).

These Kohn anomalies are relatively weak compared to those around the $\Gamma$ point, but their presence has also been observed in inelastic x-ray scattering experiments of the phonon dispersion~\cite{shukla2003}. From a computational point of view, early first-principles calculations usually adopted regular $\boldsymbol{q}$-point grids with a maximum size of $6 \times 6 \times 6$, and as a result overlooked these Kohn anomalies~\cite{yildirim2001, bohnen2001, shukla2003, baron2004,  matteo2007, alarco2014, mackinnon2017}. The Kohn anomalies along the $\Gamma-\mathrm{A}$ direction have only recently been correctly captured, requiring a regular grid of size $12 \times 12 \times 12$ or larger \cite{pevsic2019, novko2020}. We highlight that our Farey grid $\mathcal{K}^{(3)}_\text{far}\left(6\right)$ has the same sampling density on the $\Gamma-\mathrm{A}$ high-symmetry segment and is able to capture the Kohn anomalies at a lower computational cost. Finally, we note that Calandra and co-workers proposed a Wannier interpolation scheme of the dynamical matrix which also successfully describes the Kohn anomalies along the $\Gamma-\mathrm{A}$ direction~ \cite{calandra2010}. They also show that the presence of Kohn anomalies on the $E_{2g}$ branch in MgB$_2$ can affect the position of the main peak in the Eliashberg function, which is a crucial quantity for calculating the superconducting transition temperature.

%~\cite{marquina2013} 7x7
%~\cite{portal1999}~\cite{zhang2011}~\cite{gu2015}~\cite{yanagisawa2005}
%~\cite{mounet2005} 16x16
%~\cite{slotman2014}~\cite{diery2018}~\cite{dubay2003} 8x8

%~\cite{novko2020}~\cite{pevsic2019} 12x12x12
%~\cite{kong2001} 6x6x6 
~% \cite{calandra2010} wannier 
%~\cite{shukla2003} 6x6x4
%~\cite{baron2004} 18x18x6
%~\cite{yildirim2001} first DFT 2x2x2
%~\cite{bohnen2001} 6x6x6

\subsubsection{Comparison between DFPT and FD with Farey grids}
\begin{table}
\caption{\label{tab:DFPTvsFD} The computational expense of FD and DFPT calculations using regular and Farey grids. Timings are given as fractions of the time taken for the regular grid DFPT calculations. $\mathcal{K}_\text{reg}^{(\tau)}$ and $\mathcal{K}_\text{far}^{(\tau)}$ are the regular and Farey $\boldsymbol{q}$-grids necessary to obtain convergence of the Kohn anomaly.}
\begin{ruledtabular}
\begin{tabular}{ccccccc}
 & & & \multicolumn{2}{c}{FD} & \multicolumn{2}{c}{DFPT} \\
 & $\mathcal{K}^{(\tau)}_\text{reg}$& $\mathcal{K}^{(\tau)}_\text{far}$ & t$_\text{reg}$ & t$_\text{far}$ &  t$_\text{reg}$ & t$_\text{far}$ \\[2pt]
\hline \\
Graphene & $\mathcal{K}^{(2)}_\text{reg}(18)$ & $\mathcal{K}^{(2)}_\text{far}(7)$ & $1.456$ & $0.254$ & $1$ & $0.294$\\[5pt]
MgB$_2$& $\mathcal{K}^{(3)}_\text{reg}(12)$ & $\mathcal{K}^{(3)}_\text{far}(6)$ & $3.894$ & $0.560$ & $1$ & $0.233 $ \\[5pt]
\end{tabular}
\end{ruledtabular}
\end{table}

The calculation of phonons using DFPT directly samples the $\boldsymbol{q}$-point grid to construct the dynamical matrix, which can be accomplished with calculations on a primitive cell. By contrast, the calculation of phonons using FD samples the $\boldsymbol{q}$-point grid by constructing commensurate real space supercells. As a result, DFPT phonon calculations are computationally cheaper than FD phonon calculations when using regular grids. This implies that if both methods are available, DFPT is typically the method of choice. The traditional advantage of FD phonon calculations is that the formalism is simpler, which means that it is available in some contexts in which DFPT phonon calculations are not, a recent example being the calculations of phonons using DFT combined with dynamical mean-field theory to study strongly correlated materials~\cite{kocer2020}. As discussed in Sec.\,\ref{sec:theory}, Farey grids provide a computational speedup for both DFPT and FD, but the latter benefit from a larger speedup. In this section, we show how the use of Farey grids leads to comparable computational expense between DFPT and FD for the calculation of phonons, with examples in which FD is computationally advantageous compared to DFPT.

In contrast to the rest of this work, we use the {\sc castep} plane-wave DFT code \cite{CASTEP} for the comparisons carried out in this section. The reason for this is that its implementation of DFPT is more efficient than that of the {\sc vasp} package, enabling a fairer comparison. We use norm-conserving pseudopotentials, and target a converged electronic self-consistent field energy per atom below $1$\,meV. 

For graphene, this is achieved with an energy cutoff of $1250$\,eV, a Monkhorst-Pack grid with adjacent points separated by $2\pi \times 0.03$ \AA$^{-1}$, and a vacuum spacing of $9$ \AA$ $ between graphene layers. For MgB$_{2}$, the corresponding quantities are a $900$\,eV cutoff and a $\boldsymbol{k}$-point spacing of $2\pi \times 0.03$ \AA$^{-1}$.

Table \ref{tab:DFPTvsFD} compares the computational time required for different methods to converge the phonon calculations of graphene and MgB$_2$ to be able to capture the various Kohn anomalies, with the reference cost set to $1$ for DFPT calculations using a regular grid, which is the current state-of-the-art. As expected, the computational cost of FD calculations using a regular grid is significantly higher than the corresponding cost of DFPT calculations, with a slowdown by a factor of $1.5$ for graphene and of $3.9$ for MgB$_2$.

The results shown in Table \ref{tab:DFPTvsFD} demonstrate that the use of Farey grids provides a notable speedup compared to the use of regular grids. For DFPT, we find a computational speedup by a factor of $3.4$ for graphene and of $4.3$ for MgB$_2$. This is due to a reduction in the number of $\boldsymbol{q}$-points necessary to obtain a converged result with Farey grids, highlighting the extensive computational savings enabled by an appropriate choice of the BZ sampling points. With FD, the speedup is even larger, since, in addition to the number of $\boldsymbol{q}$-points, the size of the largest supercells required in the calculations is also reduced. Explicitly, the computational speedup compared to regular-grid FD is $5.7$ for graphene and $7.0$ for MgB$_{2}$.

It should be noted that the timings in Table \ref{tab:DFPTvsFD} are taken only for the converged grids in each case; as explained in Sec.\,\ref{sec:theory}, a typical convergence workflow for regular grids requires a significant amount of calculations with insufficiently dense grids, which are ultimately discarded. This implies that a realistic regular grid calculation requires more time than accounted for in the timings presented here. By contrast, the convergence workflow for the Farey grid does not discard any points, so that the timings presented here correspond to a realistic calculation. As a result, the speedups obtained with Farey grids compared to regular grids in Table \ref{tab:DFPTvsFD} are a lower bound to the expected speedups in a standard calculation that includes a convergence test.

It is remarkable that for graphene, the overall fastest method for a converged phonon calculation is FD combined with Farey grids. The likely origin of this is a favorable scaling of the supercell sizes in $2$D compared to 3D, combined with the reduction of the number of $\boldsymbol{k}$-points required to sample the electronic BZ of a supercell compared to a primitive cell. In this sense, we can conclude that Farey grids unlock the full potential of the FD method. This will prove particularly useful in calculations using electronic structure methods for which DFPT is unavailable. This has so far been a bottleneck in describing lattice dynamics with theories beyond semi-local DFT, such as meta-GGA exchange-correlation functional DFT or even DFT combined with dynamical mean-field theory, as DFPT requires mathematics specific to a given theory~\cite{giustino2017}. FD, on the other hand, is completely agnostic to the underlying theory, as long as it can calculate the forces on an atom. Our results significantly broaden the range of applicability of FD, demonstrating that FD can become competitive with DFPT. We expect this to open the doors to many new applications.

\subsection{Bandgap renormalization driven by electron-phonon coupling}
Electron-phonon coupling dominates a variety of phenomena, from transport to superconductivity. An area of recent interest is the temperature dependence of electronic structures driven by EPC~\cite{giustino2010, bhosale2012, monserrat2013, Ponce2015, Karsai2018, zhang2020}, and we use this as an example of calculating electron-phonon interactions from first principles. 

The temperature dependence of an electronic energy level $E_{n\boldsymbol{k}}$ can be understood starting from the Allen-Heine-Cardona theory~\cite{allen1976,allen1981,allen1983}, which states that
\begin{equation}
\label{Eq.{Enk(T)-quadratic}}
E_{n\boldsymbol{k}}(T)=E_{n\boldsymbol{k}}^{\text{static}}+\frac{1}{2} \sum_{ \nu \boldsymbol{q}} \frac{1}{2 \omega_{\nu \boldsymbol{q}}} \frac{\partial^{2} E_{n\boldsymbol{k}}}{\partial \mu_{\nu \boldsymbol{q}}^{2}}\left[1+2n_{\mathrm{B}}\left(\omega_{\nu \boldsymbol{q}}, T\right)\right],
\end{equation}
where $E_{n\boldsymbol{k}}^{\text{static}}$ is the static electronic eigenenergy, $\mu_{\nu \boldsymbol{q}}$ is the phonon amplitude that one can directly obtain from the phonon polarization $\boldsymbol{\xi}_{\nu \boldsymbol{q}}$~\cite{monserrat2018}, and $n_{\mathrm{B}}\left(\omega_{\nu \boldsymbol{q}}, T\right) =\left(\mathrm{e}^{\omega_{\nu \boldsymbol{q}} / k_{\mathrm{B}} T}-1\right)^{-1}$ is the Bose-Einstein factor. It is worth noting that there is a difference between the static energy band $E_{n\boldsymbol{k}}^{\text{static}}$ and the zero-temperature energy band $E_{n\boldsymbol{k}}(T=0)$, arising from zero-point quantum fluctuations and typically referred to as the zero-point correction.

The numerical implementation of Eq.\,\eqref{Eq.{Enk(T)-quadratic}} requires the solution of the lattice dynamics of the system, and the subsequent evaluation of the second derivatives of the electronic eigenenergy of interest with respect to all possible phonon displacements. In practice, the calculations are performed on a $\boldsymbol{q}$-point grid of the BZ. However, this procedure often requires a very fine $\boldsymbol{q}$-point grid whose corresponding computational cost can become prohibitive, particularly in the context of FD where the grid is sampled through commensurate real space supercells. To understand the requirement of a fine $\boldsymbol{q}$-point grid, we invoke the Hellmann-Feynman theorem to split the second derivative in Eq.\,\eqref{Eq.{Enk(T)-quadratic}} into three terms:
\begin{equation}
\label{Eq.{HF-split}}
\begin{aligned} 
\frac{\partial^{2} E_{n \boldsymbol{k}}}{\partial \mu_{\nu \boldsymbol{q}}^{2}} &=\left\langle\varphi_{n \boldsymbol{k}}\left|\frac{\partial^{2} \mathcal{H}_{\mathrm{el}}}{\partial \mu_{\nu \boldsymbol{q}}^{2}}\right| \varphi_{n \boldsymbol{k}}\right\rangle \\ &+ \left\langle\varphi_{n \boldsymbol{k}}\left|\frac{\partial\mathcal{H}_{\mathrm{el}}}{\partial \mu_{\nu \boldsymbol{q}}}\right| \frac{\partial \varphi_{n \boldsymbol{k}}}{\partial \mu_{\nu \boldsymbol{q}}}\right\rangle+\left\langle \frac{\partial \varphi_{n \boldsymbol{k}}}{\partial \mu_{\nu \boldsymbol{q}}}\left|\frac{\partial\mathcal{H}_{\mathrm{el}}}{\partial \mu_{\nu \boldsymbol{q}}}\right|\varphi_{n \boldsymbol{k}}\right\rangle.
\end{aligned}
\end{equation}
The derivative of the state appearing in Eq.\,\eqref{Eq.{HF-split}} can be expanded in the complete (unperturbed) basis $\{|\varphi_{n \boldsymbol{k}}\rangle\}$ according to perturbation theory:
\begin{equation}
\label{Eq.{dphi/dmu}}
\left| \frac{\partial \varphi_{n \boldsymbol{k}}}{\partial \mu_{\nu \boldsymbol{q}}}\right\rangle =\sum_{\tiny{\begin{array}{c}(m, \boldsymbol{k^\prime})\\ \neq  (n, \boldsymbol{k}) \end{array}}}\left|\varphi_{m \boldsymbol{k}^\prime}\right\rangle \frac{\left\langle\varphi_{m \boldsymbol{k}^\prime}\left|\frac{\partial\mathcal{H}_{\mathrm{el}}}{\partial \mu_{\nu \boldsymbol{q}}}\right| \varphi_{n \boldsymbol{k}}\right\rangle}{E_{n \boldsymbol{k}}^{(0)} -E_{m \boldsymbol{k}^\prime}^{(0)} }.
\end{equation}
Substituting Eq.\,\eqref{Eq.{HF-split}} and Eq.\,\eqref{Eq.{dphi/dmu}} into Eq.\,\eqref{Eq.{Enk(T)-quadratic}} yields
\begin{widetext}
\begin{equation} 
\label{Eq.{E_nk(T)-g}}
E_{n\boldsymbol{k}}(T)=E_{n\boldsymbol{k}}^{\text{static}} +\frac{1}{N}\sum_{\nu  \boldsymbol{q}} \left[\sum_{m} \frac{\left|g_{m n \nu}(\boldsymbol{k}, \boldsymbol{q})\right|^{2}}{E_{n\boldsymbol{k}}^{(0)} -E_{m \boldsymbol{k}+\boldsymbol{q}}^{(0)}}+g_{n n \nu \nu}^{\mathrm{DW}}(\boldsymbol{k}, \boldsymbol{q},-\boldsymbol{q})\right]\left(1+2n_{\mathrm{B}}\left(\omega_{\nu \boldsymbol{q}}, T\right)\right),
\end{equation}
\end{widetext}
where $g_{m n \nu}(\boldsymbol{k}, \boldsymbol{q})$ and $g_{n n \nu \nu}^{\mathrm{DW}}(\boldsymbol{k}, \boldsymbol{q},-\boldsymbol{q})$ are the EPC matrix elements of first and second order respectively, defined as in Ref~\cite{giustino2017}. It is worth noting that the sum excludes $\boldsymbol{q}=\boldsymbol{0}$ for the first-order term when $m=n$, but the denominator $E_{n\boldsymbol{k}}^{(0)}-E_{m\boldsymbol{k+q}}^{(0)}$  can still tend to $0$ for some $\boldsymbol{q} \neq \boldsymbol{0}$. When this happens, Equation \eqref{Eq.{Enk(T)-quadratic}} becomes an improper integral over the BZ: the integrand diverges at some $\boldsymbol{q}$-point(s), leading to a sharp peak in the EPC. A converged calculation requires that this singular behavior is correctly captured, which can only be accomplished if the distance between two adjacent sampling points is less than the width of these peaks. This implies that fine sampling grids become necessary.

In the following, we evaluate the temperature dependence of the band gaps of diamond and bismuthene.

\subsubsection{Computational details}
For diamond, the exchange-correlation functional is treated in the LDA~\cite{Perdew-Zunger-LDA}. The energy is converged to within the criterion of $10^{-4}$\,meV per cell with a plane-wave energy cutoff of $500$\,eV and by sampling the electronic BZ using an $11 \times 11 \times 11$ $\Gamma$-centered $\boldsymbol{k}$-point grid.

For bismuthene, the exchange-correlation functional is modeled using the PBE GGA parametrization~\cite{PBE-exchange-correlation}. A vacuum layer with an even larger thickness than that in the graphene calculation, $20$\,\AA $ $, is used given the buckling structure in bismuthene. The same energy convergence criterion as that in diamond is achieved with a plane-wave energy cutoff of $300$\,eV and by sampling the electronic BZ using an $11 \times 11$ $\Gamma$-centered $\boldsymbol{k}$-point grid. The spin-orbital interaction is included in the calculations \textit{via} a perturbation to the scalar relativistic Hamiltonian~\cite{Koelling1977}.

\subsubsection{Temperature dependence of the indirect band gap of diamond}
\label{sec:3D-diamond}

Diamond has been the testbed for the development of computational tools to study the electron-phonon driven temperature dependence of band structures~\cite{giustino2010, zacharias2016, Ponce2014Temperature, Ponce2015, Monserrat2014, Tomeu-nondiagonal-supercells}. It is a particularly useful test case for our method in $3$D, owing to the fact that it is possible to obtain converged results for its thermal (indirect) band gap using a regular grid with both DFPT~\cite{Ponce2014Temperature, Ponce2015} and FD~\cite{Monserrat2014, Tomeu-nondiagonal-supercells}. Therefore, we can carry out a direct comparison between the predictions from the regular and Farey grids.

\begin{figure}
\centering
\includegraphics[width=1.02\linewidth]{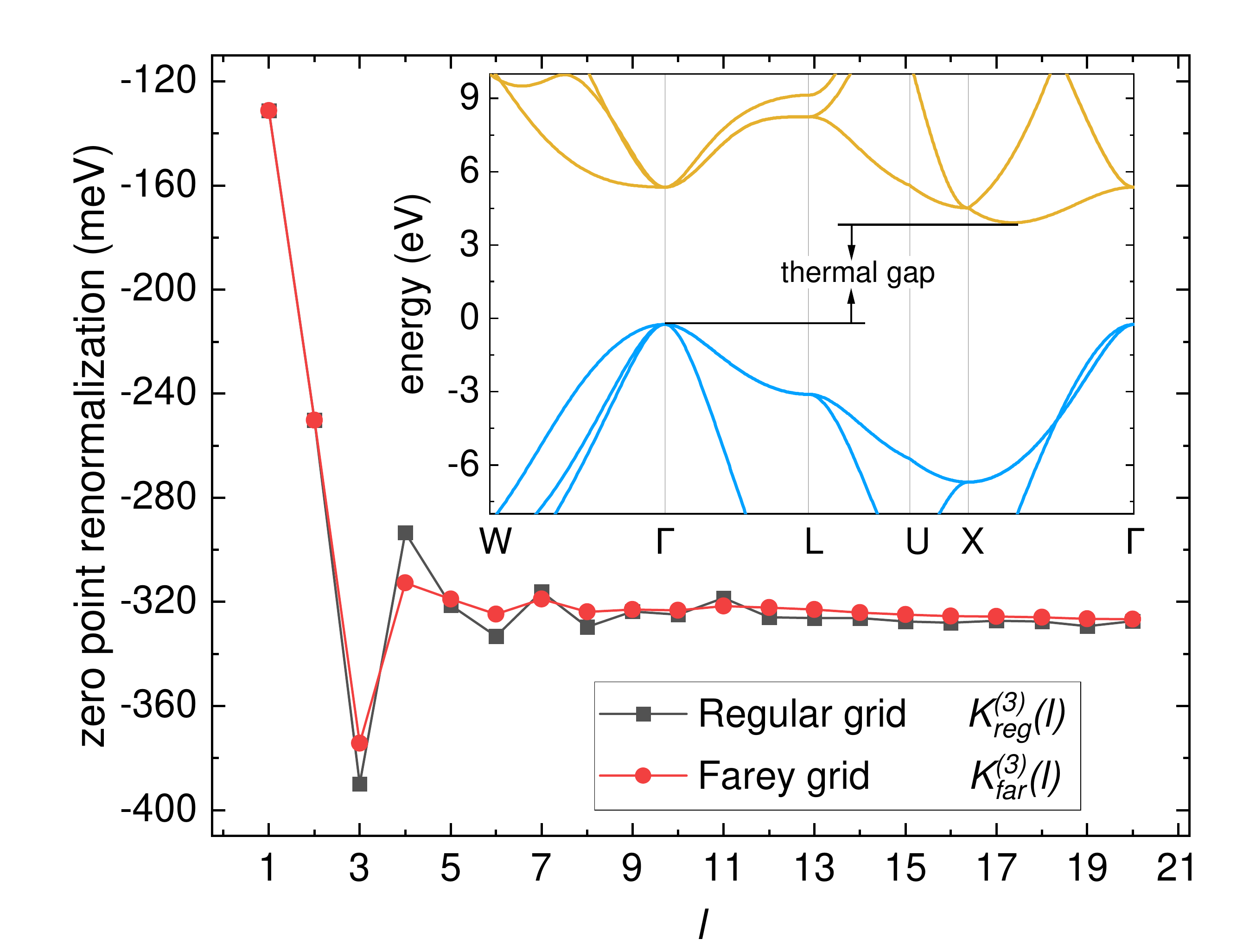}
\caption{Zero point renormalization of the thermal band gap of diamond as a function of the scale of the regular (black squares) and Farey (red circles) $\boldsymbol{q}$-point grids. Inset: static band structure of diamond along the high-symmetry path $W \rightarrow \Gamma \rightarrow \mathrm{L} \rightarrow \mathrm{U} \rightarrow \mathrm{X} \rightarrow \Gamma$.}
     \label{fig:diamond_zpe}
\end{figure}

The inset of Fig.\,\ref{fig:diamond_zpe} shows the static band structure of diamond, highlighting its wide thermal band gap, $E_{\mathrm{gap}}(\mu_{\nu \boldsymbol{q}}=0) = 4.165$\,eV. This result, in a manner typical of semi-local DFT, underestimates the experimental value of $5.450$ eV~\cite{Donnell1991}. The valence band maximum is located at the $\Gamma$ point and the conduction band minimum lies on the high-symmetry line between $\Gamma$ and $\mathrm{X}$. Figure \ref{fig:diamond_zpe} also shows the renormalization of this band gap at $0$\,K due to nuclear zero-point motion, calculated with regular grids, $\mathcal{K}^{(3)}_\text{reg}\left(l\right)$, and Farey grids, $\mathcal{K}^{(3)}_\text{far}\left(l\right)$, of different sizes: $l$ in both cases ranges from $1$ to $20$. The Farey grid of order $l$ is constructed by combining the EPC calculations of all regular grids with sizes up to $l$.

The results obtained from the Farey grid exhibit a smoother convergence compared to those obtained with the regular grid. The zero-point renormalization is converged to $-325 \pm 5$\,meV when using the Farey grid  $\mathcal{K}^{(3)}_\text{far}\left(6\right)$. This uncertainty is an order of magnitude smaller than the change to the zero-point renormalization due to the choice of pseudopotential for carbon~\cite{Ponce2014}. Using the regular grid, a $12 \times 12 \times 12$ grid is needed to achieve the same level of convergence. It is worth noting that the grid size of the Farey grid here ($|\mathcal{K}^{(3)}_\text{far}\left(6\right)|=396$) is less than a quarter of that of the regular grid ($|\mathcal{K}^{(3)}_\text{reg}\left(12\right)|=1728$), and the actual computational cost of the Farey grid is approximately a sixth that of the regular grid. This computational cost reduction comes not only from reducing the number of required sampling points, but also from the fact that the largest supercells needed for the Farey grid calculations are only 6 times larger than the primitive cell, whereas we require supercells that are 12 times larger than the primitive cell in the calculation with a $12 \times 12 \times 12$ regular grid. It is also clear that the regular and Farey grids converge to the same value as $l$ increases.

Overall, the converged result from the first-principles calculations is very close to the experimental measurement of $-340$\,meV~\cite{Cardona2001}. The small difference compared to the experimental value is likely to be due to the use of the LDA to describe the exchange-correlation energy. In principle, this can be overcome with more expensive calculations using hybrid exchange-correlation functionals or by invoking the $GW$ approximation from many-body perturbation theory. Recently, Karsai and co-workers evaluated the renormalization of the diamond thermal band gap using the $G_0W_0$ approximation, finding that $G_0W_0$ further increases the magnitude of the zero-point renormalization by 17\,meV~\cite{Karsai2018}. Adding this value to our converged DFT result, one obtains a theoretical value very close to the experimental result.

We also note that thermal expansion can also renormalize the electronic energy levels of materials, but in the case of diamond the effect is minimal compared to the changes driven by EPC, so it can be safely neglected~\cite{monserrat2013}.

\subsubsection{Topologically nontrivial band gap of bismuthene}
\begin{figure}
\includegraphics[width=0.99\linewidth]{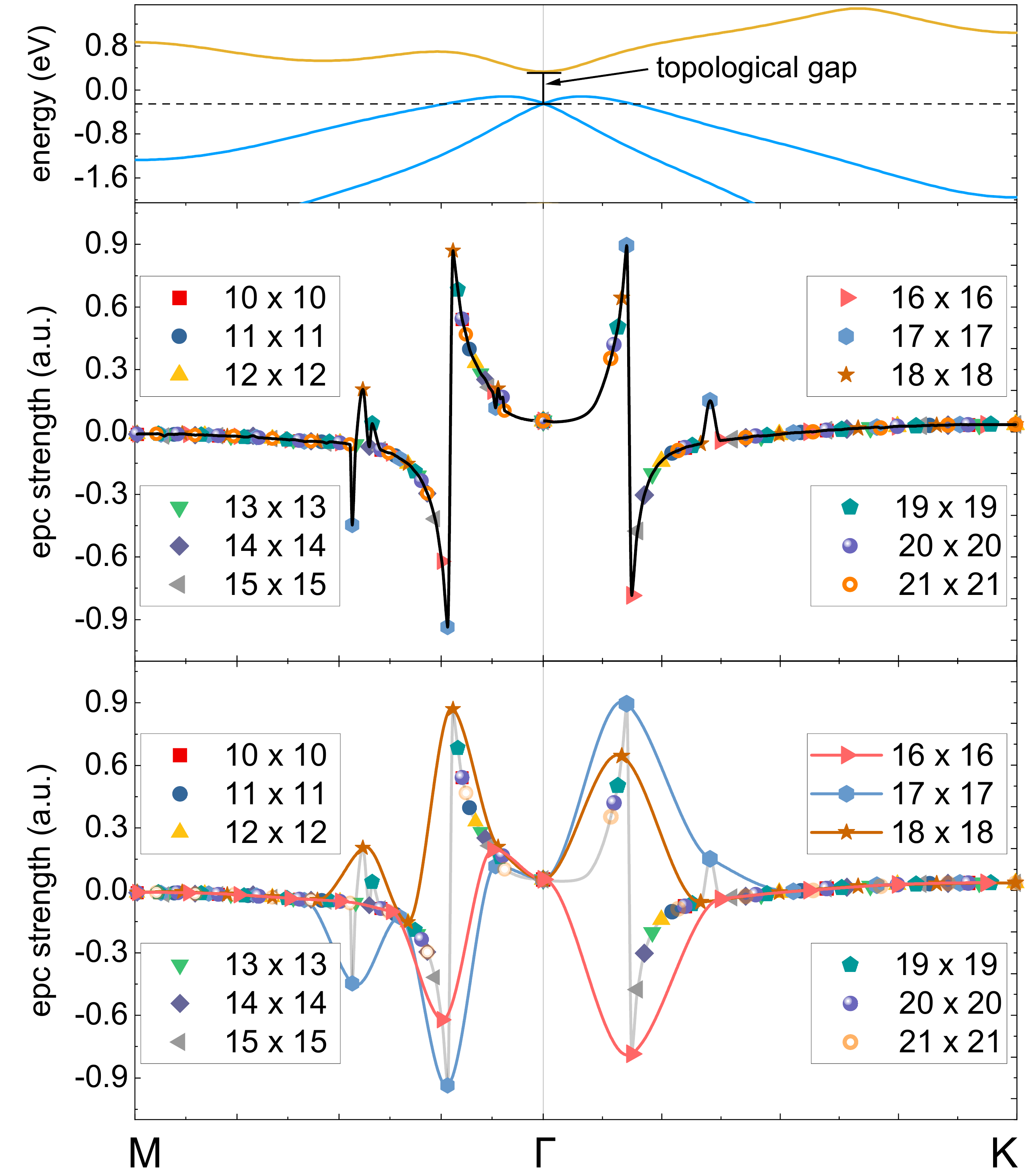}
\caption{(color online). (top panel) Static band structure of bismuthene (where the Fermi level is set to zero and the dashed line indicates the valence band energy at the $\Gamma$-point) and (middle panel) the EPC strength along the high-symmetry path $\mathrm{M} \rightarrow \Gamma \rightarrow \mathrm{K}$ (where the symbols on the line mark the $\boldsymbol{q}$-points that can be explicitly assessed by the regular $\boldsymbol{q}$-point grids of different sizes). The bottom panel highlights the EPC strength `seen' by regular $\boldsymbol{q}$-point grids of sizes $16 \times 16$, $17 \times 17$ and $18 \times 18$.}
\label{fig:bi-epc}
\end{figure}

Bismuthene is a 2D topological insulator with a topologically nontrivial band gap at the BZ center due to its intrinsic strong spin-orbital coupling~\cite{Murakami2006, Koroteev2008, Wada2011, Liu2011, Reis2017}. Motivated by the search for topological insulators which can work at ambient temperature ($T=300$ K), here we investigate the EPC-driven renormalization of this topological band gap. The top panel of Fig.\,\ref{fig:bi-epc} shows the static band structure of bismuthene near the Fermi level with the spin-orbital coupling interaction taken into account. The conduction band has conical shape in the zone center, while the highest valence band exhibits a double peak around the $\Gamma$-point, with a trough exactly at it. This results in a sizable static $\Gamma$-point band gap of $E_{\mathrm{gap}}(\mu_{\nu \boldsymbol{q}}=0) = 572 $\,meV.

\begin{figure}
\includegraphics[width=1.02\linewidth]{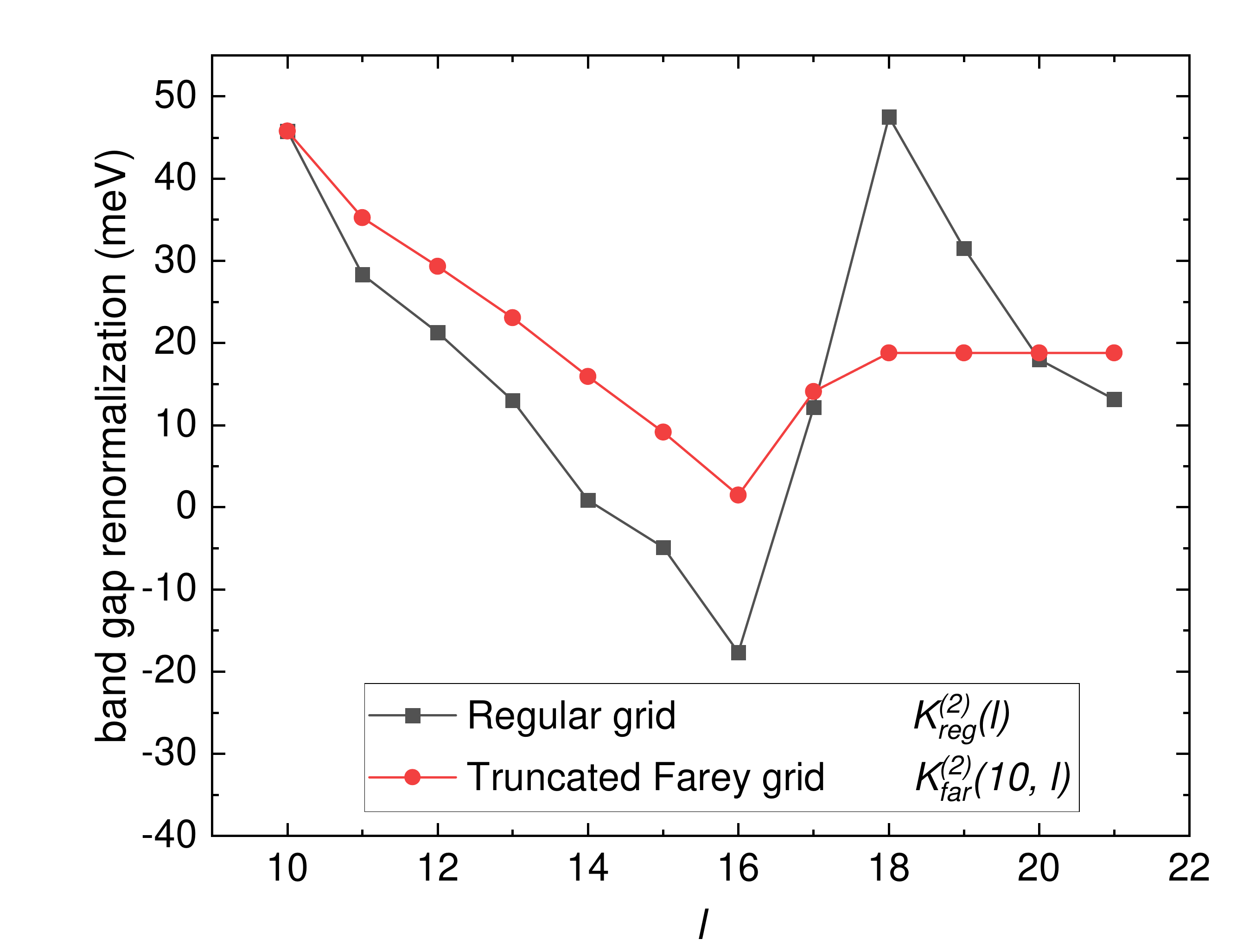}
\caption{(color online). Renormalization of the topologically nontrivial band gap of bismuthene at room temperature ($T=300$ K) as a function of the scale of the regular (black squares) and the truncated Farey (red circles) $\boldsymbol{q}$-point grids.}
\label{fig:bi-gapvst}
\end{figure}

Figure \ref{fig:bi-gapvst} shows the renormalization of this band gap at $300$\,K calculated with a regular and a truncated Farey $\boldsymbol{q}$-point grids. Its value converges to $18.8\pm 0.05$\,meV using the truncated Farey $\boldsymbol{q}$-point grid generated by $\mathcal{K}^{(2)}_\text{far}\left(10,18\right)$. This stands in stark contrast with the results calculated with the regular $\boldsymbol{q}$-point grids, which oscillate so strongly as to prevent even qualitatively consistent conclusions. For instance, while the calculation with the $14 \times 14$ regular $\boldsymbol{q}$-point grid suggests the band gap is almost independent of temperature, the calculation with the $16 \times 16$ ($18 \times 18$) regular $\boldsymbol{q}$-point grid implies that it would be significantly lower (higher) at room temperature. Even at a grid size of $21\times 21$, the largest regular grid considered here, no clear convergence trend can be observed.

To obtain more insight into these difficulties with converging the EPC calculations in bismuthene using regular grids, we define the EPC amplitude in the context of the Allen-Heine-Cardona formula [Eq.\,\eqref{Eq.{Enk(T)-quadratic}}] as 
\begin{equation}
S^{\mathrm{el-ph}}_{\nu \boldsymbol{q}} = \frac{\partial^2 E_{\mathrm{gap}}}{\partial \mu_{\nu \boldsymbol{q}}^2},
\end{equation}
and the $\boldsymbol{q}$-resolved EPC amplitude:
\begin{equation}
S^{\mathrm{el-ph}}(\boldsymbol{q}) = \sum_{\nu} S^{\mathrm{el-ph}}_{\nu \boldsymbol{q}}.
\end{equation}
The middle panel of Fig.\,\ref{fig:bi-epc} shows the $\boldsymbol{q}$-resolved EPC strength along the high-symmetry path $\mathrm{M}-\Gamma-\mathrm{K}$, where it can be clearly seen that the curve exhibits strong peaks at around $1/4$ of the $\Gamma-\mathrm{M}$ distance and $1/5$ of the $\Gamma-\mathrm{K}$ distance. This agrees with the perturbation picture provided by Eq.\,\eqref{Eq.{E_nk(T)-g}}: the double-peak shape of the valence band of bismuthene favors processes where a $\Gamma$-point electron scatters low-frequency phonons with wavevectors $|\boldsymbol{q}| \approx 1/4$ of the $\Gamma-\mathrm{M}$ distance or $|\boldsymbol{q}| \approx 1/5$ of the $\Gamma-\mathrm{K}$ distance. These phonons scatter the electron to another state at a similar energy (as shown in the top panel), but with different momentum corresponding to that contributed by the phonon. More generally, we find that the EPC amplitude is almost isotropic around $\Gamma$, which implies that the band gap renormalization is dominated by low-frequency phonons with wavevectors $|\boldsymbol{q}| \approx b/8$, where $b$ is the lattice constant of the bismuthene reciprocal lattice. 

The bottom panel of Fig.\,\ref{fig:bi-epc} shows the EPC strength `seen' by regular $\boldsymbol{q}$-point grids of a range of different sizes, which explains the qualitatively inconsistent results observed in Fig.\,\ref{fig:bi-gapvst}. The $18 \times18 $ uniform $\boldsymbol{q}$-point grid overestimates the EPC amplitude of the system as it only samples the peak points of the curve, while the $16 \times 16 $ uniform $\boldsymbol{q}$-point grid underestimates it due to only sampling the valley points. As shown in the middle panel of Fig.\,\ref{fig:bi-epc}, to capture this peak, it is necessary to use a regular grid that at least contains $\mathrm{lcm}(16,17,18) \times \mathrm{lcm}(16,17,18)$ points, that is, a $2448 \times 2448$ regular $\boldsymbol{q}$-point grid. This grid size is beyond the capabilities of modern high-performance computing. By comparison, the Farey grid leads to converged results at a moderate computational cost.

It should also be noted that, although the sharpest peaks in the middle panel of Fig.\,\ref{fig:bi-epc} agree very well with the expectation of Eq.\,\eqref{Eq.{E_nk(T)-g}}, there are two smaller peaks in the EPC amplitude appearing at around $|\boldsymbol{q} |\approx 1/2$ of the $\Gamma-\mathrm{M}$ distance and $|\boldsymbol{q} |\approx 3/8$ of the $\Gamma-\mathrm{K}$ distance. These cannot be explained by the perturbation picture, and we infer that this behavior is mostly likely subject to details of the EPC matrix element and high-order EPC interaction terms.

More generally, the temperature dependence of the band gap of bismuthene has not been reported elsewhere. Our results show that bismuthene will have a topological band of almost $600$\,meV at $300$\,K, making it an ideal material to study topological phenomena at room temperature. We also note that our calculations do not include thermal expansion, but the known dependence of the band gap of bismuthene on the lattice parameter~\cite{chen2013, liu2017} implies that the inclusion of thermal expansion should not change our conclusions.

\subsection{Discussion and outlook}
First principles calculations of the electronic structure of materials have reached a high level of sophistication, enabling the description of electrons moving in a static lattice with high accuracy. Examples of this are hybrid functional DFT, which accurately captures the exchange-correlation energy~\cite{Heyd2003, Krukau2006}, the $GW$ method~\cite{Hybertsen1986,Aryasetiawan1998}, which includes many-body effects, and the DFT+U method~\cite{Anisimov1991, Liechtenstein1995} and dynamical mean-field theory~\cite{Georges1996, Kotliar2006}, which treat strong correlation effects. These developments have laid a solid foundation for studying the coupling between phonons and electrons in a wide variety of materials from first principles. In addition, with advances in the study of other quasiparticles, such as the Heisenberg ferromagnet equation for magnons~\cite{Gebauer2000, Xu2021} and the Bethe-Salpeter equation for excitons~\cite{Salpeter1951, Rohlfing1998}, one can investigate phenomena involving magnon-phonon and exciton-phonon coupling. In most cases, the lattice dynamics and related properties predicted by the methods mentioned in this section can only be calculated using the FD method. By reducing the computational cost associated with phonon calculations dramatically, the Farey grid and the SAVT technique proposed in this paper pave the way for realistic implementations using these state-of-the-art methods to carry out electron-phonon (or magnon-phonon or exciton-phonon) coupling calculations. 

\section{Conclusion}
\label{sec:conclusion}
We have introduced two developments to carry out BZ integration. First, a nonuniform grid, called the Farey grid, which is computationally advantageous compared to the widely used regular uniform grids. Second, the symmetry-adapted Voronoi tessellation technique to evaluate the weights necessary for a numerical BZ integration using an arbitrary grid. We have demonstrated that these developments provide significant computational speedups in the calculation of phonons and EPC, irrespective of whether DFPT or finite difference methods are used. However, the most dramatic advantages are observed for finite difference methods, which have traditionally been slower than DFPT, but that now become competitive from a computational expense perspective. Specifically, we apply our methodology to the study of Kohn anomalies in graphene and MgB$_2$, and to the study of electron-phonon driven band gap renormalizations in diamond and bismuthene. The results for bismuthene reveal that this material should have a sizeable topological band gap at room temperature, making it an ideal platform to study room temperature topological phenomena.

A practical insight of our work is that the Farey grid can be constructed by a combination of differently-sized regular grids. This implies that it should be relatively straightforward to adapt modern first principles software, which is typically based on regular grids, to incorporate the Farey grid.

An important conclusion from our work is that, in many problems of physical interest, the regular grid is far from the most efficient sampling method: it samples regions in the BZ where the integrand varies slowly just as densely as those where the integrand varies rapidly. Ideally, however, one would use a few points to represent the former, while focusing more computational power on accurately representing the latter. This is a possibility that opens up as a result of the symmetry-adapted Voronoi tessellation technique. Although we have only used it in conjunction with the nonuniform Farey grid, it is generally applicable to \textit{any} grid (uniform or not) that one may conceive to evaluate the BZ integration. Thus, it provides a solid platform on which to design the most appropriate nonuniform grid for any given task, with the prospect of saving valuable computational time and allowing for new frontiers in computational condensed matter physics to be charted.

\begin{acknowledgments}
S.C. acknowledges financial support from the Cambridge Trust and from the Winton Programme for the Physics of Sustainability. P.T.S. gratefully acknowledges funding from the Department of Materials Science and Metallurgy at the University of Cambridge and from a Trinity Hall Research Studentship. P.T.S. also thanks Chris J. Pickard for helpful discussions and support. B.M. acknowledges support from a UKRI Future Leaders Fellowship (Grant No. MR/V023926/1), from the Gianna Angelopoulos Programme for Science, Technology, and Innovation, and from the Winton Programme for the Physics of Sustainability. The calculations in this work have been performed using resources provided by the Cambridge Tier-2 system (operated by the University of Cambridge Research Computing Service and funded by EPSRC [EP/P020259/1]), as well as by the UK Materials and Molecular Modelling Hub (partially funded by EPSRC [EP/P020194]), Thomas, and by the UK National Supercomputing Service, ARCHER. Access to Thomas and ARCHER was obtained via the UKCP consortium and funded by EPSRC [EP/P022561/1].
\end{acknowledgments}

\bibliography{Bibliography}

\providecommand{\noopsort}[1]{}\providecommand{\singleletter}[1]{#1}%
\begin{thebibliography}{101}%
\makeatletter
\providecommand \@ifxundefined [1]{%
 \@ifx{#1\undefined}
}%
\providecommand \@ifnum [1]{%
 \ifnum #1\expandafter \@firstoftwo
 \else \expandafter \@secondoftwo
 \fi
}%
\providecommand \@ifx [1]{%
 \ifx #1\expandafter \@firstoftwo
 \else \expandafter \@secondoftwo
 \fi
}%
\providecommand \natexlab [1]{#1}%
\providecommand \enquote  [1]{``#1''}%
\providecommand \bibnamefont  [1]{#1}%
\providecommand \bibfnamefont [1]{#1}%
\providecommand \citenamefont [1]{#1}%
\providecommand \href@noop [0]{\@secondoftwo}%
\providecommand \href [0]{\begingroup \@sanitize@url \@href}%
\providecommand \@href[1]{\@@startlink{#1}\@@href}%
\providecommand \@@href[1]{\endgroup#1\@@endlink}%
\providecommand \@sanitize@url [0]{\catcode `\\12\catcode `\$12\catcode
  `\&12\catcode `\#12\catcode `\^12\catcode `\_12\catcode `\%12\relax}%
\providecommand \@@startlink[1]{}%
\providecommand \@@endlink[0]{}%
\providecommand \url  [0]{\begingroup\@sanitize@url \@url }%
\providecommand \@url [1]{\endgroup\@href {#1}{\urlprefix }}%
\providecommand \urlprefix  [0]{URL }%
\providecommand \Eprint [0]{\href }%
\providecommand \doibase [0]{https://doi.org/}%
\providecommand \selectlanguage [0]{\@gobble}%
\providecommand \bibinfo  [0]{\@secondoftwo}%
\providecommand \bibfield  [0]{\@secondoftwo}%
\providecommand \translation [1]{[#1]}%
\providecommand \BibitemOpen [0]{}%
\providecommand \bibitemStop [0]{}%
\providecommand \bibitemNoStop [0]{.\EOS\space}%
\providecommand \EOS [0]{\spacefactor3000\relax}%
\providecommand \BibitemShut  [1]{\csname bibitem#1\endcsname}%
\let\auto@bib@innerbib\@empty
%</preamble>
\bibitem [{\citenamefont {Baldereschi}(1973)}]{Baldereschi-mean-value-point}%
  \BibitemOpen
  \bibfield  {author} {\bibinfo {author} {\bibfnamefont {A.}~\bibnamefont
  {Baldereschi}},\ }\bibfield  {title} {\bibinfo {title} {Mean-value point in
  the {B}rillouin zone},\ }\href {https://doi.org/10.1103/PhysRevB.7.5212}
  {\bibfield  {journal} {\bibinfo  {journal} {Phys. Rev. B}\ }\textbf {\bibinfo
  {volume} {7}},\ \bibinfo {pages} {5212} (\bibinfo {year} {1973})}\BibitemShut
  {NoStop}%
\bibitem [{\citenamefont {Chadi}\ and\ \citenamefont
  {Cohen}(1973)}]{Chadi-Cohen-special-points}%
  \BibitemOpen
  \bibfield  {author} {\bibinfo {author} {\bibfnamefont {D.~J.}\ \bibnamefont
  {Chadi}}\ and\ \bibinfo {author} {\bibfnamefont {M.~L.}\ \bibnamefont
  {Cohen}},\ }\bibfield  {title} {\bibinfo {title} {Special points in the
  {B}rillouin zone},\ }\href {https://doi.org/10.1103/PhysRevB.8.5747}
  {\bibfield  {journal} {\bibinfo  {journal} {Phys. Rev. B}\ }\textbf {\bibinfo
  {volume} {8}},\ \bibinfo {pages} {5747} (\bibinfo {year} {1973})}\BibitemShut
  {NoStop}%
\bibitem [{\citenamefont {Cunningham}(1974)}]{cunningham1974}%
  \BibitemOpen
  \bibfield  {author} {\bibinfo {author} {\bibfnamefont {S.~L.}\ \bibnamefont
  {Cunningham}},\ }\bibfield  {title} {\bibinfo {title} {Special points in the
  two-dimensional {B}rillouin zone},\ }\href
  {https://doi.org/10.1103/PhysRevB.10.4988} {\bibfield  {journal} {\bibinfo
  {journal} {Phys. Rev. B}\ }\textbf {\bibinfo {volume} {10}},\ \bibinfo
  {pages} {4988} (\bibinfo {year} {1974})}\BibitemShut {NoStop}%
\bibitem [{\citenamefont {Monkhorst}\ and\ \citenamefont
  {Pack}(1976)}]{Monkhorst-Pack-grids}%
  \BibitemOpen
  \bibfield  {author} {\bibinfo {author} {\bibfnamefont {H.~J.}\ \bibnamefont
  {Monkhorst}}\ and\ \bibinfo {author} {\bibfnamefont {J.~D.}\ \bibnamefont
  {Pack}},\ }\bibfield  {title} {\bibinfo {title} {Special points for
  {{B}rillouin-zone} integrations},\ }\href
  {https://link.aps.org/doi/10.1103/PhysRevB.13.5188} {\bibfield  {journal}
  {\bibinfo  {journal} {Phys. Rev. B}\ }\textbf {\bibinfo {volume} {13}},\
  \bibinfo {pages} {5188} (\bibinfo {year} {1976})}\BibitemShut {NoStop}%
\bibitem [{\citenamefont {Fehlner}\ and\ \citenamefont
  {Vosko}(1977)}]{Fehlner1977}%
  \BibitemOpen
  \bibfield  {author} {\bibinfo {author} {\bibfnamefont {W.~R.}\ \bibnamefont
  {Fehlner}}\ and\ \bibinfo {author} {\bibfnamefont {S.~H.}\ \bibnamefont
  {Vosko}},\ }\bibfield  {title} {\bibinfo {title} {Special points for the
  evaluation of form factors and fourier series representations in hexagonal
  crystals},\ }\href {https://doi.org/10.1139/p77-248} {\bibfield  {journal}
  {\bibinfo  {journal} {Can. J. Phys.}\ }\textbf {\bibinfo {volume} {55}},\
  \bibinfo {pages} {2041} (\bibinfo {year} {1977})}\BibitemShut {NoStop}%
\bibitem [{\citenamefont {Froyen}(1989)}]{Froyen1989}%
  \BibitemOpen
  \bibfield  {author} {\bibinfo {author} {\bibfnamefont {S.}~\bibnamefont
  {Froyen}},\ }\bibfield  {title} {\bibinfo {title} {{B}rillouin-zone
  integration by fourier quadrature: Special points for superlattice and
  supercell calculations},\ }\href {https://doi.org/10.1103/PhysRevB.39.3168}
  {\bibfield  {journal} {\bibinfo  {journal} {Phys. Rev. B}\ }\textbf {\bibinfo
  {volume} {39}},\ \bibinfo {pages} {3168} (\bibinfo {year}
  {1989})}\BibitemShut {NoStop}%
\bibitem [{\citenamefont {Moreno}\ and\ \citenamefont
  {Soler}(1992)}]{Moreno1992}%
  \BibitemOpen
  \bibfield  {author} {\bibinfo {author} {\bibfnamefont {J.}~\bibnamefont
  {Moreno}}\ and\ \bibinfo {author} {\bibfnamefont {J.~M.}\ \bibnamefont
  {Soler}},\ }\bibfield  {title} {\bibinfo {title} {Optimal meshes for
  integrals in real- and reciprocal-space unit cells},\ }\href
  {https://doi.org/10.1103/PhysRevB.45.13891} {\bibfield  {journal} {\bibinfo
  {journal} {Phys. Rev. B}\ }\textbf {\bibinfo {volume} {45}},\ \bibinfo
  {pages} {13891} (\bibinfo {year} {1992})}\BibitemShut {NoStop}%
\bibitem [{\citenamefont {Wisesa}\ \emph {et~al.}(2016)\citenamefont {Wisesa},
  \citenamefont {McGill},\ and\ \citenamefont {Mueller}}]{Wisesa2016}%
  \BibitemOpen
  \bibfield  {author} {\bibinfo {author} {\bibfnamefont {P.}~\bibnamefont
  {Wisesa}}, \bibinfo {author} {\bibfnamefont {K.~A.}\ \bibnamefont {McGill}},\
  and\ \bibinfo {author} {\bibfnamefont {T.}~\bibnamefont {Mueller}},\
  }\bibfield  {title} {\bibinfo {title} {Efficient generation of generalized
  monkhorst-pack grids through the use of informatics},\ }\href
  {https://doi.org/10.1103/PhysRevB.93.155109} {\bibfield  {journal} {\bibinfo
  {journal} {Phys. Rev. B}\ }\textbf {\bibinfo {volume} {93}},\ \bibinfo
  {pages} {155109} (\bibinfo {year} {2016})}\BibitemShut {NoStop}%
\bibitem [{\citenamefont {Kresse}\ and\ \citenamefont
  {Furthm{\"u}ller}(1996)}]{VASP-Original-Paper}%
  \BibitemOpen
  \bibfield  {author} {\bibinfo {author} {\bibfnamefont {G.}~\bibnamefont
  {Kresse}}\ and\ \bibinfo {author} {\bibfnamefont {J.}~\bibnamefont
  {Furthm{\"u}ller}},\ }\bibfield  {title} {\bibinfo {title} {Efficiency of
  ab-initio total energy calculations for metals and semiconductors using a
  plane-wave basis set},\ }\href
  {https://doi.org/https://doi.org/10.1016/0927-0256(96)00008-0} {\bibfield
  {journal} {\bibinfo  {journal} {Comput. Mater. Sci.}\ }\textbf {\bibinfo
  {volume} {6}},\ \bibinfo {pages} {15 } (\bibinfo {year} {1996})}\BibitemShut
  {NoStop}%
\bibitem [{\citenamefont {Alfè}(2009)}]{alfe2009}%
  \BibitemOpen
  \bibfield  {author} {\bibinfo {author} {\bibfnamefont {D.}~\bibnamefont
  {Alfè}},\ }\bibfield  {title} {\bibinfo {title} {Phon: A program to
  calculate phonons using the small displacement method},\ }\href
  {https://doi.org/https://doi.org/10.1016/j.cpc.2009.03.010} {\bibfield
  {journal} {\bibinfo  {journal} {Comput. Phys. Commun.}\ }\textbf {\bibinfo
  {volume} {180}},\ \bibinfo {pages} {2622} (\bibinfo {year}
  {2009})}\BibitemShut {NoStop}%
\bibitem [{\citenamefont {Marini}\ \emph {et~al.}(2009)\citenamefont {Marini},
  \citenamefont {Hogan}, \citenamefont {Grüning},\ and\ \citenamefont
  {Varsano}}]{marini2009}%
  \BibitemOpen
  \bibfield  {author} {\bibinfo {author} {\bibfnamefont {A.}~\bibnamefont
  {Marini}}, \bibinfo {author} {\bibfnamefont {C.}~\bibnamefont {Hogan}},
  \bibinfo {author} {\bibfnamefont {M.}~\bibnamefont {Grüning}},\ and\
  \bibinfo {author} {\bibfnamefont {D.}~\bibnamefont {Varsano}},\ }\bibfield
  {title} {\bibinfo {title} {yambo: An ab initio tool for excited state
  calculations},\ }\href
  {https://doi.org/https://doi.org/10.1016/j.cpc.2009.02.003} {\bibfield
  {journal} {\bibinfo  {journal} {Comput. Phys. Commun.}\ }\textbf {\bibinfo
  {volume} {180}},\ \bibinfo {pages} {1392} (\bibinfo {year}
  {2009})}\BibitemShut {NoStop}%
\bibitem [{\citenamefont {Tancogne-Dejean}\ \emph {et~al.}(2020)\citenamefont
  {Tancogne-Dejean}, \citenamefont {Eich},\ and\ \citenamefont
  {Rubio}}]{Tancogne2020}%
  \BibitemOpen
  \bibfield  {author} {\bibinfo {author} {\bibfnamefont {N.}~\bibnamefont
  {Tancogne-Dejean}}, \bibinfo {author} {\bibfnamefont {F.~G.}\ \bibnamefont
  {Eich}},\ and\ \bibinfo {author} {\bibfnamefont {A.}~\bibnamefont {Rubio}},\
  }\bibfield  {title} {\bibinfo {title} {Time-dependent magnons from first
  principles},\ }\href {https://doi.org/10.1021/acs.jctc.9b01064} {\bibfield
  {journal} {\bibinfo  {journal} {J. Chem. Theory Comput.}\ }\textbf {\bibinfo
  {volume} {16}},\ \bibinfo {pages} {1007} (\bibinfo {year}
  {2020})}\BibitemShut {NoStop}%
\bibitem [{\citenamefont {Sabiryanov}\ and\ \citenamefont
  {Jaswal}(1999)}]{sabiryanov1999}%
  \BibitemOpen
  \bibfield  {author} {\bibinfo {author} {\bibfnamefont {R.~F.}\ \bibnamefont
  {Sabiryanov}}\ and\ \bibinfo {author} {\bibfnamefont {S.~S.}\ \bibnamefont
  {Jaswal}},\ }\bibfield  {title} {\bibinfo {title} {Magnons and magnon-phonon
  interactions in iron},\ }\href {https://doi.org/10.1103/PhysRevLett.83.2062}
  {\bibfield  {journal} {\bibinfo  {journal} {Phys. Rev. Lett.}\ }\textbf
  {\bibinfo {volume} {83}},\ \bibinfo {pages} {2062} (\bibinfo {year}
  {1999})}\BibitemShut {NoStop}%
\bibitem [{\citenamefont {Poncé}\ \emph {et~al.}(2016)\citenamefont {Poncé},
  \citenamefont {Margine}, \citenamefont {Verdi},\ and\ \citenamefont
  {Giustino}}]{ponce2016}%
  \BibitemOpen
  \bibfield  {author} {\bibinfo {author} {\bibfnamefont {S.}~\bibnamefont
  {Poncé}}, \bibinfo {author} {\bibfnamefont {E.}~\bibnamefont {Margine}},
  \bibinfo {author} {\bibfnamefont {C.}~\bibnamefont {Verdi}},\ and\ \bibinfo
  {author} {\bibfnamefont {F.}~\bibnamefont {Giustino}},\ }\bibfield  {title}
  {\bibinfo {title} {{EPW}: Electron–phonon coupling, transport and
  superconducting properties using maximally localized wannier functions},\
  }\href {https://doi.org/https://doi.org/10.1016/j.cpc.2016.07.028} {\bibfield
   {journal} {\bibinfo  {journal} {Comput. Phys. Commun.}\ }\textbf {\bibinfo
  {volume} {209}},\ \bibinfo {pages} {116} (\bibinfo {year}
  {2016})}\BibitemShut {NoStop}%
\bibitem [{\citenamefont {Antonius}\ and\ \citenamefont
  {Louie}(2022)}]{antonius2022}%
  \BibitemOpen
  \bibfield  {author} {\bibinfo {author} {\bibfnamefont {G.}~\bibnamefont
  {Antonius}}\ and\ \bibinfo {author} {\bibfnamefont {S.~G.}\ \bibnamefont
  {Louie}},\ }\bibfield  {title} {\bibinfo {title} {Theory of exciton-phonon
  coupling},\ }\href {https://doi.org/10.1103/PhysRevB.105.085111} {\bibfield
  {journal} {\bibinfo  {journal} {Phys. Rev. B}\ }\textbf {\bibinfo {volume}
  {105}},\ \bibinfo {pages} {085111} (\bibinfo {year} {2022})}\BibitemShut
  {NoStop}%
\bibitem [{\citenamefont {Piscanec}\ \emph {et~al.}(2004)\citenamefont
  {Piscanec}, \citenamefont {Lazzeri}, \citenamefont {Mauri}, \citenamefont
  {Ferrari},\ and\ \citenamefont {Robertson}}]{piscanec2004}%
  \BibitemOpen
  \bibfield  {author} {\bibinfo {author} {\bibfnamefont {S.}~\bibnamefont
  {Piscanec}}, \bibinfo {author} {\bibfnamefont {M.}~\bibnamefont {Lazzeri}},
  \bibinfo {author} {\bibfnamefont {F.}~\bibnamefont {Mauri}}, \bibinfo
  {author} {\bibfnamefont {A.~C.}\ \bibnamefont {Ferrari}},\ and\ \bibinfo
  {author} {\bibfnamefont {J.}~\bibnamefont {Robertson}},\ }\bibfield  {title}
  {\bibinfo {title} {Kohn anomalies and electron-phonon interactions in
  graphite},\ }\href {https://doi.org/10.1103/PhysRevLett.93.185503} {\bibfield
   {journal} {\bibinfo  {journal} {Phys. Rev. Lett.}\ }\textbf {\bibinfo
  {volume} {93}},\ \bibinfo {pages} {185503} (\bibinfo {year}
  {2004})}\BibitemShut {NoStop}%
\bibitem [{\citenamefont {Calandra}\ \emph {et~al.}(2010)\citenamefont
  {Calandra}, \citenamefont {Profeta},\ and\ \citenamefont
  {Mauri}}]{calandra2010}%
  \BibitemOpen
  \bibfield  {author} {\bibinfo {author} {\bibfnamefont {M.}~\bibnamefont
  {Calandra}}, \bibinfo {author} {\bibfnamefont {G.}~\bibnamefont {Profeta}},\
  and\ \bibinfo {author} {\bibfnamefont {F.}~\bibnamefont {Mauri}},\ }\bibfield
   {title} {\bibinfo {title} {Adiabatic and nonadiabatic phonon dispersion in a
  wannier function approach},\ }\href
  {https://doi.org/10.1103/PhysRevB.82.165111} {\bibfield  {journal} {\bibinfo
  {journal} {Phys. Rev. B}\ }\textbf {\bibinfo {volume} {82}},\ \bibinfo
  {pages} {165111} (\bibinfo {year} {2010})}\BibitemShut {NoStop}%
\bibitem [{\citenamefont {Giustino}\ \emph {et~al.}(2007)\citenamefont
  {Giustino}, \citenamefont {Cohen},\ and\ \citenamefont
  {Louie}}]{giustino2007}%
  \BibitemOpen
  \bibfield  {author} {\bibinfo {author} {\bibfnamefont {F.}~\bibnamefont
  {Giustino}}, \bibinfo {author} {\bibfnamefont {M.~L.}\ \bibnamefont
  {Cohen}},\ and\ \bibinfo {author} {\bibfnamefont {S.~G.}\ \bibnamefont
  {Louie}},\ }\bibfield  {title} {\bibinfo {title} {Electron-phonon interaction
  using wannier functions},\ }\href
  {https://doi.org/10.1103/PhysRevB.76.165108} {\bibfield  {journal} {\bibinfo
  {journal} {Phys. Rev. B}\ }\textbf {\bibinfo {volume} {76}},\ \bibinfo
  {pages} {165108} (\bibinfo {year} {2007})}\BibitemShut {NoStop}%
\bibitem [{\citenamefont {Sjakste}\ \emph {et~al.}(2015)\citenamefont
  {Sjakste}, \citenamefont {Vast}, \citenamefont {Calandra},\ and\
  \citenamefont {Mauri}}]{sjakste2015}%
  \BibitemOpen
  \bibfield  {author} {\bibinfo {author} {\bibfnamefont {J.}~\bibnamefont
  {Sjakste}}, \bibinfo {author} {\bibfnamefont {N.}~\bibnamefont {Vast}},
  \bibinfo {author} {\bibfnamefont {M.}~\bibnamefont {Calandra}},\ and\
  \bibinfo {author} {\bibfnamefont {F.}~\bibnamefont {Mauri}},\ }\bibfield
  {title} {\bibinfo {title} {Wannier interpolation of the electron-phonon
  matrix elements in polar semiconductors: Polar-optical coupling in gaas},\
  }\href {https://doi.org/10.1103/PhysRevB.92.054307} {\bibfield  {journal}
  {\bibinfo  {journal} {Phys. Rev. B}\ }\textbf {\bibinfo {volume} {92}},\
  \bibinfo {pages} {054307} (\bibinfo {year} {2015})}\BibitemShut {NoStop}%
\bibitem [{\citenamefont {Verdi}\ and\ \citenamefont
  {Giustino}(2015)}]{verdi2015}%
  \BibitemOpen
  \bibfield  {author} {\bibinfo {author} {\bibfnamefont {C.}~\bibnamefont
  {Verdi}}\ and\ \bibinfo {author} {\bibfnamefont {F.}~\bibnamefont
  {Giustino}},\ }\bibfield  {title} {\bibinfo {title} {Fr\"ohlich
  electron-phonon vertex from first principles},\ }\href
  {https://doi.org/10.1103/PhysRevLett.115.176401} {\bibfield  {journal}
  {\bibinfo  {journal} {Phys. Rev. Lett.}\ }\textbf {\bibinfo {volume} {115}},\
  \bibinfo {pages} {176401} (\bibinfo {year} {2015})}\BibitemShut {NoStop}%
\bibitem [{\citenamefont {Marzari}\ \emph {et~al.}(2012)\citenamefont
  {Marzari}, \citenamefont {Mostofi}, \citenamefont {Yates}, \citenamefont
  {Souza},\ and\ \citenamefont {Vanderbilt}}]{marzari2012}%
  \BibitemOpen
  \bibfield  {author} {\bibinfo {author} {\bibfnamefont {N.}~\bibnamefont
  {Marzari}}, \bibinfo {author} {\bibfnamefont {A.~A.}\ \bibnamefont
  {Mostofi}}, \bibinfo {author} {\bibfnamefont {J.~R.}\ \bibnamefont {Yates}},
  \bibinfo {author} {\bibfnamefont {I.}~\bibnamefont {Souza}},\ and\ \bibinfo
  {author} {\bibfnamefont {D.}~\bibnamefont {Vanderbilt}},\ }\bibfield  {title}
  {\bibinfo {title} {Maximally localized wannier functions: Theory and
  applications},\ }\href {https://doi.org/10.1103/RevModPhys.84.1419}
  {\bibfield  {journal} {\bibinfo  {journal} {Rev. Mod. Phys.}\ }\textbf
  {\bibinfo {volume} {84}},\ \bibinfo {pages} {1419} (\bibinfo {year}
  {2012})}\BibitemShut {NoStop}%
\bibitem [{\citenamefont {Zhou}\ \emph {et~al.}(2021)\citenamefont {Zhou},
  \citenamefont {Park}, \citenamefont {Lu}, \citenamefont {Maliyov},
  \citenamefont {Tong},\ and\ \citenamefont {Bernardi}}]{zhou2021}%
  \BibitemOpen
  \bibfield  {author} {\bibinfo {author} {\bibfnamefont {J.-J.}\ \bibnamefont
  {Zhou}}, \bibinfo {author} {\bibfnamefont {J.}~\bibnamefont {Park}}, \bibinfo
  {author} {\bibfnamefont {I.-T.}\ \bibnamefont {Lu}}, \bibinfo {author}
  {\bibfnamefont {I.}~\bibnamefont {Maliyov}}, \bibinfo {author} {\bibfnamefont
  {X.}~\bibnamefont {Tong}},\ and\ \bibinfo {author} {\bibfnamefont
  {M.}~\bibnamefont {Bernardi}},\ }\bibfield  {title} {\bibinfo {title}
  {Perturbo: A software package for ab initio electron–phonon interactions,
  charge transport and ultrafast dynamics},\ }\href
  {https://doi.org/https://doi.org/10.1016/j.cpc.2021.107970} {\bibfield
  {journal} {\bibinfo  {journal} {Comput. Phys. Commun.}\ }\textbf {\bibinfo
  {volume} {264}},\ \bibinfo {pages} {107970} (\bibinfo {year}
  {2021})}\BibitemShut {NoStop}%
\bibitem [{\citenamefont {Cepellotti}\ \emph {et~al.}(2021)\citenamefont
  {Cepellotti}, \citenamefont {Coulter}, \citenamefont {Johansson},
  \citenamefont {Fedorova},\ and\ \citenamefont {Kozinsky}}]{cepellotti2021}%
  \BibitemOpen
  \bibfield  {author} {\bibinfo {author} {\bibfnamefont {A.}~\bibnamefont
  {Cepellotti}}, \bibinfo {author} {\bibfnamefont {J.}~\bibnamefont {Coulter}},
  \bibinfo {author} {\bibfnamefont {A.}~\bibnamefont {Johansson}}, \bibinfo
  {author} {\bibfnamefont {N.~S.}\ \bibnamefont {Fedorova}},\ and\ \bibinfo
  {author} {\bibfnamefont {B.}~\bibnamefont {Kozinsky}},\ }\bibfield  {title}
  {\bibinfo {title} {Phoebe: a collection of phonon and electron boltzmann
  equation solvers},\ }\href {https://doi.org/10.48550/arXiv.2111.14999}
  {\bibfield  {journal} {\bibinfo  {journal} {arXiv:2111.14999}\ } (\bibinfo
  {year} {2021})}\BibitemShut {NoStop}%
\bibitem [{\citenamefont {Lloyd-Williams}\ and\ \citenamefont
  {Monserrat}(2015)}]{Tomeu-nondiagonal-supercells}%
  \BibitemOpen
  \bibfield  {author} {\bibinfo {author} {\bibfnamefont {J.~H.}\ \bibnamefont
  {Lloyd-Williams}}\ and\ \bibinfo {author} {\bibfnamefont {B.}~\bibnamefont
  {Monserrat}},\ }\bibfield  {title} {\bibinfo {title} {Lattice dynamics and
  electron-phonon coupling calculations using nondiagonal supercells},\ }\href
  {https://doi.org/10.1103/PhysRevB.92.184301} {\bibfield  {journal} {\bibinfo
  {journal} {Phys. Rev. B}\ }\textbf {\bibinfo {volume} {92}},\ \bibinfo
  {pages} {184301} (\bibinfo {year} {2015})}\BibitemShut {NoStop}%
\bibitem [{\citenamefont {Niven}\ \emph {et~al.}(1991)\citenamefont {Niven},
  \citenamefont {Zuckerman},\ and\ \citenamefont {Montgomery}}]{niven1991}%
  \BibitemOpen
  \bibfield  {author} {\bibinfo {author} {\bibfnamefont {I.}~\bibnamefont
  {Niven}}, \bibinfo {author} {\bibfnamefont {H.~S.}\ \bibnamefont
  {Zuckerman}},\ and\ \bibinfo {author} {\bibfnamefont {H.~L.}\ \bibnamefont
  {Montgomery}},\ }\href@noop {} {\emph {\bibinfo {title} {An introduction to
  the theory of numbers}}}\ (\bibinfo  {publisher} {John Wiley \& Sons},\
  \bibinfo {year} {1991})\ pp.\ \bibinfo {pages} {297--300}\BibitemShut
  {NoStop}%
\bibitem [{\citenamefont {Ledoan}(2018)}]{ledoan2018}%
  \BibitemOpen
  \bibfield  {author} {\bibinfo {author} {\bibfnamefont {A.}~\bibnamefont
  {Ledoan}},\ }\bibfield  {title} {\bibinfo {title} {The discrepancy of farey
  series},\ }\href {https://doi.org/10.1007/s10474-018-0868-x} {\bibfield
  {journal} {\bibinfo  {journal} {Acta Math. Hungar}\ }\textbf {\bibinfo
  {volume} {156}},\ \bibinfo {pages} {465} (\bibinfo {year}
  {2018})}\BibitemShut {NoStop}%
\bibitem [{\citenamefont {Kleinman}(1983)}]{Kleinman1983}%
  \BibitemOpen
  \bibfield  {author} {\bibinfo {author} {\bibfnamefont {L.}~\bibnamefont
  {Kleinman}},\ }\bibfield  {title} {\bibinfo {title} {Error in the tetrahedron
  integration scheme},\ }\href {https://doi.org/10.1103/PhysRevB.28.1139}
  {\bibfield  {journal} {\bibinfo  {journal} {Phys. Rev. B}\ }\textbf {\bibinfo
  {volume} {28}},\ \bibinfo {pages} {1139} (\bibinfo {year}
  {1983})}\BibitemShut {NoStop}%
\bibitem [{\citenamefont {Jepsen}\ and\ \citenamefont
  {Andersen}(1984)}]{Jepsen1984}%
  \BibitemOpen
  \bibfield  {author} {\bibinfo {author} {\bibfnamefont {O.}~\bibnamefont
  {Jepsen}}\ and\ \bibinfo {author} {\bibfnamefont {O.~K.}\ \bibnamefont
  {Andersen}},\ }\bibfield  {title} {\bibinfo {title} {No error in the
  tetrahedron integration scheme},\ }\href
  {https://doi.org/10.1103/PhysRevB.29.5965} {\bibfield  {journal} {\bibinfo
  {journal} {Phys. Rev. B}\ }\textbf {\bibinfo {volume} {29}},\ \bibinfo
  {pages} {5965} (\bibinfo {year} {1984})}\BibitemShut {NoStop}%
\bibitem [{\citenamefont {Bl\"ochl}\ \emph {et~al.}(1994)\citenamefont
  {Bl\"ochl}, \citenamefont {Jepsen},\ and\ \citenamefont
  {Andersen}}]{Blochl1994}%
  \BibitemOpen
  \bibfield  {author} {\bibinfo {author} {\bibfnamefont {P.~E.}\ \bibnamefont
  {Bl\"ochl}}, \bibinfo {author} {\bibfnamefont {O.}~\bibnamefont {Jepsen}},\
  and\ \bibinfo {author} {\bibfnamefont {O.~K.}\ \bibnamefont {Andersen}},\
  }\bibfield  {title} {\bibinfo {title} {Improved tetrahedron method for
  {B}rillouin-zone integrations},\ }\href
  {https://doi.org/10.1103/PhysRevB.49.16223} {\bibfield  {journal} {\bibinfo
  {journal} {Phys. Rev. B}\ }\textbf {\bibinfo {volume} {49}},\ \bibinfo
  {pages} {16223} (\bibinfo {year} {1994})}\BibitemShut {NoStop}%
\bibitem [{\citenamefont {Sibson}(1981)}]{sibson1981}%
  \BibitemOpen
  \bibfield  {author} {\bibinfo {author} {\bibfnamefont {R.}~\bibnamefont
  {Sibson}},\ }\bibfield  {title} {\bibinfo {title} {A brief description of
  natural neighbour interpolation},\ }\href@noop {} {\bibfield  {journal}
  {\bibinfo  {journal} {Interpreting multivariate data}\ } (\bibinfo {year}
  {1981})}\BibitemShut {NoStop}%
\bibitem [{\citenamefont {Hohenberg}\ and\ \citenamefont
  {Kohn}(1964)}]{DFT-Hohenberg-Kohn}%
  \BibitemOpen
  \bibfield  {author} {\bibinfo {author} {\bibfnamefont {P.}~\bibnamefont
  {Hohenberg}}\ and\ \bibinfo {author} {\bibfnamefont {W.}~\bibnamefont
  {Kohn}},\ }\bibfield  {title} {\bibinfo {title} {Inhomogeneous electron
  gas},\ }\href {https://doi.org/10.1103/PhysRev.136.B864} {\bibfield
  {journal} {\bibinfo  {journal} {Phys. Rev.}\ }\textbf {\bibinfo {volume}
  {136}},\ \bibinfo {pages} {B864} (\bibinfo {year} {1964})}\BibitemShut
  {NoStop}%
\bibitem [{\citenamefont {Kohn}\ and\ \citenamefont
  {Sham}(1965)}]{DFT-Kohn-Sham}%
  \BibitemOpen
  \bibfield  {author} {\bibinfo {author} {\bibfnamefont {W.}~\bibnamefont
  {Kohn}}\ and\ \bibinfo {author} {\bibfnamefont {L.~J.}\ \bibnamefont
  {Sham}},\ }\bibfield  {title} {\bibinfo {title} {Self-consistent equations
  including exchange and correlation effects},\ }\href
  {https://doi.org/10.1103/PhysRev.140.A1133} {\bibfield  {journal} {\bibinfo
  {journal} {Phys. Rev.}\ }\textbf {\bibinfo {volume} {140}},\ \bibinfo {pages}
  {A1133} (\bibinfo {year} {1965})}\BibitemShut {NoStop}%
\bibitem [{\citenamefont {Bl{\"o}chl}(1994)}]{VASP-PAW-One}%
  \BibitemOpen
  \bibfield  {author} {\bibinfo {author} {\bibfnamefont {P.~E.}\ \bibnamefont
  {Bl{\"o}chl}},\ }\bibfield  {title} {\bibinfo {title} {Projector
  augmented-wave method},\ }\href {https://doi.org/10.1103/PhysRevB.50.17953}
  {\bibfield  {journal} {\bibinfo  {journal} {Phys. Rev. B}\ }\textbf {\bibinfo
  {volume} {50}},\ \bibinfo {pages} {17953} (\bibinfo {year}
  {1994})}\BibitemShut {NoStop}%
\bibitem [{\citenamefont {Kresse}\ and\ \citenamefont
  {Joubert}(1999)}]{VASP-PAW-Two}%
  \BibitemOpen
  \bibfield  {author} {\bibinfo {author} {\bibfnamefont {G.}~\bibnamefont
  {Kresse}}\ and\ \bibinfo {author} {\bibfnamefont {D.}~\bibnamefont
  {Joubert}},\ }\bibfield  {title} {\bibinfo {title} {From ultrasoft
  pseudopotentials to the projector augmented-wave method},\ }\href
  {https://doi.org/10.1103/PhysRevB.59.1758} {\bibfield  {journal} {\bibinfo
  {journal} {Phys. Rev. B}\ }\textbf {\bibinfo {volume} {59}},\ \bibinfo
  {pages} {1758} (\bibinfo {year} {1999})}\BibitemShut {NoStop}%
\bibitem [{\citenamefont {Clark}\ \emph {et~al.}(2005)\citenamefont {Clark},
  \citenamefont {Segall}, \citenamefont {Pickard}, \citenamefont {Hasnip},
  \citenamefont {Probert}, \citenamefont {Refson},\ and\ \citenamefont
  {Payne}}]{CASTEP}%
  \BibitemOpen
  \bibfield  {author} {\bibinfo {author} {\bibfnamefont {S.~J.}\ \bibnamefont
  {Clark}}, \bibinfo {author} {\bibfnamefont {M.~D.}\ \bibnamefont {Segall}},
  \bibinfo {author} {\bibfnamefont {C.~J.}\ \bibnamefont {Pickard}}, \bibinfo
  {author} {\bibfnamefont {P.~J.}\ \bibnamefont {Hasnip}}, \bibinfo {author}
  {\bibfnamefont {M.~I.~J.}\ \bibnamefont {Probert}}, \bibinfo {author}
  {\bibfnamefont {K.}~\bibnamefont {Refson}},\ and\ \bibinfo {author}
  {\bibfnamefont {M.~C.}\ \bibnamefont {Payne}},\ }\bibfield  {title} {\bibinfo
  {title} {First principles methods using {CASTEP}},\ }\href
  {https://doi.org/https://doi.org/10.1524/zkri.220.5.567.65075} {\bibfield
  {journal} {\bibinfo  {journal} {Z. Kristallogr. Cryst. Mater.}\ }\textbf
  {\bibinfo {volume} {220}},\ \bibinfo {pages} {567 } (\bibinfo {year}
  {2005})}\BibitemShut {NoStop}%
\bibitem [{\citenamefont {Hamann}\ \emph {et~al.}(1979)\citenamefont {Hamann},
  \citenamefont {Schl\"uter},\ and\ \citenamefont {Chiang}}]{nc_pseudo}%
  \BibitemOpen
  \bibfield  {author} {\bibinfo {author} {\bibfnamefont {D.~R.}\ \bibnamefont
  {Hamann}}, \bibinfo {author} {\bibfnamefont {M.}~\bibnamefont {Schl\"uter}},\
  and\ \bibinfo {author} {\bibfnamefont {C.}~\bibnamefont {Chiang}},\
  }\bibfield  {title} {\bibinfo {title} {Norm-conserving pseudopotentials},\
  }\href {http://link.aps.org/doi/10.1103/PhysRevLett.43.1494} {\bibfield
  {journal} {\bibinfo  {journal} {Phys. Rev. Lett.}\ }\textbf {\bibinfo
  {volume} {43}},\ \bibinfo {pages} {1494} (\bibinfo {year}
  {1979})}\BibitemShut {NoStop}%
\bibitem [{\citenamefont {Parlinski}\ \emph {et~al.}(1997)\citenamefont
  {Parlinski}, \citenamefont {Li},\ and\ \citenamefont
  {Kawazoe}}]{Parlinski1997}%
  \BibitemOpen
  \bibfield  {author} {\bibinfo {author} {\bibfnamefont {K.}~\bibnamefont
  {Parlinski}}, \bibinfo {author} {\bibfnamefont {Z.~Q.}\ \bibnamefont {Li}},\
  and\ \bibinfo {author} {\bibfnamefont {Y.}~\bibnamefont {Kawazoe}},\
  }\bibfield  {title} {\bibinfo {title} {First-principles determination of the
  soft mode in cubic ${\mathrm{zro}}_{2}$},\ }\href
  {https://doi.org/10.1103/PhysRevLett.78.4063} {\bibfield  {journal} {\bibinfo
   {journal} {Phys. Rev. Lett.}\ }\textbf {\bibinfo {volume} {78}},\ \bibinfo
  {pages} {4063} (\bibinfo {year} {1997})}\BibitemShut {NoStop}%
\bibitem [{\citenamefont {Baroni}\ \emph {et~al.}(2001)\citenamefont {Baroni},
  \citenamefont {de~Gironcoli}, \citenamefont {Dal~Corso},\ and\ \citenamefont
  {Giannozzi}}]{baroni2001}%
  \BibitemOpen
  \bibfield  {author} {\bibinfo {author} {\bibfnamefont {S.}~\bibnamefont
  {Baroni}}, \bibinfo {author} {\bibfnamefont {S.}~\bibnamefont
  {de~Gironcoli}}, \bibinfo {author} {\bibfnamefont {A.}~\bibnamefont
  {Dal~Corso}},\ and\ \bibinfo {author} {\bibfnamefont {P.}~\bibnamefont
  {Giannozzi}},\ }\bibfield  {title} {\bibinfo {title} {Phonons and related
  crystal properties from density-functional perturbation theory},\ }\href
  {https://doi.org/10.1103/RevModPhys.73.515} {\bibfield  {journal} {\bibinfo
  {journal} {Rev. Mod. Phys.}\ }\textbf {\bibinfo {volume} {73}},\ \bibinfo
  {pages} {515} (\bibinfo {year} {2001})}\BibitemShut {NoStop}%
\bibitem [{\citenamefont {Perdew}\ and\ \citenamefont
  {Zunger}(1981)}]{Perdew-Zunger-LDA}%
  \BibitemOpen
  \bibfield  {author} {\bibinfo {author} {\bibfnamefont {J.~P.}\ \bibnamefont
  {Perdew}}\ and\ \bibinfo {author} {\bibfnamefont {A.}~\bibnamefont
  {Zunger}},\ }\bibfield  {title} {\bibinfo {title} {Self-interaction
  correction to density-functional approximations for many-electron systems},\
  }\href {https://link.aps.org/doi/10.1103/PhysRevB.23.5048} {\bibfield
  {journal} {\bibinfo  {journal} {Phys. Rev. B}\ }\textbf {\bibinfo {volume}
  {23}},\ \bibinfo {pages} {5048} (\bibinfo {year} {1981})}\BibitemShut
  {NoStop}%
\bibitem [{\citenamefont {Perdew}\ \emph {et~al.}(1996)\citenamefont {Perdew},
  \citenamefont {Burke},\ and\ \citenamefont
  {Ernzerhof}}]{PBE-exchange-correlation}%
  \BibitemOpen
  \bibfield  {author} {\bibinfo {author} {\bibfnamefont {J.~P.}\ \bibnamefont
  {Perdew}}, \bibinfo {author} {\bibfnamefont {K.}~\bibnamefont {Burke}},\ and\
  \bibinfo {author} {\bibfnamefont {M.}~\bibnamefont {Ernzerhof}},\ }\bibfield
  {title} {\bibinfo {title} {Generalized gradient approximation made simple},\
  }\href {https://doi.org/10.1103/PhysRevLett.77.3865} {\bibfield  {journal}
  {\bibinfo  {journal} {Phys. Rev. Lett.}\ }\textbf {\bibinfo {volume} {77}},\
  \bibinfo {pages} {3865} (\bibinfo {year} {1996})}\BibitemShut {NoStop}%
\bibitem [{\citenamefont {Kohn}(1959)}]{kohn1959}%
  \BibitemOpen
  \bibfield  {author} {\bibinfo {author} {\bibfnamefont {W.}~\bibnamefont
  {Kohn}},\ }\bibfield  {title} {\bibinfo {title} {Image of the fermi surface
  in the vibration spectrum of a metal},\ }\href
  {https://doi.org/10.1103/PhysRevLett.2.393} {\bibfield  {journal} {\bibinfo
  {journal} {Phys. Rev. Lett.}\ }\textbf {\bibinfo {volume} {2}},\ \bibinfo
  {pages} {393} (\bibinfo {year} {1959})}\BibitemShut {NoStop}%
\bibitem [{\citenamefont {Maultzsch}\ \emph {et~al.}(2004)\citenamefont
  {Maultzsch}, \citenamefont {Reich}, \citenamefont {Thomsen}, \citenamefont
  {Requardt},\ and\ \citenamefont {Ordej\'on}}]{maultzsch2004}%
  \BibitemOpen
  \bibfield  {author} {\bibinfo {author} {\bibfnamefont {J.}~\bibnamefont
  {Maultzsch}}, \bibinfo {author} {\bibfnamefont {S.}~\bibnamefont {Reich}},
  \bibinfo {author} {\bibfnamefont {C.}~\bibnamefont {Thomsen}}, \bibinfo
  {author} {\bibfnamefont {H.}~\bibnamefont {Requardt}},\ and\ \bibinfo
  {author} {\bibfnamefont {P.}~\bibnamefont {Ordej\'on}},\ }\bibfield  {title}
  {\bibinfo {title} {Phonon dispersion in graphite},\ }\href
  {https://doi.org/10.1103/PhysRevLett.92.075501} {\bibfield  {journal}
  {\bibinfo  {journal} {Phys. Rev. Lett.}\ }\textbf {\bibinfo {volume} {92}},\
  \bibinfo {pages} {075501} (\bibinfo {year} {2004})}\BibitemShut {NoStop}%
\bibitem [{\citenamefont {Mohr}\ \emph {et~al.}(2007)\citenamefont {Mohr},
  \citenamefont {Maultzsch}, \citenamefont {Dobard\ifmmode \check{z}\else
  \v{z}\fi{}i\ifmmode~\acute{c}\else \'{c}\fi{}}, \citenamefont {Reich},
  \citenamefont {Milo\ifmmode \check{s}\else
  \v{s}\fi{}evi\ifmmode~\acute{c}\else \'{c}\fi{}}, \citenamefont
  {Damnjanovi\ifmmode~\acute{c}\else \'{c}\fi{}}, \citenamefont {Bosak},
  \citenamefont {Krisch},\ and\ \citenamefont {Thomsen}}]{mohr2007}%
  \BibitemOpen
  \bibfield  {author} {\bibinfo {author} {\bibfnamefont {M.}~\bibnamefont
  {Mohr}}, \bibinfo {author} {\bibfnamefont {J.}~\bibnamefont {Maultzsch}},
  \bibinfo {author} {\bibfnamefont {E.}~\bibnamefont {Dobard\ifmmode
  \check{z}\else \v{z}\fi{}i\ifmmode~\acute{c}\else \'{c}\fi{}}}, \bibinfo
  {author} {\bibfnamefont {S.}~\bibnamefont {Reich}}, \bibinfo {author}
  {\bibfnamefont {I.}~\bibnamefont {Milo\ifmmode \check{s}\else
  \v{s}\fi{}evi\ifmmode~\acute{c}\else \'{c}\fi{}}}, \bibinfo {author}
  {\bibfnamefont {M.}~\bibnamefont {Damnjanovi\ifmmode~\acute{c}\else
  \'{c}\fi{}}}, \bibinfo {author} {\bibfnamefont {A.}~\bibnamefont {Bosak}},
  \bibinfo {author} {\bibfnamefont {M.}~\bibnamefont {Krisch}},\ and\ \bibinfo
  {author} {\bibfnamefont {C.}~\bibnamefont {Thomsen}},\ }\bibfield  {title}
  {\bibinfo {title} {Phonon dispersion of graphite by inelastic {X}-ray
  scattering},\ }\href {https://doi.org/10.1103/PhysRevB.76.035439} {\bibfield
  {journal} {\bibinfo  {journal} {Phys. Rev. B}\ }\textbf {\bibinfo {volume}
  {76}},\ \bibinfo {pages} {035439} (\bibinfo {year} {2007})}\BibitemShut
  {NoStop}%
\bibitem [{\citenamefont {S\'anchez-Portal}\ \emph {et~al.}(1999)\citenamefont
  {S\'anchez-Portal}, \citenamefont {Artacho}, \citenamefont {Soler},
  \citenamefont {Rubio},\ and\ \citenamefont {Ordej\'on}}]{portal1999}%
  \BibitemOpen
  \bibfield  {author} {\bibinfo {author} {\bibfnamefont {D.}~\bibnamefont
  {S\'anchez-Portal}}, \bibinfo {author} {\bibfnamefont {E.}~\bibnamefont
  {Artacho}}, \bibinfo {author} {\bibfnamefont {J.~M.}\ \bibnamefont {Soler}},
  \bibinfo {author} {\bibfnamefont {A.}~\bibnamefont {Rubio}},\ and\ \bibinfo
  {author} {\bibfnamefont {P.}~\bibnamefont {Ordej\'on}},\ }\bibfield  {title}
  {\bibinfo {title} {Ab initio structural, elastic, and vibrational properties
  of carbon nanotubes},\ }\href {https://doi.org/10.1103/PhysRevB.59.12678}
  {\bibfield  {journal} {\bibinfo  {journal} {Phys. Rev. B}\ }\textbf {\bibinfo
  {volume} {59}},\ \bibinfo {pages} {12678} (\bibinfo {year}
  {1999})}\BibitemShut {NoStop}%
\bibitem [{\citenamefont {Dubay}\ and\ \citenamefont
  {Kresse}(2003)}]{dubay2003}%
  \BibitemOpen
  \bibfield  {author} {\bibinfo {author} {\bibfnamefont {O.}~\bibnamefont
  {Dubay}}\ and\ \bibinfo {author} {\bibfnamefont {G.}~\bibnamefont {Kresse}},\
  }\bibfield  {title} {\bibinfo {title} {Accurate density functional
  calculations for the phonon dispersion relations of graphite layer and carbon
  nanotubes},\ }\href {https://doi.org/10.1103/PhysRevB.67.035401} {\bibfield
  {journal} {\bibinfo  {journal} {Phys. Rev. B}\ }\textbf {\bibinfo {volume}
  {67}},\ \bibinfo {pages} {035401} (\bibinfo {year} {2003})}\BibitemShut
  {NoStop}%
\bibitem [{\citenamefont {Yanagisawa}\ \emph {et~al.}(2005)\citenamefont
  {Yanagisawa}, \citenamefont {Tanaka}, \citenamefont {Ishida}, \citenamefont
  {Matsue}, \citenamefont {Rokuta}, \citenamefont {Otani},\ and\ \citenamefont
  {Oshima}}]{yanagisawa2005}%
  \BibitemOpen
  \bibfield  {author} {\bibinfo {author} {\bibfnamefont {H.}~\bibnamefont
  {Yanagisawa}}, \bibinfo {author} {\bibfnamefont {T.}~\bibnamefont {Tanaka}},
  \bibinfo {author} {\bibfnamefont {Y.}~\bibnamefont {Ishida}}, \bibinfo
  {author} {\bibfnamefont {M.}~\bibnamefont {Matsue}}, \bibinfo {author}
  {\bibfnamefont {E.}~\bibnamefont {Rokuta}}, \bibinfo {author} {\bibfnamefont
  {S.}~\bibnamefont {Otani}},\ and\ \bibinfo {author} {\bibfnamefont
  {C.}~\bibnamefont {Oshima}},\ }\bibfield  {title} {\bibinfo {title} {Analysis
  of phonons in graphene sheets by means of hreels measurement and ab initio
  calculation},\ }\href {https://doi.org/https://doi.org/10.1002/sia.1948}
  {\bibfield  {journal} {\bibinfo  {journal} {Surf. Interface Anal.}\ }\textbf
  {\bibinfo {volume} {37}},\ \bibinfo {pages} {133} (\bibinfo {year}
  {2005})}\BibitemShut {NoStop}%
\bibitem [{\citenamefont {Mounet}\ and\ \citenamefont
  {Marzari}(2005)}]{mounet2005}%
  \BibitemOpen
  \bibfield  {author} {\bibinfo {author} {\bibfnamefont {N.}~\bibnamefont
  {Mounet}}\ and\ \bibinfo {author} {\bibfnamefont {N.}~\bibnamefont
  {Marzari}},\ }\bibfield  {title} {\bibinfo {title} {First-principles
  determination of the structural, vibrational and thermodynamic properties of
  diamond, graphite, and derivatives},\ }\href
  {https://doi.org/10.1103/PhysRevB.71.205214} {\bibfield  {journal} {\bibinfo
  {journal} {Phys. Rev. B}\ }\textbf {\bibinfo {volume} {71}},\ \bibinfo
  {pages} {205214} (\bibinfo {year} {2005})}\BibitemShut {NoStop}%
\bibitem [{\citenamefont {Zhang}\ \emph {et~al.}(2011)\citenamefont {Zhang},
  \citenamefont {Lee},\ and\ \citenamefont {Cho}}]{zhang2011}%
  \BibitemOpen
  \bibfield  {author} {\bibinfo {author} {\bibfnamefont {H.}~\bibnamefont
  {Zhang}}, \bibinfo {author} {\bibfnamefont {G.}~\bibnamefont {Lee}},\ and\
  \bibinfo {author} {\bibfnamefont {K.}~\bibnamefont {Cho}},\ }\bibfield
  {title} {\bibinfo {title} {Thermal transport in graphene and effects of
  vacancy defects},\ }\href {https://doi.org/10.1103/PhysRevB.84.115460}
  {\bibfield  {journal} {\bibinfo  {journal} {Phys. Rev. B}\ }\textbf {\bibinfo
  {volume} {84}},\ \bibinfo {pages} {115460} (\bibinfo {year}
  {2011})}\BibitemShut {NoStop}%
\bibitem [{\citenamefont {Marquina}\ \emph {et~al.}(2013)\citenamefont
  {Marquina}, \citenamefont {Power}, \citenamefont {Gonzalez},\ and\
  \citenamefont {Broto}}]{marquina2013}%
  \BibitemOpen
  \bibfield  {author} {\bibinfo {author} {\bibfnamefont {J.}~\bibnamefont
  {Marquina}}, \bibinfo {author} {\bibfnamefont {C.}~\bibnamefont {Power}},
  \bibinfo {author} {\bibfnamefont {J.}~\bibnamefont {Gonzalez}},\ and\
  \bibinfo {author} {\bibfnamefont {J.-M.}\ \bibnamefont {Broto}},\ }\bibfield
  {title} {\bibinfo {title} {Ab initio study of the electronic and vibrational
  properties of 1-nm-diameter single-walled nanotubes},\ }\href
  {https://doi.org/10.4236/ampc.2013.32025} {\bibfield  {journal} {\bibinfo
  {journal} {Adv. Mater. Phys. Chem.}\ }\textbf {\bibinfo {volume} {03}},\
  \bibinfo {pages} {178} (\bibinfo {year} {2013})}\BibitemShut {NoStop}%
\bibitem [{\citenamefont {Slotman}\ \emph {et~al.}(2014)\citenamefont
  {Slotman}, \citenamefont {de~Wijs}, \citenamefont {Fasolino},\ and\
  \citenamefont {Katsnelson}}]{slotman2014}%
  \BibitemOpen
  \bibfield  {author} {\bibinfo {author} {\bibfnamefont {G.~J.}\ \bibnamefont
  {Slotman}}, \bibinfo {author} {\bibfnamefont {G.~A.}\ \bibnamefont
  {de~Wijs}}, \bibinfo {author} {\bibfnamefont {A.}~\bibnamefont {Fasolino}},\
  and\ \bibinfo {author} {\bibfnamefont {M.~I.}\ \bibnamefont {Katsnelson}},\
  }\bibfield  {title} {\bibinfo {title} {Phonons and electron-phonon coupling
  in graphene-{h-BN} heterostructures},\ }\href
  {https://doi.org/https://doi.org/10.1002/andp.201400155} {\bibfield
  {journal} {\bibinfo  {journal} {Ann. Phys.}\ }\textbf {\bibinfo {volume}
  {526}},\ \bibinfo {pages} {381} (\bibinfo {year} {2014})}\BibitemShut
  {NoStop}%
\bibitem [{\citenamefont {Gu}\ and\ \citenamefont {Yang}(2015)}]{gu2015}%
  \BibitemOpen
  \bibfield  {author} {\bibinfo {author} {\bibfnamefont {X.}~\bibnamefont
  {Gu}}\ and\ \bibinfo {author} {\bibfnamefont {R.}~\bibnamefont {Yang}},\
  }\bibfield  {title} {\bibinfo {title} {First-principles prediction of
  phononic thermal conductivity of silicene: A comparison with graphene},\
  }\href {https://doi.org/10.1063/1.4905540} {\bibfield  {journal} {\bibinfo
  {journal} {J. Appl. Phys.}\ }\textbf {\bibinfo {volume} {117}},\ \bibinfo
  {pages} {025102} (\bibinfo {year} {2015})}\BibitemShut {NoStop}%
\bibitem [{\citenamefont {Diery}\ \emph {et~al.}(2018)\citenamefont {Diery},
  \citenamefont {Moujaes},\ and\ \citenamefont {Nunes}}]{diery2018}%
  \BibitemOpen
  \bibfield  {author} {\bibinfo {author} {\bibfnamefont {W.}~\bibnamefont
  {Diery}}, \bibinfo {author} {\bibfnamefont {E.~A.}\ \bibnamefont {Moujaes}},\
  and\ \bibinfo {author} {\bibfnamefont {R.}~\bibnamefont {Nunes}},\ }\bibfield
   {title} {\bibinfo {title} {Nature of localized phonon modes of tilt grain
  boundaries in graphene},\ }\href
  {https://doi.org/https://doi.org/10.1016/j.carbon.2018.08.045} {\bibfield
  {journal} {\bibinfo  {journal} {Carbon}\ }\textbf {\bibinfo {volume} {140}},\
  \bibinfo {pages} {250} (\bibinfo {year} {2018})}\BibitemShut {NoStop}%
\bibitem [{\citenamefont {Nagamatsu}\ \emph {et~al.}(2001)\citenamefont
  {Nagamatsu}, \citenamefont {Nakagawa}, \citenamefont {Muranaka},
  \citenamefont {Zenitani},\ and\ \citenamefont {Akimitsu}}]{nagamatsu2001}%
  \BibitemOpen
  \bibfield  {author} {\bibinfo {author} {\bibfnamefont {J.}~\bibnamefont
  {Nagamatsu}}, \bibinfo {author} {\bibfnamefont {N.}~\bibnamefont {Nakagawa}},
  \bibinfo {author} {\bibfnamefont {T.}~\bibnamefont {Muranaka}}, \bibinfo
  {author} {\bibfnamefont {Y.}~\bibnamefont {Zenitani}},\ and\ \bibinfo
  {author} {\bibfnamefont {J.}~\bibnamefont {Akimitsu}},\ }\bibfield  {title}
  {\bibinfo {title} {Superconductivity at 39 {K} in magnesium diboride},\
  }\href {https://doi.org/10.1038/35065039} {\bibfield  {journal} {\bibinfo
  {journal} {Nature}\ }\textbf {\bibinfo {volume} {410}},\ \bibinfo {pages}
  {63} (\bibinfo {year} {2001})}\BibitemShut {NoStop}%
\bibitem [{\citenamefont {Kong}\ \emph {et~al.}(2001)\citenamefont {Kong},
  \citenamefont {Dolgov}, \citenamefont {Jepsen},\ and\ \citenamefont
  {Andersen}}]{kong2001}%
  \BibitemOpen
  \bibfield  {author} {\bibinfo {author} {\bibfnamefont {Y.}~\bibnamefont
  {Kong}}, \bibinfo {author} {\bibfnamefont {O.~V.}\ \bibnamefont {Dolgov}},
  \bibinfo {author} {\bibfnamefont {O.}~\bibnamefont {Jepsen}},\ and\ \bibinfo
  {author} {\bibfnamefont {O.~K.}\ \bibnamefont {Andersen}},\ }\bibfield
  {title} {\bibinfo {title} {Electron-phonon interaction in the normal and
  superconducting states of {MgB}$_{2}$},\ }\href
  {https://doi.org/10.1103/PhysRevB.64.020501} {\bibfield  {journal} {\bibinfo
  {journal} {Phys. Rev. B}\ }\textbf {\bibinfo {volume} {64}},\ \bibinfo
  {pages} {020501(R)} (\bibinfo {year} {2001})}\BibitemShut {NoStop}%
\bibitem [{\citenamefont {Bohnen}\ \emph {et~al.}(2001)\citenamefont {Bohnen},
  \citenamefont {Heid},\ and\ \citenamefont {Renker}}]{bohnen2001}%
  \BibitemOpen
  \bibfield  {author} {\bibinfo {author} {\bibfnamefont {K.-P.}\ \bibnamefont
  {Bohnen}}, \bibinfo {author} {\bibfnamefont {R.}~\bibnamefont {Heid}},\ and\
  \bibinfo {author} {\bibfnamefont {B.}~\bibnamefont {Renker}},\ }\bibfield
  {title} {\bibinfo {title} {Phonon dispersion and electron-phonon coupling in
  {M}g{B}$_{2}$ and {A}l{B}$_{2}$},\ }\href
  {https://doi.org/10.1103/PhysRevLett.86.5771} {\bibfield  {journal} {\bibinfo
   {journal} {Phys. Rev. Lett.}\ }\textbf {\bibinfo {volume} {86}},\ \bibinfo
  {pages} {5771} (\bibinfo {year} {2001})}\BibitemShut {NoStop}%
\bibitem [{\citenamefont {d'Astuto}\ \emph {et~al.}(2007)\citenamefont
  {d'Astuto}, \citenamefont {Calandra}, \citenamefont {Reich}, \citenamefont
  {Shukla}, \citenamefont {Lazzeri}, \citenamefont {Mauri}, \citenamefont
  {Karpinski}, \citenamefont {Zhigadlo}, \citenamefont {Bossak},\ and\
  \citenamefont {Krisch}}]{matteo2007}%
  \BibitemOpen
  \bibfield  {author} {\bibinfo {author} {\bibfnamefont {M.}~\bibnamefont
  {d'Astuto}}, \bibinfo {author} {\bibfnamefont {M.}~\bibnamefont {Calandra}},
  \bibinfo {author} {\bibfnamefont {S.}~\bibnamefont {Reich}}, \bibinfo
  {author} {\bibfnamefont {A.}~\bibnamefont {Shukla}}, \bibinfo {author}
  {\bibfnamefont {M.}~\bibnamefont {Lazzeri}}, \bibinfo {author} {\bibfnamefont
  {F.}~\bibnamefont {Mauri}}, \bibinfo {author} {\bibfnamefont
  {J.}~\bibnamefont {Karpinski}}, \bibinfo {author} {\bibfnamefont {N.~D.}\
  \bibnamefont {Zhigadlo}}, \bibinfo {author} {\bibfnamefont {A.}~\bibnamefont
  {Bossak}},\ and\ \bibinfo {author} {\bibfnamefont {M.}~\bibnamefont
  {Krisch}},\ }\bibfield  {title} {\bibinfo {title} {Weak anharmonic effects in
  $\mathrm{Mg}{\mathrm{b}}_{2}$: A comparative inelastic {X}-ray scattering and
  raman study},\ }\href {https://doi.org/10.1103/PhysRevB.75.174508} {\bibfield
   {journal} {\bibinfo  {journal} {Phys. Rev. B}\ }\textbf {\bibinfo {volume}
  {75}},\ \bibinfo {pages} {174508} (\bibinfo {year} {2007})}\BibitemShut
  {NoStop}%
\bibitem [{\citenamefont {Alarco}\ \emph {et~al.}(2014)\citenamefont {Alarco},
  \citenamefont {Chou}, \citenamefont {Talbot},\ and\ \citenamefont
  {Mackinnon}}]{alarco2014}%
  \BibitemOpen
  \bibfield  {author} {\bibinfo {author} {\bibfnamefont {J.~A.}\ \bibnamefont
  {Alarco}}, \bibinfo {author} {\bibfnamefont {A.}~\bibnamefont {Chou}},
  \bibinfo {author} {\bibfnamefont {P.~C.}\ \bibnamefont {Talbot}},\ and\
  \bibinfo {author} {\bibfnamefont {I.~D.}\ \bibnamefont {Mackinnon}},\
  }\bibfield  {title} {\bibinfo {title} {Phonon modes of {MgB}$_{2}$:
  super-lattice structures and spectral response},\ }\href
  {https://doi.org/10.1039/C4CP03449J} {\bibfield  {journal} {\bibinfo
  {journal} {Phys. Chem. Chem. Phys.}\ }\textbf {\bibinfo {volume} {16}},\
  \bibinfo {pages} {24443} (\bibinfo {year} {2014})}\BibitemShut {NoStop}%
\bibitem [{\citenamefont {Mackinnon}\ \emph {et~al.}(2017)\citenamefont
  {Mackinnon}, \citenamefont {Talbot},\ and\ \citenamefont
  {Alarco}}]{mackinnon2017}%
  \BibitemOpen
  \bibfield  {author} {\bibinfo {author} {\bibfnamefont {I.~D.}\ \bibnamefont
  {Mackinnon}}, \bibinfo {author} {\bibfnamefont {P.~C.}\ \bibnamefont
  {Talbot}},\ and\ \bibinfo {author} {\bibfnamefont {J.~A.}\ \bibnamefont
  {Alarco}},\ }\bibfield  {title} {\bibinfo {title} {Phonon dispersion
  anomalies and superconductivity in metal substituted {MgB}$_{2}$},\ }\href
  {https://doi.org/https://doi.org/10.1016/j.commatsci.2017.01.011} {\bibfield
  {journal} {\bibinfo  {journal} {Comput. Mater. Sci.}\ }\textbf {\bibinfo
  {volume} {130}},\ \bibinfo {pages} {191} (\bibinfo {year}
  {2017})}\BibitemShut {NoStop}%
\bibitem [{\citenamefont {Shukla}\ \emph {et~al.}(2003)\citenamefont {Shukla},
  \citenamefont {Calandra}, \citenamefont {d'Astuto}, \citenamefont {Lazzeri},
  \citenamefont {Mauri}, \citenamefont {Bellin}, \citenamefont {Krisch},
  \citenamefont {Karpinski}, \citenamefont {Kazakov}, \citenamefont {Jun},
  \citenamefont {Daghero},\ and\ \citenamefont {Parlinski}}]{shukla2003}%
  \BibitemOpen
  \bibfield  {author} {\bibinfo {author} {\bibfnamefont {A.}~\bibnamefont
  {Shukla}}, \bibinfo {author} {\bibfnamefont {M.}~\bibnamefont {Calandra}},
  \bibinfo {author} {\bibfnamefont {M.}~\bibnamefont {d'Astuto}}, \bibinfo
  {author} {\bibfnamefont {M.}~\bibnamefont {Lazzeri}}, \bibinfo {author}
  {\bibfnamefont {F.}~\bibnamefont {Mauri}}, \bibinfo {author} {\bibfnamefont
  {C.}~\bibnamefont {Bellin}}, \bibinfo {author} {\bibfnamefont
  {M.}~\bibnamefont {Krisch}}, \bibinfo {author} {\bibfnamefont
  {J.}~\bibnamefont {Karpinski}}, \bibinfo {author} {\bibfnamefont {S.~M.}\
  \bibnamefont {Kazakov}}, \bibinfo {author} {\bibfnamefont {J.}~\bibnamefont
  {Jun}}, \bibinfo {author} {\bibfnamefont {D.}~\bibnamefont {Daghero}},\ and\
  \bibinfo {author} {\bibfnamefont {K.}~\bibnamefont {Parlinski}},\ }\bibfield
  {title} {\bibinfo {title} {Phonon dispersion and lifetimes in {MgB}$_2$},\
  }\href {https://doi.org/10.1103/PhysRevLett.90.095506} {\bibfield  {journal}
  {\bibinfo  {journal} {Phys. Rev. Lett.}\ }\textbf {\bibinfo {volume} {90}},\
  \bibinfo {pages} {095506} (\bibinfo {year} {2003})}\BibitemShut {NoStop}%
\bibitem [{\citenamefont {Yildirim}\ \emph {et~al.}(2001)\citenamefont
  {Yildirim}, \citenamefont {G\"ulseren}, \citenamefont {Lynn}, \citenamefont
  {Brown}, \citenamefont {Udovic}, \citenamefont {Huang}, \citenamefont
  {Rogado}, \citenamefont {Regan}, \citenamefont {Hayward}, \citenamefont
  {Slusky}, \citenamefont {He}, \citenamefont {Haas}, \citenamefont {Khalifah},
  \citenamefont {Inumaru},\ and\ \citenamefont {Cava}}]{yildirim2001}%
  \BibitemOpen
  \bibfield  {author} {\bibinfo {author} {\bibfnamefont {T.}~\bibnamefont
  {Yildirim}}, \bibinfo {author} {\bibfnamefont {O.}~\bibnamefont
  {G\"ulseren}}, \bibinfo {author} {\bibfnamefont {J.~W.}\ \bibnamefont
  {Lynn}}, \bibinfo {author} {\bibfnamefont {C.~M.}\ \bibnamefont {Brown}},
  \bibinfo {author} {\bibfnamefont {T.~J.}\ \bibnamefont {Udovic}}, \bibinfo
  {author} {\bibfnamefont {Q.}~\bibnamefont {Huang}}, \bibinfo {author}
  {\bibfnamefont {N.}~\bibnamefont {Rogado}}, \bibinfo {author} {\bibfnamefont
  {K.~A.}\ \bibnamefont {Regan}}, \bibinfo {author} {\bibfnamefont {M.~A.}\
  \bibnamefont {Hayward}}, \bibinfo {author} {\bibfnamefont {J.~S.}\
  \bibnamefont {Slusky}}, \bibinfo {author} {\bibfnamefont {T.}~\bibnamefont
  {He}}, \bibinfo {author} {\bibfnamefont {M.~K.}\ \bibnamefont {Haas}},
  \bibinfo {author} {\bibfnamefont {P.}~\bibnamefont {Khalifah}}, \bibinfo
  {author} {\bibfnamefont {K.}~\bibnamefont {Inumaru}},\ and\ \bibinfo {author}
  {\bibfnamefont {R.~J.}\ \bibnamefont {Cava}},\ }\bibfield  {title} {\bibinfo
  {title} {Giant anharmonicity and nonlinear electron-phonon coupling in
  {MgB}$_{2}$: A combined first-principles calculation and neutron scattering
  study},\ }\href {https://doi.org/10.1103/PhysRevLett.87.037001} {\bibfield
  {journal} {\bibinfo  {journal} {Phys. Rev. Lett.}\ }\textbf {\bibinfo
  {volume} {87}},\ \bibinfo {pages} {037001} (\bibinfo {year}
  {2001})}\BibitemShut {NoStop}%
\bibitem [{\citenamefont {Baron}\ \emph {et~al.}(2004)\citenamefont {Baron},
  \citenamefont {Uchiyama}, \citenamefont {Tanaka}, \citenamefont {Tsutsui},
  \citenamefont {Ishikawa}, \citenamefont {Lee}, \citenamefont {Heid},
  \citenamefont {Bohnen}, \citenamefont {Tajima},\ and\ \citenamefont
  {Ishikawa}}]{baron2004}%
  \BibitemOpen
  \bibfield  {author} {\bibinfo {author} {\bibfnamefont {A.~Q.~R.}\
  \bibnamefont {Baron}}, \bibinfo {author} {\bibfnamefont {H.}~\bibnamefont
  {Uchiyama}}, \bibinfo {author} {\bibfnamefont {Y.}~\bibnamefont {Tanaka}},
  \bibinfo {author} {\bibfnamefont {S.}~\bibnamefont {Tsutsui}}, \bibinfo
  {author} {\bibfnamefont {D.}~\bibnamefont {Ishikawa}}, \bibinfo {author}
  {\bibfnamefont {S.}~\bibnamefont {Lee}}, \bibinfo {author} {\bibfnamefont
  {R.}~\bibnamefont {Heid}}, \bibinfo {author} {\bibfnamefont {K.-P.}\
  \bibnamefont {Bohnen}}, \bibinfo {author} {\bibfnamefont {S.}~\bibnamefont
  {Tajima}},\ and\ \bibinfo {author} {\bibfnamefont {T.}~\bibnamefont
  {Ishikawa}},\ }\bibfield  {title} {\bibinfo {title} {Kohn anomaly in
  {MgB}$_2$ by inelastic {X}-ray scattering},\ }\href
  {https://doi.org/10.1103/PhysRevLett.92.197004} {\bibfield  {journal}
  {\bibinfo  {journal} {Phys. Rev. Lett.}\ }\textbf {\bibinfo {volume} {92}},\
  \bibinfo {pages} {197004} (\bibinfo {year} {2004})}\BibitemShut {NoStop}%
\bibitem [{\citenamefont {Pe{\v{s}}i{\'c}}\ \emph {et~al.}(2019)\citenamefont
  {Pe{\v{s}}i{\'c}}, \citenamefont {Popov}, \citenamefont {{\v{S}}olaji{\'c}},
  \citenamefont {Damljanovi{\'c}}, \citenamefont {Hingerl}, \citenamefont
  {Beli{\'c}},\ and\ \citenamefont {Gaji{\'c}}}]{pevsic2019}%
  \BibitemOpen
  \bibfield  {author} {\bibinfo {author} {\bibfnamefont {J.}~\bibnamefont
  {Pe{\v{s}}i{\'c}}}, \bibinfo {author} {\bibfnamefont {I.}~\bibnamefont
  {Popov}}, \bibinfo {author} {\bibfnamefont {A.}~\bibnamefont
  {{\v{S}}olaji{\'c}}}, \bibinfo {author} {\bibfnamefont {V.}~\bibnamefont
  {Damljanovi{\'c}}}, \bibinfo {author} {\bibfnamefont {K.}~\bibnamefont
  {Hingerl}}, \bibinfo {author} {\bibfnamefont {M.}~\bibnamefont {Beli{\'c}}},\
  and\ \bibinfo {author} {\bibfnamefont {R.}~\bibnamefont {Gaji{\'c}}},\
  }\bibfield  {title} {\bibinfo {title} {Ab initio study of the electronic,
  vibrational, and mechanical properties of the magnesium diboride monolayer},\
  }\href {https://doi.org/10.3390/condmat4020037} {\bibfield  {journal}
  {\bibinfo  {journal} {Condens. Matter}\ }\textbf {\bibinfo {volume} {4}},\
  \bibinfo {pages} {37} (\bibinfo {year} {2019})}\BibitemShut {NoStop}%
\bibitem [{\citenamefont {Novko}\ \emph {et~al.}(2020)\citenamefont {Novko},
  \citenamefont {Caruso}, \citenamefont {Draxl},\ and\ \citenamefont
  {Cappelluti}}]{novko2020}%
  \BibitemOpen
  \bibfield  {author} {\bibinfo {author} {\bibfnamefont {D.}~\bibnamefont
  {Novko}}, \bibinfo {author} {\bibfnamefont {F.}~\bibnamefont {Caruso}},
  \bibinfo {author} {\bibfnamefont {C.}~\bibnamefont {Draxl}},\ and\ \bibinfo
  {author} {\bibfnamefont {E.}~\bibnamefont {Cappelluti}},\ }\bibfield  {title}
  {\bibinfo {title} {Ultrafast hot phonon dynamics in ${\mathrm{mgb}}_{2}$
  driven by anisotropic electron-phonon coupling},\ }\href
  {https://doi.org/10.1103/PhysRevLett.124.077001} {\bibfield  {journal}
  {\bibinfo  {journal} {Phys. Rev. Lett.}\ }\textbf {\bibinfo {volume} {124}},\
  \bibinfo {pages} {077001} (\bibinfo {year} {2020})}\BibitemShut {NoStop}%
\bibitem [{\citenamefont {Ko\ifmmode~\mbox{\c{c}}\else \c{c}\fi{}er}\ \emph
  {et~al.}(2020)\citenamefont {Ko\ifmmode~\mbox{\c{c}}\else \c{c}\fi{}er},
  \citenamefont {Haule}, \citenamefont {Pascut},\ and\ \citenamefont
  {Monserrat}}]{kocer2020}%
  \BibitemOpen
  \bibfield  {author} {\bibinfo {author} {\bibfnamefont {C.~P.}\ \bibnamefont
  {Ko\ifmmode~\mbox{\c{c}}\else \c{c}\fi{}er}}, \bibinfo {author}
  {\bibfnamefont {K.}~\bibnamefont {Haule}}, \bibinfo {author} {\bibfnamefont
  {G.~L.}\ \bibnamefont {Pascut}},\ and\ \bibinfo {author} {\bibfnamefont
  {B.}~\bibnamefont {Monserrat}},\ }\bibfield  {title} {\bibinfo {title}
  {Efficient lattice dynamics calculations for correlated materials with
  $\mathrm{DFT}+\mathrm{DMFT}$},\ }\href
  {https://doi.org/10.1103/PhysRevB.102.245104} {\bibfield  {journal} {\bibinfo
   {journal} {Phys. Rev. B}\ }\textbf {\bibinfo {volume} {102}},\ \bibinfo
  {pages} {245104} (\bibinfo {year} {2020})}\BibitemShut {NoStop}%
\bibitem [{\citenamefont {Giustino}(2017)}]{giustino2017}%
  \BibitemOpen
  \bibfield  {author} {\bibinfo {author} {\bibfnamefont {F.}~\bibnamefont
  {Giustino}},\ }\bibfield  {title} {\bibinfo {title} {Electron-phonon
  interactions from first principles},\ }\href
  {https://doi.org/10.1103/RevModPhys.89.015003} {\bibfield  {journal}
  {\bibinfo  {journal} {Rev. Mod. Phys.}\ }\textbf {\bibinfo {volume} {89}},\
  \bibinfo {pages} {015003} (\bibinfo {year} {2017})}\BibitemShut {NoStop}%
\bibitem [{\citenamefont {Giustino}\ \emph {et~al.}(2010)\citenamefont
  {Giustino}, \citenamefont {Louie},\ and\ \citenamefont
  {Cohen}}]{giustino2010}%
  \BibitemOpen
  \bibfield  {author} {\bibinfo {author} {\bibfnamefont {F.}~\bibnamefont
  {Giustino}}, \bibinfo {author} {\bibfnamefont {S.~G.}\ \bibnamefont
  {Louie}},\ and\ \bibinfo {author} {\bibfnamefont {M.~L.}\ \bibnamefont
  {Cohen}},\ }\bibfield  {title} {\bibinfo {title} {Electron-phonon
  renormalization of the direct band gap of diamond},\ }\href
  {https://doi.org/10.1103/PhysRevLett.105.265501} {\bibfield  {journal}
  {\bibinfo  {journal} {Phys. Rev. Lett.}\ }\textbf {\bibinfo {volume} {105}},\
  \bibinfo {pages} {265501} (\bibinfo {year} {2010})}\BibitemShut {NoStop}%
\bibitem [{\citenamefont {Bhosale}\ \emph {et~al.}(2012)\citenamefont
  {Bhosale}, \citenamefont {Ramdas}, \citenamefont {Burger}, \citenamefont
  {Mu\~noz}, \citenamefont {Romero}, \citenamefont {Cardona}, \citenamefont
  {Lauck},\ and\ \citenamefont {Kremer}}]{bhosale2012}%
  \BibitemOpen
  \bibfield  {author} {\bibinfo {author} {\bibfnamefont {J.}~\bibnamefont
  {Bhosale}}, \bibinfo {author} {\bibfnamefont {A.~K.}\ \bibnamefont {Ramdas}},
  \bibinfo {author} {\bibfnamefont {A.}~\bibnamefont {Burger}}, \bibinfo
  {author} {\bibfnamefont {A.}~\bibnamefont {Mu\~noz}}, \bibinfo {author}
  {\bibfnamefont {A.~H.}\ \bibnamefont {Romero}}, \bibinfo {author}
  {\bibfnamefont {M.}~\bibnamefont {Cardona}}, \bibinfo {author} {\bibfnamefont
  {R.}~\bibnamefont {Lauck}},\ and\ \bibinfo {author} {\bibfnamefont {R.~K.}\
  \bibnamefont {Kremer}},\ }\bibfield  {title} {\bibinfo {title} {Temperature
  dependence of band gaps in semiconductors: Electron-phonon interaction},\
  }\href {https://doi.org/10.1103/PhysRevB.86.195208} {\bibfield  {journal}
  {\bibinfo  {journal} {Phys. Rev. B}\ }\textbf {\bibinfo {volume} {86}},\
  \bibinfo {pages} {195208} (\bibinfo {year} {2012})}\BibitemShut {NoStop}%
\bibitem [{\citenamefont {Monserrat}\ \emph {et~al.}(2013)\citenamefont
  {Monserrat}, \citenamefont {Drummond},\ and\ \citenamefont
  {Needs}}]{monserrat2013}%
  \BibitemOpen
  \bibfield  {author} {\bibinfo {author} {\bibfnamefont {B.}~\bibnamefont
  {Monserrat}}, \bibinfo {author} {\bibfnamefont {N.~D.}\ \bibnamefont
  {Drummond}},\ and\ \bibinfo {author} {\bibfnamefont {R.~J.}\ \bibnamefont
  {Needs}},\ }\bibfield  {title} {\bibinfo {title} {Anharmonic vibrational
  properties in periodic systems: energy, electron-phonon coupling, and
  stress},\ }\href {https://doi.org/10.1103/PhysRevB.87.144302} {\bibfield
  {journal} {\bibinfo  {journal} {Phys. Rev. B}\ }\textbf {\bibinfo {volume}
  {87}},\ \bibinfo {pages} {144302} (\bibinfo {year} {2013})}\BibitemShut
  {NoStop}%
\bibitem [{\citenamefont {Poncé}\ \emph {et~al.}(2015)\citenamefont {Poncé},
  \citenamefont {Gillet}, \citenamefont {Laflamme~Janssen}, \citenamefont
  {Marini}, \citenamefont {Verstraete},\ and\ \citenamefont
  {Gonze}}]{Ponce2015}%
  \BibitemOpen
  \bibfield  {author} {\bibinfo {author} {\bibfnamefont {S.}~\bibnamefont
  {Poncé}}, \bibinfo {author} {\bibfnamefont {Y.}~\bibnamefont {Gillet}},
  \bibinfo {author} {\bibfnamefont {J.}~\bibnamefont {Laflamme~Janssen}},
  \bibinfo {author} {\bibfnamefont {A.}~\bibnamefont {Marini}}, \bibinfo
  {author} {\bibfnamefont {M.}~\bibnamefont {Verstraete}},\ and\ \bibinfo
  {author} {\bibfnamefont {X.}~\bibnamefont {Gonze}},\ }\bibfield  {title}
  {\bibinfo {title} {Temperature dependence of the electronic structure of
  semiconductors and insulators},\ }\href {https://doi.org/10.1063/1.4927081}
  {\bibfield  {journal} {\bibinfo  {journal} {J. Chem. Phys.}\ }\textbf
  {\bibinfo {volume} {143}},\ \bibinfo {pages} {102813} (\bibinfo {year}
  {2015})}\BibitemShut {NoStop}%
\bibitem [{\citenamefont {Karsai}\ \emph {et~al.}(2018)\citenamefont {Karsai},
  \citenamefont {Engel}, \citenamefont {Flage-Larsen},\ and\ \citenamefont
  {Kresse}}]{Karsai2018}%
  \BibitemOpen
  \bibfield  {author} {\bibinfo {author} {\bibfnamefont {F.}~\bibnamefont
  {Karsai}}, \bibinfo {author} {\bibfnamefont {M.}~\bibnamefont {Engel}},
  \bibinfo {author} {\bibfnamefont {E.}~\bibnamefont {Flage-Larsen}},\ and\
  \bibinfo {author} {\bibfnamefont {G.}~\bibnamefont {Kresse}},\ }\bibfield
  {title} {\bibinfo {title} {Electron{\textendash}phonon coupling in
  semiconductors within the {GW} approximation},\ }\href
  {https://doi.org/10.1088/1367-2630/aaf53f} {\bibfield  {journal} {\bibinfo
  {journal} {New J. Phys.}\ }\textbf {\bibinfo {volume} {20}},\ \bibinfo
  {pages} {123008} (\bibinfo {year} {2018})}\BibitemShut {NoStop}%
\bibitem [{\citenamefont {Zhang}\ \emph {et~al.}(2020)\citenamefont {Zhang},
  \citenamefont {Wang}, \citenamefont {Xi},\ and\ \citenamefont
  {Yang}}]{zhang2020}%
  \BibitemOpen
  \bibfield  {author} {\bibinfo {author} {\bibfnamefont {Y.}~\bibnamefont
  {Zhang}}, \bibinfo {author} {\bibfnamefont {Z.}~\bibnamefont {Wang}},
  \bibinfo {author} {\bibfnamefont {J.}~\bibnamefont {Xi}},\ and\ \bibinfo
  {author} {\bibfnamefont {J.}~\bibnamefont {Yang}},\ }\bibfield  {title}
  {\bibinfo {title} {Temperature-dependent band gaps in several semiconductors:
  from the role of electron{\textendash}phonon renormalization},\ }\href
  {https://doi.org/10.1088/1361-648x/aba45d} {\bibfield  {journal} {\bibinfo
  {journal} {J. Condens. Matter Phys.}\ }\textbf {\bibinfo {volume} {32}},\
  \bibinfo {pages} {475503} (\bibinfo {year} {2020})}\BibitemShut {NoStop}%
\bibitem [{\citenamefont {Allen}\ and\ \citenamefont
  {Heine}(1976)}]{allen1976}%
  \BibitemOpen
  \bibfield  {author} {\bibinfo {author} {\bibfnamefont {P.}~\bibnamefont
  {Allen}}\ and\ \bibinfo {author} {\bibfnamefont {V.}~\bibnamefont {Heine}},\
  }\bibfield  {title} {\bibinfo {title} {Theory of the temperature dependence
  of electronic band structures},\ }\href
  {https://doi.org/10.1088/0022-3719/9/12/013} {\bibfield  {journal} {\bibinfo
  {journal} {J. Phys. C: Solid State Phys.}\ }\textbf {\bibinfo {volume} {9}},\
  \bibinfo {pages} {2305} (\bibinfo {year} {1976})}\BibitemShut {NoStop}%
\bibitem [{\citenamefont {Allen}\ and\ \citenamefont
  {Cardona}(1981)}]{allen1981}%
  \BibitemOpen
  \bibfield  {author} {\bibinfo {author} {\bibfnamefont {P.~B.}\ \bibnamefont
  {Allen}}\ and\ \bibinfo {author} {\bibfnamefont {M.}~\bibnamefont
  {Cardona}},\ }\bibfield  {title} {\bibinfo {title} {Theory of the temperature
  dependence of the direct gap of germanium},\ }\href
  {https://doi.org/10.1103/PhysRevB.23.1495} {\bibfield  {journal} {\bibinfo
  {journal} {Phys. Rev. B}\ }\textbf {\bibinfo {volume} {23}},\ \bibinfo
  {pages} {1495} (\bibinfo {year} {1981})}\BibitemShut {NoStop}%
\bibitem [{\citenamefont {Allen}\ and\ \citenamefont
  {Cardona}(1983)}]{allen1983}%
  \BibitemOpen
  \bibfield  {author} {\bibinfo {author} {\bibfnamefont {P.~B.}\ \bibnamefont
  {Allen}}\ and\ \bibinfo {author} {\bibfnamefont {M.}~\bibnamefont
  {Cardona}},\ }\bibfield  {title} {\bibinfo {title} {Temperature dependence of
  the direct gap of {Si} and {Ge}},\ }\href
  {https://doi.org/10.1103/PhysRevB.27.4760} {\bibfield  {journal} {\bibinfo
  {journal} {Phys. Rev. B}\ }\textbf {\bibinfo {volume} {27}},\ \bibinfo
  {pages} {4760} (\bibinfo {year} {1983})}\BibitemShut {NoStop}%
\bibitem [{\citenamefont {Monserrat}(2018)}]{monserrat2018}%
  \BibitemOpen
  \bibfield  {author} {\bibinfo {author} {\bibfnamefont {B.}~\bibnamefont
  {Monserrat}},\ }\bibfield  {title} {\bibinfo {title}
  {Electron{\textendash}phonon coupling from finite differences},\ }\href
  {https://doi.org/10.1088/1361-648x/aaa737} {\bibfield  {journal} {\bibinfo
  {journal} {J. Condens. Matter Phys.}\ }\textbf {\bibinfo {volume} {30}},\
  \bibinfo {pages} {083001} (\bibinfo {year} {2018})}\BibitemShut {NoStop}%
\bibitem [{\citenamefont {Koelling}\ and\ \citenamefont
  {Harmon}(1977)}]{Koelling1977}%
  \BibitemOpen
  \bibfield  {author} {\bibinfo {author} {\bibfnamefont {D.~D.}\ \bibnamefont
  {Koelling}}\ and\ \bibinfo {author} {\bibfnamefont {B.~N.}\ \bibnamefont
  {Harmon}},\ }\bibfield  {title} {\bibinfo {title} {A technique for
  relativistic spin-polarised calculations},\ }\href
  {https://doi.org/10.1088/0022-3719/10/16/019} {\bibfield  {journal} {\bibinfo
   {journal} {J. Phys. C: Solid State Phys.}\ }\textbf {\bibinfo {volume}
  {10}},\ \bibinfo {pages} {3107} (\bibinfo {year} {1977})}\BibitemShut
  {NoStop}%
\bibitem [{\citenamefont {Zacharias}\ and\ \citenamefont
  {Giustino}(2016)}]{zacharias2016}%
  \BibitemOpen
  \bibfield  {author} {\bibinfo {author} {\bibfnamefont {M.}~\bibnamefont
  {Zacharias}}\ and\ \bibinfo {author} {\bibfnamefont {F.}~\bibnamefont
  {Giustino}},\ }\bibfield  {title} {\bibinfo {title} {One-shot calculation of
  temperature-dependent optical spectra and phonon-induced band-gap
  renormalization},\ }\href {https://doi.org/10.1103/PhysRevB.94.075125}
  {\bibfield  {journal} {\bibinfo  {journal} {Phys. Rev. B}\ }\textbf {\bibinfo
  {volume} {94}},\ \bibinfo {pages} {075125} (\bibinfo {year}
  {2016})}\BibitemShut {NoStop}%
\bibitem [{\citenamefont {Ponc\'e}\ \emph {et~al.}(2014)\citenamefont
  {Ponc\'e}, \citenamefont {Antonius}, \citenamefont {Gillet}, \citenamefont
  {Boulanger}, \citenamefont {Laflamme~Janssen}, \citenamefont {Marini},
  \citenamefont {C\^ot\'e},\ and\ \citenamefont
  {Gonze}}]{Ponce2014Temperature}%
  \BibitemOpen
  \bibfield  {author} {\bibinfo {author} {\bibfnamefont {S.}~\bibnamefont
  {Ponc\'e}}, \bibinfo {author} {\bibfnamefont {G.}~\bibnamefont {Antonius}},
  \bibinfo {author} {\bibfnamefont {Y.}~\bibnamefont {Gillet}}, \bibinfo
  {author} {\bibfnamefont {P.}~\bibnamefont {Boulanger}}, \bibinfo {author}
  {\bibfnamefont {J.}~\bibnamefont {Laflamme~Janssen}}, \bibinfo {author}
  {\bibfnamefont {A.}~\bibnamefont {Marini}}, \bibinfo {author} {\bibfnamefont
  {M.}~\bibnamefont {C\^ot\'e}},\ and\ \bibinfo {author} {\bibfnamefont
  {X.}~\bibnamefont {Gonze}},\ }\bibfield  {title} {\bibinfo {title}
  {Temperature dependence of electronic eigenenergies in the adiabatic harmonic
  approximation},\ }\href {https://doi.org/10.1103/PhysRevB.90.214304}
  {\bibfield  {journal} {\bibinfo  {journal} {Phys. Rev. B}\ }\textbf {\bibinfo
  {volume} {90}},\ \bibinfo {pages} {214304} (\bibinfo {year}
  {2014})}\BibitemShut {NoStop}%
\bibitem [{\citenamefont {Monserrat}\ and\ \citenamefont
  {Needs}(2014)}]{Monserrat2014}%
  \BibitemOpen
  \bibfield  {author} {\bibinfo {author} {\bibfnamefont {B.}~\bibnamefont
  {Monserrat}}\ and\ \bibinfo {author} {\bibfnamefont {R.~J.}\ \bibnamefont
  {Needs}},\ }\bibfield  {title} {\bibinfo {title} {Comparing electron-phonon
  coupling strength in diamond, silicon, and silicon carbide: First-principles
  study},\ }\href {https://doi.org/10.1103/PhysRevB.89.214304} {\bibfield
  {journal} {\bibinfo  {journal} {Phys. Rev. B}\ }\textbf {\bibinfo {volume}
  {89}},\ \bibinfo {pages} {214304} (\bibinfo {year} {2014})}\BibitemShut
  {NoStop}%
\bibitem [{\citenamefont {O’Donnell}\ and\ \citenamefont
  {Chen}(1991)}]{Donnell1991}%
  \BibitemOpen
  \bibfield  {author} {\bibinfo {author} {\bibfnamefont {K.~P.}\ \bibnamefont
  {O’Donnell}}\ and\ \bibinfo {author} {\bibfnamefont {X.}~\bibnamefont
  {Chen}},\ }\bibfield  {title} {\bibinfo {title} {Temperature dependence of
  semiconductor band gaps},\ }\href {https://doi.org/10.1063/1.104723}
  {\bibfield  {journal} {\bibinfo  {journal} {Appl. Phys. Lett.}\ }\textbf
  {\bibinfo {volume} {58}},\ \bibinfo {pages} {2924} (\bibinfo {year}
  {1991})}\BibitemShut {NoStop}%
\bibitem [{\citenamefont {Poncé}\ \emph {et~al.}(2014)\citenamefont {Poncé},
  \citenamefont {Antonius}, \citenamefont {Boulanger}, \citenamefont
  {Cannuccia}, \citenamefont {Marini}, \citenamefont {Côté},\ and\
  \citenamefont {Gonze}}]{Ponce2014}%
  \BibitemOpen
  \bibfield  {author} {\bibinfo {author} {\bibfnamefont {S.}~\bibnamefont
  {Poncé}}, \bibinfo {author} {\bibfnamefont {G.}~\bibnamefont {Antonius}},
  \bibinfo {author} {\bibfnamefont {P.}~\bibnamefont {Boulanger}}, \bibinfo
  {author} {\bibfnamefont {E.}~\bibnamefont {Cannuccia}}, \bibinfo {author}
  {\bibfnamefont {A.}~\bibnamefont {Marini}}, \bibinfo {author} {\bibfnamefont
  {M.}~\bibnamefont {Côté}},\ and\ \bibinfo {author} {\bibfnamefont
  {X.}~\bibnamefont {Gonze}},\ }\bibfield  {title} {\bibinfo {title}
  {Verification of first-principles codes: Comparison of total energies, phonon
  frequencies, electron–phonon coupling and zero-point motion correction to
  the gap between {ABINIT} and {QE/Yambo}},\ }\href
  {https://doi.org/https://doi.org/10.1016/j.commatsci.2013.11.031} {\bibfield
  {journal} {\bibinfo  {journal} {Comput. Mater. Sci.}\ }\textbf {\bibinfo
  {volume} {83}},\ \bibinfo {pages} {341} (\bibinfo {year} {2014})}\BibitemShut
  {NoStop}%
\bibitem [{\citenamefont {Cardona}(2001)}]{Cardona2001}%
  \BibitemOpen
  \bibfield  {author} {\bibinfo {author} {\bibfnamefont {M.}~\bibnamefont
  {Cardona}},\ }\bibfield  {title} {\bibinfo {title} {Renormalization of the
  optical response of semiconductors by electron–phonon interaction},\ }\href
  {https://doi.org/https://doi.org/10.1002/1521-396X(200112)188:4<1209::AID-PSSA1209>3.0.CO;2-2}
  {\bibfield  {journal} {\bibinfo  {journal} {Phys. Status Solidi A}\ }\textbf
  {\bibinfo {volume} {188}},\ \bibinfo {pages} {1209} (\bibinfo {year}
  {2001})}\BibitemShut {NoStop}%
\bibitem [{\citenamefont {Murakami}(2006)}]{Murakami2006}%
  \BibitemOpen
  \bibfield  {author} {\bibinfo {author} {\bibfnamefont {S.}~\bibnamefont
  {Murakami}},\ }\bibfield  {title} {\bibinfo {title} {Quantum spin hall effect
  and enhanced magnetic response by spin-orbit coupling},\ }\href
  {https://doi.org/10.1103/PhysRevLett.97.236805} {\bibfield  {journal}
  {\bibinfo  {journal} {Phys. Rev. Lett.}\ }\textbf {\bibinfo {volume} {97}},\
  \bibinfo {pages} {236805} (\bibinfo {year} {2006})}\BibitemShut {NoStop}%
\bibitem [{\citenamefont {Koroteev}\ \emph {et~al.}(2008)\citenamefont
  {Koroteev}, \citenamefont {Bihlmayer}, \citenamefont {Chulkov},\ and\
  \citenamefont {Bl\"ugel}}]{Koroteev2008}%
  \BibitemOpen
  \bibfield  {author} {\bibinfo {author} {\bibfnamefont {Y.~M.}\ \bibnamefont
  {Koroteev}}, \bibinfo {author} {\bibfnamefont {G.}~\bibnamefont {Bihlmayer}},
  \bibinfo {author} {\bibfnamefont {E.~V.}\ \bibnamefont {Chulkov}},\ and\
  \bibinfo {author} {\bibfnamefont {S.}~\bibnamefont {Bl\"ugel}},\ }\bibfield
  {title} {\bibinfo {title} {First-principles investigation of structural and
  electronic properties of ultrathin {Bi} films},\ }\href
  {https://doi.org/10.1103/PhysRevB.77.045428} {\bibfield  {journal} {\bibinfo
  {journal} {Phys. Rev. B}\ }\textbf {\bibinfo {volume} {77}},\ \bibinfo
  {pages} {045428} (\bibinfo {year} {2008})}\BibitemShut {NoStop}%
\bibitem [{\citenamefont {Wada}\ \emph {et~al.}(2011)\citenamefont {Wada},
  \citenamefont {Murakami}, \citenamefont {Freimuth},\ and\ \citenamefont
  {Bihlmayer}}]{Wada2011}%
  \BibitemOpen
  \bibfield  {author} {\bibinfo {author} {\bibfnamefont {M.}~\bibnamefont
  {Wada}}, \bibinfo {author} {\bibfnamefont {S.}~\bibnamefont {Murakami}},
  \bibinfo {author} {\bibfnamefont {F.}~\bibnamefont {Freimuth}},\ and\
  \bibinfo {author} {\bibfnamefont {G.}~\bibnamefont {Bihlmayer}},\ }\bibfield
  {title} {\bibinfo {title} {Localized edge states in two-dimensional
  topological insulators: Ultrathin {Bi} films},\ }\href
  {https://doi.org/10.1103/PhysRevB.83.121310} {\bibfield  {journal} {\bibinfo
  {journal} {Phys. Rev. B}\ }\textbf {\bibinfo {volume} {83}},\ \bibinfo
  {pages} {121310(R)} (\bibinfo {year} {2011})}\BibitemShut {NoStop}%
\bibitem [{\citenamefont {Liu}\ \emph {et~al.}(2011)\citenamefont {Liu},
  \citenamefont {Liu}, \citenamefont {Wu}, \citenamefont {Duan}, \citenamefont
  {Liu},\ and\ \citenamefont {Wu}}]{Liu2011}%
  \BibitemOpen
  \bibfield  {author} {\bibinfo {author} {\bibfnamefont {Z.}~\bibnamefont
  {Liu}}, \bibinfo {author} {\bibfnamefont {C.-X.}\ \bibnamefont {Liu}},
  \bibinfo {author} {\bibfnamefont {Y.-S.}\ \bibnamefont {Wu}}, \bibinfo
  {author} {\bibfnamefont {W.-H.}\ \bibnamefont {Duan}}, \bibinfo {author}
  {\bibfnamefont {F.}~\bibnamefont {Liu}},\ and\ \bibinfo {author}
  {\bibfnamefont {J.}~\bibnamefont {Wu}},\ }\bibfield  {title} {\bibinfo
  {title} {Stable nontrivial ${Z}_{2}$ topology in ultrathin {Bi} (111) films:
  A first-principles study},\ }\href
  {https://doi.org/10.1103/PhysRevLett.107.136805} {\bibfield  {journal}
  {\bibinfo  {journal} {Phys. Rev. Lett.}\ }\textbf {\bibinfo {volume} {107}},\
  \bibinfo {pages} {136805} (\bibinfo {year} {2011})}\BibitemShut {NoStop}%
\bibitem [{\citenamefont {Reis}\ \emph {et~al.}(2017)\citenamefont {Reis},
  \citenamefont {Li}, \citenamefont {Dudy}, \citenamefont {Bauernfeind},
  \citenamefont {Glass}, \citenamefont {Hanke}, \citenamefont {Thomale},
  \citenamefont {Schäfer},\ and\ \citenamefont {Claessen}}]{Reis2017}%
  \BibitemOpen
  \bibfield  {author} {\bibinfo {author} {\bibfnamefont {F.}~\bibnamefont
  {Reis}}, \bibinfo {author} {\bibfnamefont {G.}~\bibnamefont {Li}}, \bibinfo
  {author} {\bibfnamefont {L.}~\bibnamefont {Dudy}}, \bibinfo {author}
  {\bibfnamefont {M.}~\bibnamefont {Bauernfeind}}, \bibinfo {author}
  {\bibfnamefont {S.}~\bibnamefont {Glass}}, \bibinfo {author} {\bibfnamefont
  {W.}~\bibnamefont {Hanke}}, \bibinfo {author} {\bibfnamefont
  {R.}~\bibnamefont {Thomale}}, \bibinfo {author} {\bibfnamefont
  {J.}~\bibnamefont {Schäfer}},\ and\ \bibinfo {author} {\bibfnamefont
  {R.}~\bibnamefont {Claessen}},\ }\bibfield  {title} {\bibinfo {title}
  {Bismuthene on a {SiC} substrate: A candidate for a high-temperature quantum
  spin hall material},\ }\href {https://doi.org/10.1126/science.aai8142}
  {\bibfield  {journal} {\bibinfo  {journal} {Science}\ }\textbf {\bibinfo
  {volume} {357}},\ \bibinfo {pages} {287} (\bibinfo {year}
  {2017})}\BibitemShut {NoStop}%
\bibitem [{\citenamefont {Chen}\ \emph {et~al.}(2013)\citenamefont {Chen},
  \citenamefont {Wang},\ and\ \citenamefont {Liu}}]{chen2013}%
  \BibitemOpen
  \bibfield  {author} {\bibinfo {author} {\bibfnamefont {L.}~\bibnamefont
  {Chen}}, \bibinfo {author} {\bibfnamefont {Z.~F.}\ \bibnamefont {Wang}},\
  and\ \bibinfo {author} {\bibfnamefont {F.}~\bibnamefont {Liu}},\ }\bibfield
  {title} {\bibinfo {title} {Robustness of two-dimensional topological
  insulator states in bilayer bismuth against strain and electrical field},\
  }\href {https://doi.org/10.1103/PhysRevB.87.235420} {\bibfield  {journal}
  {\bibinfo  {journal} {Phys. Rev. B}\ }\textbf {\bibinfo {volume} {87}},\
  \bibinfo {pages} {235420} (\bibinfo {year} {2013})}\BibitemShut {NoStop}%
\bibitem [{\citenamefont {Liu}\ \emph {et~al.}(2017)\citenamefont {Liu},
  \citenamefont {Huang}, \citenamefont {Chen}, \citenamefont {Li},
  \citenamefont {Cao},\ and\ \citenamefont {He}}]{liu2017}%
  \BibitemOpen
  \bibfield  {author} {\bibinfo {author} {\bibfnamefont {M.-Y.}\ \bibnamefont
  {Liu}}, \bibinfo {author} {\bibfnamefont {Y.}~\bibnamefont {Huang}}, \bibinfo
  {author} {\bibfnamefont {Q.-Y.}\ \bibnamefont {Chen}}, \bibinfo {author}
  {\bibfnamefont {Z.-Y.}\ \bibnamefont {Li}}, \bibinfo {author} {\bibfnamefont
  {C.}~\bibnamefont {Cao}},\ and\ \bibinfo {author} {\bibfnamefont
  {Y.}~\bibnamefont {He}},\ }\bibfield  {title} {\bibinfo {title} {Strain and
  electric field tunable electronic structure of buckled bismuthene},\ }\href
  {https://doi.org/10.1039/C7RA05787C} {\bibfield  {journal} {\bibinfo
  {journal} {RSC Adv.}\ }\textbf {\bibinfo {volume} {7}},\ \bibinfo {pages}
  {39546} (\bibinfo {year} {2017})}\BibitemShut {NoStop}%
\bibitem [{\citenamefont {Heyd}\ \emph {et~al.}(2003)\citenamefont {Heyd},
  \citenamefont {Scuseria},\ and\ \citenamefont {Ernzerhof}}]{Heyd2003}%
  \BibitemOpen
  \bibfield  {author} {\bibinfo {author} {\bibfnamefont {J.}~\bibnamefont
  {Heyd}}, \bibinfo {author} {\bibfnamefont {G.~E.}\ \bibnamefont {Scuseria}},\
  and\ \bibinfo {author} {\bibfnamefont {M.}~\bibnamefont {Ernzerhof}},\
  }\bibfield  {title} {\bibinfo {title} {Hybrid functionals based on a screened
  coulomb potential},\ }\href {https://doi.org/10.1063/1.1564060} {\bibfield
  {journal} {\bibinfo  {journal} {J. Chem. Phys.}\ }\textbf {\bibinfo {volume}
  {118}},\ \bibinfo {pages} {8207} (\bibinfo {year} {2003})}\BibitemShut
  {NoStop}%
\bibitem [{\citenamefont {Krukau}\ \emph {et~al.}(2006)\citenamefont {Krukau},
  \citenamefont {Vydrov}, \citenamefont {Izmaylov},\ and\ \citenamefont
  {Scuseria}}]{Krukau2006}%
  \BibitemOpen
  \bibfield  {author} {\bibinfo {author} {\bibfnamefont {A.~V.}\ \bibnamefont
  {Krukau}}, \bibinfo {author} {\bibfnamefont {O.~A.}\ \bibnamefont {Vydrov}},
  \bibinfo {author} {\bibfnamefont {A.~F.}\ \bibnamefont {Izmaylov}},\ and\
  \bibinfo {author} {\bibfnamefont {G.~E.}\ \bibnamefont {Scuseria}},\
  }\bibfield  {title} {\bibinfo {title} {Influence of the exchange screening
  parameter on the performance of screened hybrid functionals},\ }\href
  {https://doi.org/10.1063/1.2404663} {\bibfield  {journal} {\bibinfo
  {journal} {J. Chem. Phys.}\ }\textbf {\bibinfo {volume} {125}},\ \bibinfo
  {pages} {224106} (\bibinfo {year} {2006})}\BibitemShut {NoStop}%
\bibitem [{\citenamefont {Hybertsen}\ and\ \citenamefont
  {Louie}(1986)}]{Hybertsen1986}%
  \BibitemOpen
  \bibfield  {author} {\bibinfo {author} {\bibfnamefont {M.~S.}\ \bibnamefont
  {Hybertsen}}\ and\ \bibinfo {author} {\bibfnamefont {S.~G.}\ \bibnamefont
  {Louie}},\ }\bibfield  {title} {\bibinfo {title} {Electron correlation in
  semiconductors and insulators: Band gaps and quasiparticle energies},\ }\href
  {https://doi.org/10.1103/PhysRevB.34.5390} {\bibfield  {journal} {\bibinfo
  {journal} {Phys. Rev. B}\ }\textbf {\bibinfo {volume} {34}},\ \bibinfo
  {pages} {5390} (\bibinfo {year} {1986})}\BibitemShut {NoStop}%
\bibitem [{\citenamefont {Aryasetiawan}\ and\ \citenamefont
  {Gunnarsson}(1998)}]{Aryasetiawan1998}%
  \BibitemOpen
  \bibfield  {author} {\bibinfo {author} {\bibfnamefont {F.}~\bibnamefont
  {Aryasetiawan}}\ and\ \bibinfo {author} {\bibfnamefont {O.}~\bibnamefont
  {Gunnarsson}},\ }\bibfield  {title} {\bibinfo {title} {The gw method},\
  }\href {https://doi.org/10.1088/0034-4885/61/3/002} {\bibfield  {journal}
  {\bibinfo  {journal} {Rep. Prog. Phys.}\ }\textbf {\bibinfo {volume} {61}},\
  \bibinfo {pages} {237} (\bibinfo {year} {1998})}\BibitemShut {NoStop}%
\bibitem [{\citenamefont {Anisimov}\ and\ \citenamefont
  {Gunnarsson}(1991)}]{Anisimov1991}%
  \BibitemOpen
  \bibfield  {author} {\bibinfo {author} {\bibfnamefont {V.~I.}\ \bibnamefont
  {Anisimov}}\ and\ \bibinfo {author} {\bibfnamefont {O.}~\bibnamefont
  {Gunnarsson}},\ }\bibfield  {title} {\bibinfo {title} {Density-functional
  calculation of effective coulomb interactions in metals},\ }\href
  {https://doi.org/10.1103/physrevb.43.7570} {\bibfield  {journal} {\bibinfo
  {journal} {Phys. Rev. B}\ }\textbf {\bibinfo {volume} {43}},\ \bibinfo
  {pages} {7570} (\bibinfo {year} {1991})}\BibitemShut {NoStop}%
\bibitem [{\citenamefont {Liechtenstein}\ \emph {et~al.}(1995)\citenamefont
  {Liechtenstein}, \citenamefont {Anisimov},\ and\ \citenamefont
  {Zaanen}}]{Liechtenstein1995}%
  \BibitemOpen
  \bibfield  {author} {\bibinfo {author} {\bibfnamefont {A.~I.}\ \bibnamefont
  {Liechtenstein}}, \bibinfo {author} {\bibfnamefont {V.~I.}\ \bibnamefont
  {Anisimov}},\ and\ \bibinfo {author} {\bibfnamefont {J.}~\bibnamefont
  {Zaanen}},\ }\bibfield  {title} {\bibinfo {title} {Density-functional theory
  and strong interactions: {Orbital} ordering in mott-hubbard insulators},\
  }\href {https://doi.org/10.1103/physrevb.52.r5467} {\bibfield  {journal}
  {\bibinfo  {journal} {Phys. Rev. B}\ }\textbf {\bibinfo {volume} {52}},\
  \bibinfo {pages} {R5467} (\bibinfo {year} {1995})}\BibitemShut {NoStop}%
\bibitem [{\citenamefont {Georges}\ \emph {et~al.}(1996)\citenamefont
  {Georges}, \citenamefont {Kotliar}, \citenamefont {Krauth},\ and\
  \citenamefont {Rozenberg}}]{Georges1996}%
  \BibitemOpen
  \bibfield  {author} {\bibinfo {author} {\bibfnamefont {A.}~\bibnamefont
  {Georges}}, \bibinfo {author} {\bibfnamefont {G.}~\bibnamefont {Kotliar}},
  \bibinfo {author} {\bibfnamefont {W.}~\bibnamefont {Krauth}},\ and\ \bibinfo
  {author} {\bibfnamefont {M.~J.}\ \bibnamefont {Rozenberg}},\ }\bibfield
  {title} {\bibinfo {title} {Dynamical mean-field theory of strongly correlated
  fermion systems and the limit of infinite dimensions},\ }\href
  {https://doi.org/10.1103/revmodphys.68.13} {\bibfield  {journal} {\bibinfo
  {journal} {Rev. Mod. Phys.}\ }\textbf {\bibinfo {volume} {68}},\ \bibinfo
  {pages} {13} (\bibinfo {year} {1996})}\BibitemShut {NoStop}%
\bibitem [{\citenamefont {Kotliar}\ \emph {et~al.}(2006)\citenamefont
  {Kotliar}, \citenamefont {Savrasov}, \citenamefont {Haule}, \citenamefont
  {Oudovenko}, \citenamefont {Parcollet},\ and\ \citenamefont
  {Marianetti}}]{Kotliar2006}%
  \BibitemOpen
  \bibfield  {author} {\bibinfo {author} {\bibfnamefont {G.}~\bibnamefont
  {Kotliar}}, \bibinfo {author} {\bibfnamefont {S.~Y.}\ \bibnamefont
  {Savrasov}}, \bibinfo {author} {\bibfnamefont {K.}~\bibnamefont {Haule}},
  \bibinfo {author} {\bibfnamefont {V.~S.}\ \bibnamefont {Oudovenko}}, \bibinfo
  {author} {\bibfnamefont {O.}~\bibnamefont {Parcollet}},\ and\ \bibinfo
  {author} {\bibfnamefont {C.~A.}\ \bibnamefont {Marianetti}},\ }\bibfield
  {title} {\bibinfo {title} {Electronic structure calculations with dynamical
  mean-field theory},\ }\href {https://doi.org/10.1103/revmodphys.78.865}
  {\bibfield  {journal} {\bibinfo  {journal} {Rev. Mod. Phys.}\ }\textbf
  {\bibinfo {volume} {78}},\ \bibinfo {pages} {865} (\bibinfo {year}
  {2006})}\BibitemShut {NoStop}%
\bibitem [{\citenamefont {Gebauer}\ and\ \citenamefont
  {Baroni}(2000)}]{Gebauer2000}%
  \BibitemOpen
  \bibfield  {author} {\bibinfo {author} {\bibfnamefont {R.}~\bibnamefont
  {Gebauer}}\ and\ \bibinfo {author} {\bibfnamefont {S.}~\bibnamefont
  {Baroni}},\ }\bibfield  {title} {\bibinfo {title} {Magnons in real materials
  from density-functional theory},\ }\href
  {https://doi.org/10.1103/PhysRevB.61.R6459} {\bibfield  {journal} {\bibinfo
  {journal} {Phys. Rev. B}\ }\textbf {\bibinfo {volume} {61}},\ \bibinfo
  {pages} {R6459} (\bibinfo {year} {2000})}\BibitemShut {NoStop}%
\bibitem [{\citenamefont {He}\ \emph {et~al.}(2021)\citenamefont {He},
  \citenamefont {Helbig}, \citenamefont {Verstraete},\ and\ \citenamefont
  {Bousquet}}]{Xu2021}%
  \BibitemOpen
  \bibfield  {author} {\bibinfo {author} {\bibfnamefont {X.}~\bibnamefont
  {He}}, \bibinfo {author} {\bibfnamefont {N.}~\bibnamefont {Helbig}}, \bibinfo
  {author} {\bibfnamefont {M.~J.}\ \bibnamefont {Verstraete}},\ and\ \bibinfo
  {author} {\bibfnamefont {E.}~\bibnamefont {Bousquet}},\ }\bibfield  {title}
  {\bibinfo {title} {Tb2{J}: A python package for computing magnetic
  interaction parameters},\ }\href
  {https://doi.org/https://doi.org/10.1016/j.cpc.2021.107938} {\bibfield
  {journal} {\bibinfo  {journal} {Comput. Phys. Commun.}\ }\textbf {\bibinfo
  {volume} {264}},\ \bibinfo {pages} {107938} (\bibinfo {year}
  {2021})}\BibitemShut {NoStop}%
\bibitem [{\citenamefont {Salpeter}\ and\ \citenamefont
  {Bethe}(1951)}]{Salpeter1951}%
  \BibitemOpen
  \bibfield  {author} {\bibinfo {author} {\bibfnamefont {E.~E.}\ \bibnamefont
  {Salpeter}}\ and\ \bibinfo {author} {\bibfnamefont {H.~A.}\ \bibnamefont
  {Bethe}},\ }\bibfield  {title} {\bibinfo {title} {A relativistic equation for
  bound-state problems},\ }\href {https://doi.org/10.1103/PhysRev.84.1232}
  {\bibfield  {journal} {\bibinfo  {journal} {Phys. Rev.}\ }\textbf {\bibinfo
  {volume} {84}},\ \bibinfo {pages} {1232} (\bibinfo {year}
  {1951})}\BibitemShut {NoStop}%
\bibitem [{\citenamefont {Rohlfing}\ and\ \citenamefont
  {Louie}(1998)}]{Rohlfing1998}%
  \BibitemOpen
  \bibfield  {author} {\bibinfo {author} {\bibfnamefont {M.}~\bibnamefont
  {Rohlfing}}\ and\ \bibinfo {author} {\bibfnamefont {S.~G.}\ \bibnamefont
  {Louie}},\ }\bibfield  {title} {\bibinfo {title} {Electron-hole excitations
  in semiconductors and insulators},\ }\href
  {https://doi.org/10.1103/PhysRevLett.81.2312} {\bibfield  {journal} {\bibinfo
   {journal} {Phys. Rev. Lett.}\ }\textbf {\bibinfo {volume} {81}},\ \bibinfo
  {pages} {2312} (\bibinfo {year} {1998})}\BibitemShut {NoStop}%
\end{thebibliography}%

\end{document}